\documentclass[12pt]{article}
\pdfoutput=1
\usepackage{pstricks}
\usepackage{color}
\usepackage{amssymb,amsmath,bm,bbm}
\usepackage{epsf}
\usepackage{epsfig}
\usepackage{afterpage}
\usepackage{longtable}
\usepackage{appendix}
\usepackage{cite}
\usepackage{latexsym,mathrsfs,dsfont}
\usepackage{graphics}
\usepackage{url}
\usepackage{paralist}
\usepackage{bbold}
\usepackage[colorlinks=true, urlcolor=blue]{hyperref}
\newcommand{\nn}{\nonumber}

\newcommand{\be}{\begin{equation}}
\newcommand{\ee}{\end{equation}}
\newcommand{\bea}{\begin{eqnarray}}
\newcommand{\eea}{\end{eqnarray}}
\newcommand{\dd}{\displaystyle}
\begin{document}

\begin{flushright}
    {BARI-TH/25-772}
\end{flushright}

\medskip

\begin{center}
{\Large\bf
  \boldmath{Correlating lepton flavour violating \\ \vskip 0.2cm $b \to s$  and leptonic decay modes \\ \vskip 0.3cm in a minimal abelian extension of the Standard Model }}
\\[0.8 cm]
{\large P.~Colangelo$^{a}$, F.~De~Fazio$^{a}$ and D.~Milillo$^{a,b}$
 \\[0.5 cm]}
{\small
$^a$
Istituto Nazionale di Fisica Nucleare, Sezione di Bari, Via Orabona 4,
70126 Bari, Italy\\
$^b$ Dipartimento Interateneo di Fisica "Michelangelo Merlin", Universit\`a degli Studi di Bari, via Orabona 4, 70126 Bari, Italy}
\end{center}

\vskip0.5cm

%{\em \red revised version of \today}

\begin{abstract}
\noindent
We consider an abelian extension of the Standard Model (SM) comprising a new gauge group $U(1)^\prime$, with the neutral gauge boson $Z^\prime$ having flavour violating couplings to quarks and leptons. The fermion content  is the same as in  SM except for the addition of three right-handed neutrinos. The model,  proposed in \cite{Aebischer:2019blw}, describes the couplings of $Z^\prime$ to fermions  
in terms of three rational parameters $\epsilon_{1,2,3}$ that sum to zero imposing the cancellation of the  gauge anomalies. Each  $\epsilon_i$ is  common to all fermions in a  generation, a feature   producing correlations among quark and lepton observables.
We focus on $b \to s \ell_1^- \ell_2^+$ transitions for the lepton flavour conserving  $\ell_1=\ell_2$ and lepton flavour violating case $\ell_1 \neq \ell_2$.  Small deviations with respect to the SM predictions are found in the first case, which reflects a feature of the model where  quark and lepton sectors prevent each other to manifest  large discrepancies with respect to  SM. 
We investigate the correlations between rare $B$ and $B_s$ decays and the leptonic processes $\tau^- \to \mu^- \mu^+ \mu^-$, $\mu^- \to e^- \gamma$, $\mu^- \to e^- e^+ e^-$ and the $ \mu^- \to e^-$ conversion in nuclei. 
We show that the current experimental upper bounds on these four channels play an increasingly important role in constraining the branching fractions of lepton flavour violating $B$ and $B_s$ decays. While the present bound on $\tau^- \to \mu^- \mu^+ \mu^-$ does not impose significant restrictions,  the other three modes set progressively more stringent limits, an important information for the planned new experimental facilities.

\end{abstract}

\thispagestyle{empty}
\newpage
\section{Introduction}
The Standard Model of particle physics (SM) has been directly probed up to the energies accessible by the Large Hadron Collider and, within the present theoretical and experimental accuracy, the tests have been successful: The model provides us with a careful description of the interactions among the basic constituents of matter, excluding gravity. 
However, there are well known and broadly shared arguments to assert that this is not the ultimate theory for the fundamental interactions. Apart from the missing description of gravity, this model does not give any insight, for example, on the large number of independent parameters and  the hierarchical structure of some of them,
the regularities in the fermion sector, e.g. the organisation in  generations,  the texture of the fermion mixing matrices,  and so on.   A theory extended at higher energies, able to face at least some of  the unsolved  issues is actively sought for. The Standard Model  would be  its  low-energy realisation. 

The perspective for physics beyond the Standard Model (BSM)  is reinforced by tensions emerged  between some SM predictions and measurements. At present  they do not reach the   needed significance to claim the failure of the theory, but  all together they could  hint  new physics (NP).
Some  tensions show up  in  the flavour sector (the {\it flavour anomalies}), identifying  flavour physics as a path to disclose the BSM dynamics. Indeed, in flavour processes energy scales much higher that those directly accessed at colliders can be probed, in modes where  heavy  particles  contribute as virtual states.
To fully exploit this possibility, progress is required in the theoretical precision, as well as  the  increase in luminosity of the experimental facilities. 

The flavour changing neutral current (FCNC) modes are of particular interest. In the SM they are forbidden at tree-level, as a consequence of  the universality of the weak interactions
and of the unitarity of the Cabibbo-Kobayashi-Maskawa matrix (CKM)  rotating quark flavour eigenstates to  mass eigenstates. 
\footnote{A complete discussion is in \cite{Buras:2020xsm}.} Among the tensions,  there are  indeed anomalies  in observables  related to the FCNC $b \to s$ transition. Moreover,  
there are difficulties concerning the value of some CKM matrix elements.  Hints of violation of  lepton flavour universality have also been spotted.
A more detailed list of discrepancies includes \cite{Colangelo:2024ped,Capdevila:2023yhq}:   
\begin{itemize}
\item 
Tensions related to  the CKM  matrix elements $|V_{cb}|$ and $|V_{ub}|$. In both cases the value measured from exclusive decay modes (namely  $B \to D^* \mu \bar \nu_\mu$ and $B \to \pi \mu \bar \nu_\mu$)  is smaller than the value obtained from the inclusive  $B \to X_{c,u} \mu \bar \nu_\mu$ modes \cite{HeavyFlavorAveragingGroupHFLAV:2024ctg}. This long standing issue is still unsolved
 despite the numerous proposed explanations.\footnote{For $|V_{cb}|$ a solution based on $\mu-e$ universality violation has been proposed in \cite{Colangelo:2016ymy}.}
 In addition, there is the so-called  Cabibbo-angle anomaly,   a $ \sim 3\sigma$  
  tension among the determinations of $|V_{us}|$, together with  a  deficit from unitarity in the first row of the  CKM matrix   \cite{Kitahara:2024azt}.
\item
Tensions in observables related to the  $b \to s$  transition.  The branching ratios ${\cal B}(B \to K \mu^+ \mu^-)$ and ${\cal B}(B_s \to \phi  \mu^+ \mu^-)$ are below the SM estimates. For the  mode $B \to  K^*  \mu^+ \mu^-$  a number of
angular observables can be defined,  which depend on  $q^2$, the squared dilepton invariant mass \cite{Altmannshofer:2008dz}. For low $q^2$  some observables exceed the SM prediction  by about $3\sigma$  \cite{HeavyFlavorAveragingGroupHFLAV:2024ctg}. 
\item 
Anomalies in semileptonic decays induced by $b \to c \ell \bar \nu_\ell$ transition ($\ell=e,\,\mu,\,\tau$). The measured  ratios of branching fractions
$R(D^{(*)})={\cal B}(B \to D^{(*)} \tau \bar \nu_\tau)/{\cal B}(B \to D^{(*)} \ell \bar \nu_\ell)$,  with $\ell=e,\,\mu$,  exceed  the SM result: The current  average of
BaBar, Belle and LHCb Collaborations measurements deviates by about  $3\sigma$ from the SM prediction  \cite{HeavyFlavorAveragingGroupHFLAV:2024ctg}. This  issue challenges the SM property of  lepton flavour universality.
\end{itemize}
 In addition to the above tensions,  a  discrepancy appeared in $(g-2)_\mu$, the anomalous magnetic moment of the muon,
 in particular with the measurements at BNL \cite{Muong-2:2006rrc} and 
  Fermilab \cite{Muong-2:2021ojo,Muong-2:2023cdq}.  In this observable the main uncertainty in the SM prediction is related to one of the hadronic contributions, the leading-order hadronic-vacuum-polarisation contribution (LO HVP),  a nonperturbative QCD effect.  A lattice QCD calculation of this  contribution  by the BMW Collaboration \cite{Borsanyi:2020mff} differs by about  $2.2\sigma$  from the one obtained by a data-driven  dispersive method
employing the hadronic $e^+ e^-$ cross section, used for the SM prediction  in the White Paper 2020 of the $g-2$ Theory Initiative \cite{Aoyama:2020ynm}.  In this respect, the measurement of  $\sigma (e^+ e^- \to \pi^+ \pi^-)$ by the CMD-3 experiment is also in tension with  previous results \cite{CMD-3:2023alj}. An  analysis which excludes results by data-driven dispersive methods and  only uses lattice QCD determinations of the LO HVP contribution  shows that measurements and SM expectations  agree within the errors, which are dominated by the theory uncertainties \cite{Aliberti:2025beg}. The agreement persists for the full data set  of the Muon $g-2$  Collaboration and for the combination with all previous measurements  \cite{Muong-2:2025xyk}.
Nevertheless,  non SM contributions to $(g-2)_\mu$  still deserve to be scrutinized. 

A single above mentioned deviation is not large enough to unambiguously signal NP.  In general, the precision analysis  is a complex task, since  most observables involve  SM parameters and nonperturbative  quantities whose  uncertainty  need to be reliably assessed, as demonstrated by $(g-2)_\mu$. Methods have been proposed to get rid, to a large extent, of the errors  of  some SM inputs in a class of rare $K$ and $B$ decays \cite{Buras:2021nns,Buras:2022wpw}.
Taken all together the tensions could hint a BSM scenario, which   however is severely constrained by the unreasonable effectiveness of the Standard Model.

A minimal extension of SM has been investigated in \cite{Aebischer:2019blw}, based on the additional $U(1)^\prime$ gauge group and comprising right-handed neutrinos. The focus was on a solution found for the set of equations enabling the cancellation of the gauge anomalies induced by the new $U(1)^\prime$. In this solution the $U(1)^\prime$  charges of the fermions  are
generation dependent, and are expressed in terms of the SM hypercharges and of two additional rational parameters. One consequence of this assignment is a set of  correlations  arising among quark observables, among lepton observables  and, remarkably,  in the quark-lepton sectors together.  Some correlations have been investigated in \cite{Aebischer:2019blw}. The  conclusion was that, within the model,  quark and lepton sectors mutually act to  prevent large deviations from  SM,  justifying  the small discrepancy with  SM in observables where the tensions have shown up. However, the path to new physics is  still viable: it relies on lepton flavour violating processes  forbidden in  SM,  the observation of which would be a smoking gun for  BSM physics. Such a kind of processes are the main topic of the present study.
Within the considered framework (from now on denoted  as the ABCD model), we specifically investigate possible correlations among charged lepton flavour violating (LFV) processes. We first analyse the interplay between rare beauty meson decays that conserve lepton flavour (LFC) and those exhibiting LFV. Subsequently, we explore potential correlations between processes induced by the $b \to s \ell_1^- \ell_2^+$ transition and purely leptonic LFV modes.
\footnote{Recent analyses of lepton flavour violating $b \to s$ modes are in \cite{Crivellin:2015mga,Crivellin:2015lwa,Crivellin:2015era,Cornella:2021sby,Bordone:2021usz,Panda:2024ygr,Becirevic:2024vwy}.}

The plan of the paper is as follows:
In Sec.~\ref{abcd} we describe the main features of the ABCD model,  summarising the discussion  in \cite{Aebischer:2019blw}. In  Sec.~\ref{obs} we select the set of observables to study within the model, and identify  the correlations among them to investigate.   The results of the numerical analysis are   presented  in Sec.~\ref{numerics}. The conclusions are in  the last section. Some details needed in the main text are included in the appendices: Appendix  \ref{appA} collects the definition of the  $B \to K^*$  matrix elements of the relevant local operators in terms of form factors.
Appendix \ref{appAngularCoeff}  contains the expressions  of the angular coefficient functions  of  the $B \to K^* (K \pi)  \ell^-_1 \ell^+_2$ fully differential  decay distribution.  Appendix \ref{AppCKM} includes the parametrization and the values of the CKM and the Pontecorvo-Maki-Nakagawa-Sakata (PMNS) mixing matrices used in  the study.
\section{The ABCD model}\label{abcd}
The ABCD model  is a minimal  extension of the SM gauge group  with an additional $\text{U}(1)^\prime$ gauge group \cite{Aebischer:2019blw}.  A new neutral gauge boson $Z^\prime$ is predicted, with
 gauge coupling  $g_{Z^\prime}$. The charge associated to the $\text{U}(1)^\prime$ symmetry  is named
$z$-hypercharge.
Many NP models are based on similar simple extensions, each  one with specific $z$-hypercharge assignments \cite{Leike:1998wr,Langacker:2000ju,Appelquist:2002mw,Rizzo:2006nw}. 
From the experimental side, the assumption on the hypercharges determines the 
  exclusion plots in the plane of  the $Z^\prime$ production cross section versus  the  mass   $M_{Z^\prime}$. \!\!\!\!\!\!
\footnote{See, e.g.,  the review: B.A.~Dobrescu and S.~Willocq, "$Z^\prime$-boson searches", in \cite{Navas:PDG}. }

 $M_{Z^\prime}$  sets the NP scale, and  it  is assumed  to be acquired after spontaneous  breaking  of the new symmetry. In the model 
it is not  specified how the SSB happens,  the only assumption is  that it occurs at a  much higher scale than the SM Higgs vacuum expectation value. 
We neglect the mixing  with  other neutral gauge bosons.
The mass mixing vanishes at three-level if the SM Higgs is a singlet under $U(1)^\prime$. As for the 
kinetic mixing,  it can be dealt with by recasting  the gauge bosons kinetic terms in canonical form redefining  the new gauge coupling  \cite{Holdom:1985ag,Babu:1996vt,DelAguila:1996fw}. Mixing generated by fermion loops  can be neglected  for processes mediated by  $Z^\prime$ at  tree-level.

Besides $Z^\prime$,  the model comprises heavy right-handed neutrinos considered as Dirac particles.
The quark sector is the same as in the SM:  $q_L^i$ are  $\text{SU}(2)_L$ left-handed quark doublets,  $u_R^i,\,d_R^i$ the right-handed singlets,    $i = \{1,2, 3\}$  is a generation index. 
In the lepton sector,  $\ell_L^i$  denote the left-handed lepton doublets,   $\nu_R^i$ and $\,e_R^i$ the right-handed singlets.
Before the electroweak SSB  the $Z^\prime$ couplings to fermions are flavour conserving:   for a generic fermion $\psi$  the interaction Lagrangian with $Z^\prime$ reads 
\be
{\cal L}_{\rm int}^{Z^\prime} (\psi) = g_{Z^\prime} \, z_\psi \,\bar \psi \, \gamma^\mu \, \psi  \,Z_\mu^\prime \;, \label{Lint}
\ee
 with  $Z_\mu^\prime$  the gauge boson field, $g_{Z^\prime}$ the $\text{U}(1)^\prime$ gauge coupling and $z_\psi$  the $z$-hypercharge  of $\psi$.
 Hence, in terms of  left- and right-handed  fields $\psi_{L(R)}$  the fermion-$Z^\prime$  interaction Lagrangian   is
\be
{\cal L}_{\rm int}^{Z^\prime}  = \sum_{i,j} \big[ (\Delta_L^\psi)^{ij} \,\bar \psi_L^i \, \gamma^\mu \, \psi_L^j  +(\Delta_R^\psi)^{ij} \,\bar \psi_R^i \, \gamma^\mu \, \psi_R^j \big] \, Z_\mu^\prime \;, \label{LintLR}
\ee
where  
\be
(\Delta_{L (R)}^\psi)^{ij} = g_{Z^\prime} \, z_{\psi_{L (R)}} \, \delta^{ij} \;. \label{delta-couplings}
\ee
After  the enlargement of the gauge group,  the SM remains free of gauge anomalies  imposing  anomaly cancellation conditions.
The $z$-hypercharges are required to satisfy  a set of six anomaly cancellation equations  (ACE)    \cite{Carena:2004xs}, suitably   written  defining 
\be\label{charge1}
z_q=\sum_{i=1}^{3}z_{q_L^i} \,\, ,\qquad z_u=\sum_{i=1}^{3}z_{u_R^i} \,\, ,\qquad z_d=\sum_{i=1}^{3}z_{d_R^i} \,\, ,
\ee
\be\label{charge2}
z_\ell=\sum_{i=1}^{3}z_{\ell_L^i} \,\, ,\qquad z_\nu=\sum_{i=1}^{3}z_{\nu_R^i} \,\, ,\qquad z_e=\sum_{i=1}^{3}z_{e_R^i} \,\, ,
\ee
and 
\be
z^{(2)}_f=\sum_{i=1,2,3} z^2_{f^i} \,\,,\qquad z^{(3)}_f=\sum_{i=1,2,3} z^3_{f^i} \,\,.
\ee
 $f^i$  are the fermions  $f^i=q_L^i,\,u_R^i,\,d_R^i,\,\ell_L^i,\,e_R^i,\,\nu_R^i$ for the generations  $i=1,2,3$.
In a notation where  groups indicate the gauge bosons in triangular graphs,  the anomaly cancellation equations  read:
\begin{enumerate}
\item 
$[SU(3)_C]^2\text{U(1)}^\prime$:
\be
A_{33z}=2z_q-z_u-z_d=0 \,\, ;\label{A33z}
\ee
\item
$[SU(2)]^2\text{U(1)}^\prime$:
\be
A_{22z}=3z_q+z_l=0 \,\, ;\label{A22z}
\ee
\item
$[\text{U(1)}_Y]^2\text{U(1)}^\prime$:
\be
\label{A11z}
A_{11z}=\frac{1}{6}z_q-\frac{4}{3}z_u-\frac{1}{3}z_d+\frac{1}{2}z_l-z_e=0 \,\, ;
\ee
\item
triangular graphs involving two gravitons and $Z^\prime$:
\be
A_{GGz}=3[2z_q-z_u-z_d]+2z_l-z_e-z_\nu=2z_l-z_e-z_\nu=0   \,\, ,         \label{AGGza}
\ee
(the second equation  is obtained after imposing \eqref{A33z});
\item 
$\text{U(1)}_Y [\text{U(1)}^\prime]^2$:
\be
A_{1zz}=[z^{(2)}_q -  2 z^{(2)}_u + z^{(2)}_d]- [z^{(2)}_l - z^{(2)}_e]=0 \,\, ;\label{A1zz}
\ee
\item
$[\text{U(1)}^\prime]^3$:
\be
A_{zzz}=3[2z^{(3)}_q -  z^{(3)}_u - z^{(3)}_d]+ [2z^{(3)}_l -  z^{(3)}_\nu - z^{(3)}_e]=0 \,\, .\label{Azzz}
\ee
\end{enumerate}

Several studies have been devoted to the rational solutions of Eqs.~(\ref{A33z})-(\ref{Azzz}) \cite{Allanach:2018vjg}.
The solution proposed  in the ABCD model  consists in writing the $z$-hypercharge of each fermion  $f_i$ in terms of the SM hypercharge $y_{f_i}=y_f$ for all $i=1,2,3$  (hence generation universal)  plus a  rational parameter $\epsilon_i$  which depends on the generation $i$ and is  the same for all fermions in the generation:
\be
z_{f_i}=y_f+\epsilon_i \, .  \label{ziass}
\ee
This choice implies
\be
z_f=\sum_{i=1}^3 z_{f_i}=3y_f +\epsilon \,\, , 
\label{zf}
\ee
with 
\be
\epsilon=\sum_{i=1}^3 \epsilon_i\,. \label{epsilondef}
\ee
Since the SM hypercharges $y_f$  solve the anomaly equations, the condition for  $\epsilon_i$
 to solve  Eqs.~\eqref{A33z}-\eqref{Azzz}   is simply \cite{Aebischer:2019blw}
\be
\epsilon=\sum_{i=1}^3 \epsilon_i=0 \,\,. \label{epsilon}
\ee

At this stage the ABCD model has  four new parameters:
$M_{Z^\prime},\, g_{Z}$, $\epsilon_1$ and $\epsilon_2$ (using Eq.~\eqref{epsilon} to get rid of $\epsilon_3$).
Other parameters are involved in  the fermion rotation from  flavour to  mass eigenstates.
We  denote by ${\hat V}_{L(R)}^\psi$ the matrix  rotating left (right)-handed components of the vector of  fermions $(\psi_{L(R)}^1,\,\psi_{L(R)}^2\,,\psi_{L(R)}^3)$ with the same quantum numbers in the three generations, to obtain a diagonal  mass matrix $(\hat M_\psi)_D$:
\be
(\hat M_\psi)_D=({\hat V}_L^\psi)^\dagger \hat M_\psi {\hat V}_R^\psi\,. \label{rotation}
\ee
In terms of the rotated fields, Eq.~\eqref{LintLR} reads:
\be
{\cal L}_{\rm int}^{Z^\prime} = \sum_{i,j} g_{Z^\prime} \Big[{\bar \psi}_L^i \gamma^\mu \, ({\hat V})^\dagger_L {\hat Z}^\psi_L {\hat V}_L \, \psi_L^j +
{\bar \psi}_R^i \gamma^\mu \,({\hat V})^\dagger_R {\hat Z}^\psi_R {\hat V}_R \, \psi_R^j \Big]\, Z^\prime_\mu \, ,
\label{int-new}
\ee
where 
\be
{\hat Z}^\psi_{L(R)}=diag(z_{\psi_{L(R)}^1},\,z_{\psi_{L(R)}^2},\,z_{\psi_{L(R)}^3}) \,. 
\ee
The Lagrangian in Eq.~\eqref{int-new} has the same form  as in \eqref{LintLR} defining
\be\label{Zprimecouplings}
(\Delta^\psi_L)^{ij}=g_{Z^\prime} [(V_L^\psi)^\dagger \hat Z^{\psi}_L  V_L^\psi]^{ij}\,\, , \quad
(\Delta^\psi_R)^{ij}=g_{Z^\prime} [(V_R^\psi)^\dagger  \hat Z^{\psi}_R  V_R^\psi]^{ij}\,.
\ee
In  SM the product $\dd ({\hat V}_L^u)^\dagger {\hat V}_L^d$ defines the CKM matrix:  $\dd ({\hat V}_L^u)^\dagger {\hat V}_L^d={\hat V}_{CKM}$, and it  is possible to choose $\dd {\hat V}_L^u=1$ and 
$\dd {\hat V}_L^d={\hat V}_{CKM}$. The same  choice is  made in the ABCD model.
 The absence of right-handed charged currents and the flavour universality of the neutral currents allows also  to  choose ${\hat V}_R^u=1$ in SM. This choice is not automatically justified in the ABCD model, due to the flavour nonuniversality of the $Z^\prime$ mediated current:  the choice ${\hat V}_R^u=1$ is retained as a model assumption.
Hence,  the quark sector only involves the matrices ${\hat V}_{CKM}$ and ${\hat V}_R^d$.

In  the lepton sector the matrix ${\hat V}_R^\nu$ can be ignored since no  phenomenological observable involving $\nu_R$ is considered.  
In analogy to the quark sector, the matrix ${\hat V}_L^\nu$ is set to the unit matrix  and $({\hat V}_L^\ell)^\dagger$ defines the  PMNS matrix:
 $({\hat V}_L^\ell)^\dagger={\hat U}_{PMNS}$. 
Hence, the remaining matrices in this sector are ${\hat U}_{PMNS}$ and ${\hat V}_R^e$.

 The parametrization of the rotation matrices can be found in \cite{Aebischer:2019blw}. Appendix \ref{AppCKM} contains the choice for the parameters of  the CKM and PMNS matrices. After the fermion rotations,
flavour violating $Z^\prime$ couplings are generated, as indicated below. 
\begin{itemize}
\item Couplings  of left-handed down quarks to $Z^\prime$:
\bea
\Delta_L^{ij}&=&g_{Z^\prime} \big[(2\epsilon_1+\epsilon_2)V_{uj}V^*_{ui}+(\epsilon_1+2\epsilon_2)V_{cj}V^*_{ci}\big] \qquad \qquad    (i\neq j) \,\, , 
  \label{LHQD} \\
 \Delta_L^{ii}&=&g_{Z^\prime} \Big[\frac{1}{6}+(2\epsilon_1+\epsilon_2)|V_{ui}|^2+(\epsilon_1+2\epsilon_2)|V_{ci}|^2-\epsilon_1-\epsilon_2\Big] \, .  \label{LHQDa}
 \eea
The generation index  $i=d,\,s,\,b$ is used to make the notation more transparent.  $V_{ui}$ and $V_{ci}$ are  CKM matrix elements.
\item 
Couplings of right-handed down quarks  to $Z^\prime$:
\bea
\Delta_R^{sd}&=&g_{Z^\prime} \,e^{i\phi_K} \tilde s_{12}\,(\epsilon_1-\epsilon_2) \, , \label{ZprimesdR}\\
\Delta_R^{bd}&=&g_{Z^\prime} \,e^{i\phi_d} \tilde s_{13}\,(2\epsilon_1+\epsilon_2) \, , \\
\Delta_R^{bs}&=&g_{Z^\prime} \,e^{i\phi_s} \tilde s_{23}\,(\epsilon_1+2\epsilon_2) \, , \\
 \Delta_R^{ii}&=&g_{Z^\prime} [-\frac{1}{3}+ \epsilon_i] \, .
 \eea
The real numbers ${\tilde s}_{12},\,{\tilde s}_{13},\,{\tilde s}_{23}$ and the phases $\phi_K,\,\phi_d,\,\phi_s$ appear in the parametrization of the rotation matrices  \cite{Aebischer:2019blw}.
 $(\tilde s_{12},\,\phi_K)$ only enter in the kaon sector,
$(\tilde s_{13},\,\phi_d)$  only in the $B_d$ sector, $(\tilde s_{23},\,\phi_s)$ only in the $B_s$ sector.
\item Couplings of  up-type quarks  to $Z^\prime$:
\be
 \Delta_L^{ii}= g_{Z^\prime} [\frac{1}{6}+\epsilon_i] \,\, ,\qquad  \Delta_R^{ii}= g_{Z^\prime} [\frac{2}{3}+\epsilon_i] \,\, , \qquad \qquad (i=u,c,t)\,.
\ee
\item Couplings of left-handed charged leptons to $Z^\prime$:
\bea\label{LHQD1}
 \Delta_L^{ij}&=&g_{Z^\prime}\,[(2\epsilon_1+\epsilon_2)U_{i 1}U^*_{j1}+(\epsilon_1+2\epsilon_2)U_{i2}U^*_{j2}] \qquad   (i\neq j) \,\, , 
  \\\label{LHQD2}
  \Delta_L^{ii}&=&g_{Z^\prime}\,[-\frac{1}{2}+(2\epsilon_1+\epsilon_2)|U_{i1}|^2+(\epsilon_1+2\epsilon_2)|U_{i2}|^2-\epsilon_1-\epsilon_2]\,,
 \eea
 where  $i=e,\,\mu,\,\tau$, and $U_{ij}$ are elements of the PMNS matrix. Lepton flavour violating couplings arise for non-vanishing $\epsilon_{1,2}$.
 \item Couplings  of right-handed charged leptons to $Z^\prime$:
 \bea
\Delta_R^{\mu e}&=&g_{Z^\prime}\,e^{i\phi_{12}} \tilde t_{12}\,(\epsilon_1-\epsilon_2)\, , \\
\Delta_R^{\tau e}&=&g_{Z^\prime}\,e^{i\phi_{13}} \tilde t_{13}\,(2\epsilon_1+\epsilon_2)\, , \\
\Delta_R^{\tau\mu}&=&g_{Z^\prime}\,e^{i\phi_{23}} \tilde t_{23}\,(\epsilon_1+2\epsilon_2)\, , \\
 \Delta_R^{ii}&=&g_{Z^\prime}\,[-1 + \epsilon_i]\, .
 \eea
The real numbers  $\tilde t_{ij}$  and the phases $\phi_{ij}$ are the analogous of $\tilde s_{ij}$  and $\phi_{K,d,s}$ in the quark sector: 
$(\tilde t_{12},\,\phi_{12})$ enter in  $\mu\to e$ transitions,
$(\tilde t_{13},\,\phi_{13})$ in   $\tau\to e$ transitions,
 $(\tilde t_{23},\,\phi_{23})$ in   $\tau\to \mu$ transitions. Also in this case $\epsilon_{1,2}\neq 0$ produces lepton flavour violating $Z^\prime$ interactions.
 \item Neutrino couplings to $Z^\prime$:
 \be\label{neutrinos}
 \Delta_L^{\nu_i\bar\nu_i}= g_{Z^\prime}\,\Big[-\frac{1}{2}+\epsilon_i \Big] \qquad \qquad (i=\nu_1,\nu_2, \nu_3)\,.
\ee
\end{itemize}
Hence, after the  fermion rotation from flavour to mass eigenstates, the free parameters in the model comprise  
\be
g_{Z^\prime}\,\, , \qquad M_{Z^\prime}\,\, , \qquad \epsilon_{1}\,\, , \qquad \epsilon_{2} \,\, , \label{4par}
\ee
with $\epsilon_{1,2}$ rational numbers, 
and the mixing angles and phases  in the  $Z^\prime$ couplings to right-handed fermions: 
\be\label{quarks}
 (\tilde s_{12},\,\phi_K)\, , \qquad  (\tilde s_{13},\,\phi_d)\, ,\qquad
(\tilde s_{23},\,\phi_s)\, 
\ee
for quarks, and
\be\label{leptons}
 (\tilde t_{12},\,\phi_{12})\, , \qquad  (\tilde t_{13},\,\phi_{13})\, ,\qquad
(\tilde t_{23},\,\phi_{23})
\ee
for leptons.
The $Z^\prime$ couplings to left-handed fermions involve the parameters of the  CKM and  PMNS matrices, respectively  in the quark and lepton sectors. 

Since the number of parameters, even in this minimal extension of the Standard Model, is not small, a  strategy for their treatment   needs to be envisaged.  
 The general  case  
where all $\tilde s_{ij}$ and $\tilde t_{ij}$  in Eqs.\eqref{quarks},\eqref{leptons} are taken to be different from zero was considered,  allowing for $Z^\prime$ couplings  to all right-handed fermions  \cite{Aebischer:2019blw}. This possibility,    denoted as the scenario C,  appeared  not efficiently constrained. \footnote{  In particular, the efficient constraints imposed by the $\Delta F=2$ observables  discussed in the following were not imposed.}  
 Another  strategy was adopted in two other  cases.  First,  all  $\tilde t_{ij}$  in \eqref{leptons}  were assumed to vanish, enabling  flavour violating couplings of $Z^\prime$ to right-handed fermions only for quarks, a case denoted as scenario B. Finally,  all  
$\tilde s_{ij}$ and $\tilde t_{ij}$  in \eqref{quarks},\eqref{leptons} were set to zero,  enabling  no flavour violating couplings of $Z^\prime$ to right-handed fermions  \cite{Aebischer:2019blw}. In this last case,  the scenario A,  the new parameters are  only those in \eqref{4par} together with the elements of the CKM and PMNS matrices. This   most economic scenario  is assumed in our study. In the following, we select a set of relevant observables, we discuss how the parameter space is constrained and the implications in  a few lepton flavour conserving  (LFC) and  lepton flavour violating processes (LFV). Since all $Z^\prime$ couplings to fermions involve the two parameters $\epsilon_{1,2}$, correlations among different modes can be established: we investigate some of them,  in particular the  correlations between quark and lepton observables. 

\section{Observables}\label{obs}
The couplings  \eqref{LHQD}-\eqref{neutrinos} show that in the ABCD model flavour changing neutral current transitions  occur at tree-level through $Z^\prime$ exchanges, and that lepton flavour violating processes are possible.
For this reason,  we study the $b \to s \ell_1^- \ell_2^+$ transition, both for $\ell_1=\ell_2$ and $\ell_1 \neq \ell_2$, focusing on the exclusive modes $B_s \to  \ell_1^- \ell_2^+$, $B \to K^*  \ell_1^- \ell_2^+$ and  
$B_s \to \phi \ell_1^- \ell_2^+$ for which a wealth of experimental information is available and some tensions are spotted. Moreover, we describe the leptonic LFV modes  
 $\mu^- \to e^- \gamma$, $ \tau^- \to \mu^- \mu^+ \mu^-$, $\mu^- \to e^- e^+ e^- $ and the $\mu^- \to e^-$ conversion in nuclei. Finally, we consider  the $Z^\prime$ contribution to $(g-2)_\mu$. 

\subsection{Effective Hamiltonian for $b \to s  \ell_1^- \ell_2^+$}\label{section-hamil}
The  low-energy  $ \Delta B =-1$, $\Delta S = 1$ Hamiltonian governing   $b \to s \ell_1^- \ell_2^+$,  comprising  the SM operators for $\ell_1= \ell_2$,  and operators that can be generated in BSM models with $Z^\prime$,
has the form
\bea
H^{\rm eff}=-\,4\,\frac{G_F }{\sqrt{2}} V_{tb} V_{ts}^* \, &&\Big\{C_1 O_1+C_2 O_2 
+\sum_{i=3,..,6}C_i O_i+\sum_{i=7,..,10} \left[ C_i O_i +C_i^{\prime  } O_i^{\prime  } \right]\Big\}\,\, . \label{hamil}
\eea
 $G_F$ is the Fermi constant,  $V_{ij}$ are
elements of the CKM  matrix, and  doubly Cabibbo suppressed terms proportional to $V_{ub} V_{us}^*$ are neglected.\footnote{The complete basis of  operators including (pseudo-)scalar and tensor operators  can be found  in \cite{Altmannshofer:2008dz}.}
 The effective Hamiltonian includes  the current-current operators $O_{1,2}$ 
\be
O_1 = ({\bar c}_\alpha \gamma_\mu P_L \, b_\beta)({\bar s}_\beta \gamma^\mu P_L \, c_\alpha) \,\, , \quad
O_2 = ({\bar c} \gamma_\mu P_L \, b)({\bar s} \gamma^\mu P_L \, c) \,\, , \label{O1}
\ee
the QCD penguin operators  $O_{3,\dots,6}$ 
\bea
O_3&=&({\bar s} \gamma^\mu P_L \, b) \sum_q ({\bar q} \gamma^\mu P_L \, q) \,\, , \quad 
O_4=({\bar s}_\alpha \gamma^\mu P_L \, b_\beta) \sum_q ({\bar q}_\beta \gamma^\mu P_L \, q_\alpha) \,\, ,  \label{O34} \\
O_5&=&({\bar s} \gamma^\mu P_L \, b) \sum_q ({\bar q} \gamma_\mu P_R \, q) \,\, ,  \quad
O_6=({\bar s}_\alpha \gamma^\mu P_L \, b_\beta) \sum_q ({\bar q}_\beta \gamma_\mu P_R \, q_\alpha) \,\, , \label{O56}
\eea
with $P_{R,L}=\displaystyle\frac{1 \pm \gamma_5}{2}$,   $\alpha$,$\beta$ colour indices, and the sum in (\ref{O34})-(\ref{O56}) running over the flavours $q=u,d,s,c,b$.
The other operators in  \eqref{hamil} are  the magnetic penguin operators 
\bea
O_7&=&\frac{e}{16 \pi^2} \Big({\bar s}\sigma^{\mu \nu}(m_s P_L + m_b P_R)\,b \Big) F_{\mu \nu} \,\, , \label{O7} \\
O_7^\prime &=& \frac{e}{16 \pi^2} \Big({\bar s}\sigma^{\mu \nu} (m_s P_R + m_b P_L) \, b\Big)  F_{\mu \nu}  \,\, ,  \label{O7p} \\
O_8&=&\frac{g_s}{16 \pi^2}\Bigg({\bar s}_{ \alpha} \sigma^{\mu \nu} \Big({\lambda^a \over 2}\Big)_{\alpha \beta} (m_s P_L + m_b P_R)  b_{ \beta}\Bigg) 
      G^a_{\mu \nu}  \,\,  ,  \label{O8} \\
      O_8^\prime&=&\frac{g_s}{16 \pi^2}  \Bigg({\bar s}_{ \alpha} \sigma^{\mu \nu} \Big({\lambda^a \over 2}\Big)_{\alpha \beta} (m_s P_R + m_b P_L)  b_{ \beta}\Bigg)
      G^a_{\mu \nu} \,\,  , \label{O8p}
      \eea
and  the semileptonic electroweak penguin operators
\bea
O_9&=&{e^2 \over 16 \pi^2}  \Big({\bar s} \gamma^\mu P_L \, b\Big) \; {\bar \ell}_1 \gamma_\mu \ell_2  \,\, ,\hskip 1.4 cm
O_9^\prime ={e^2 \over 16 \pi^2}  \Big({\bar s} \gamma^\mu P_R \, b\Big) \; {\bar \ell}_1 \gamma_\mu \ell_2  \,\, , \label{O9} \\
O_{10}&=&{e^2 \over 16 \pi^2}  \Big({\bar s} \gamma^\mu P_L \, b\Big) \; {\bar \ell}_1 \gamma_\mu \gamma_5 \ell_2  \,\, ,\hskip 1 cm 
O_{10}^\prime={e^2 \over 16 \pi^2}  \Big({\bar s} \gamma^\mu P_R \, b\Big) \; {\bar \ell}_1 \gamma_\mu \gamma_5 \ell_2\,\,.\label{O10}
\eea
Primed operators have opposite chirality with respect to the unprimed ones. 
$\lambda^a$ are the Gell-Mann matrices,  $F_{\mu \nu}$ and $G^a_{\mu \nu}$   the electromagnetic and  gluonic field strengths, $e$ and $g_s$  the
electromagnetic and strong coupling constant. The  $b$ and $s$ quark mass $m_b$ and  $m_s$ are defined in the ${\rm \overline{MS}}$ scheme at the scale $m_b$.

In SM,  at the leading order in $\alpha_s$  the only operators to be considered are $O_7$ and  $O_{9,10}$  with $\ell_1 = \ell_2$. At this order, 
$O_7$   is the only operator contributing to $b \to s \gamma$.
The renormalization group evolution to the scale $\mu_b \simeq {\cal O}(m_b)$ also involves the magnetic gluon penguin operator $O_8$ and the operators $O_{1,\dots 6}$. Their mixing into $O_7$ generates large logarithms  producing a strong enhancement of the rates. The anomalous dimension matrix  governing the mixing is regularization scheme dependent. One can get rid of such a dependence   defining an effective coefficient
  $C_7^{(0)\rm eff}(\mu_b)$ which includes contributions of $O_{1,\dots 6}$ \cite{Buras:1993xp}. Non factorizable contributions to the matrix elements of $O_1$ and $O_2$ must also be taken into account. Their effects
  produce  a further shift in the coefficient of $O_7$: $C_7^{\rm eff}=1.33\, C_7^{(0)\rm eff}$ \cite{Beneke:2001at}.
In this way,  $O_7$ turns out to give  the dominant contribution to $b \to s \gamma$, with the  SM Wilson coefficients known at NNLO in QCD \cite{Misiak:2018cec,Buras:2020xsm}.
$O_7$ contributes to  $b \to s \ell^+_1 \ell^-_2$  for   $\ell_1 = \ell_2$, due to the photon conversion into the lepton pair.

In SM, the contribution to $b \to s \ell^+ \ell^-$ comes  from the operators $O_9$ and $O_{10}$  and also from other operators in \eqref{hamil}, in particular  $O_1$ and $O_2$ due to charm re-scattering. This contribution overwhelms the contributions of the other operators in the region of $q^2$, the dilepton invariant mass,  corresponding to  the narrow charmonium resonances $J/\psi$, $\psi(2S)$, \dots
The contribution of charm re-scattering in the other intervals of $q^2$, the role of which has been pointed out  long ago \cite{Colangelo:1989gi}, has  theoretical uncertainties difficult to evaluate.
 A  possibility to account for it is to replace  $C_9$ with $C_9^{\rm eff}=C_9+Y(q^2)$ \cite{Lim:1988yu,Deshpande:1988bd,ODonnell:1991cdx,Paver:1991tn,Colangelo:1995jv},  introducing a  function $Y(q^2)$
which comprises the terms described in \cite{Bordone:2024hui}.   This could obscure  genuine new short-distance contributions to $C_9$.  Charm re-scattering contributions can be described through non-local form factors 
 determined, e.g., by dispersion relations  employing information from the nonleptonic modes $B \to  K^* (J/\psi, \psi(2S))$  \cite{Khodjamirian:2010vf,Khodjamirian:2012rm,Bobeth:2017vxj,Bordone:2024hui},  with an
 uncertainty  difficult to estimate \cite{Ciuchini:2022wbq}.  In the comparison with measurements it is customary to exclude the $q^2$ region corresponding to the resonances  $J/\psi$, $\psi(2S)$,  which are cut in the experimental analyses  to isolate the
 short-distance part of the $q^2$ spectrum  \cite{LHCb:2024onj}.
Since charm re-scattering does not affect the LFV processes with  $\ell_1 \neq \ell_2$, of prime interest in our study, for the $\ell^+ \ell^-$ mode we also cut the resonance region and omit to include charm rescattering effects in the remaining $q^2$ interval.

Summarizing,  we only consider  the operators $O_7$, $O_{9,10}$ and $O^\prime_{9,10}$ in the effective Hamiltonian \eqref{hamil} to analyse  the exclusive  modes $B_s \to  \ell_1^- \ell_2^+$,  $B \to K^*  \ell_1^- \ell_2^+$
and $B_s \to \phi  \ell_1^- \ell_2^+$.
In  SM the Wilson coefficients are flavour universal, in the ABCD  extension they  depend on the lepton flavour.  
\subsubsection{$B_s \to  \ell_1^- \ell_2^+$}
For the decay $B_s (p) \to  \ell_1^-(k_1)\ell_2^+(k_2)$, with $p$ and $k_{1,2}$  the  particle four-momenta, a single hadronic quantity is required,  the  decay constant $f_{B_s}$  defined by 
\be
\langle 0|{\bar s}\gamma_\mu \gamma_5 b| B_s(p) \rangle=i\,f_{B_s} p_\mu \,\,.
\ee
The branching fraction is given by
\bea
{\bar {\cal B}}(B_s \to \ell^-_1 \ell_2^+)&=&\frac{\tau_{B_s}}{(1-y_s)}\frac{G_F^2 \alpha^2 |\lambda_{ts}^*|^2 f_{B_s}^2}{64 \pi^3 m_{B_s}^3 }\lambda^{1/2}(m_{B_s}^2,\, m_{\ell_1}^2,\, m_{\ell_2}^2)
\nn \\
&\times&\Big\{|C_9-C_9^\prime|^2 (m_{\ell_1}-m_{\ell_2})^2 \, [m_{B_s}^2-(m_{\ell_1}+m_{\ell_2})^2]
\nn \\
&+&|C_{10}-C_{10}^\prime|^2 (m_{\ell_1}+m_{\ell_2})^2 \, [m_{B_s}^2-(m_{\ell_1}-m_{\ell_2})^2]\Big\} \,\, , \label{Bslept}
\eea
with $\lambda_{ts}^*=V_{tb}V_{ts}^*$, $\lambda(x,y,z)=x^2+y^2+z^2-2xy-2xz-2yz$  the K\"{a}llen function,   $\tau_{B_s}$ the average $B_s$ lifetime. The notation ${\bar {\cal B}}$ indicates that the effects of the $B_s-\bar B_s$ oscillations are taken into account through the factor $1/(1-y_s)$ \cite{DeBruyn:2012wj}, with $y_s=0.061 \pm 0.009$  \cite{LHCb:2014iah}.
In  SM with $\ell_1=\ell_2$  Eq.~\eqref{Bslept} shows that only the operator  $O_{10}$ contributes to this decay.

\subsubsection{$\bar B^0 \to \bar K^{*0} \ell_1^- \ell_2^+$ and  $B_s \to \phi \ell_1^- \ell_2^+$}\label{Kstar}
Let us consider   $\bar B^0(p)  \to \bar K^{*0}(p^\prime,\,\epsilon) \ell_1^-(k_1) \ell_2^+(k_2)$ (the description of  $B_s \to \phi \ell_1^- \ell_2^+$ is analogous).
$\epsilon$ is the $K^*$  polarization four-vector,    $q=k_1+k_2$.
 The local form factors parametrising the matrix elements of the operators in  \eqref{hamil} are defined  in appendix \ref{appA}.
\begin{figure}[t]
\begin{center}
\includegraphics[width = 0.35\textwidth]{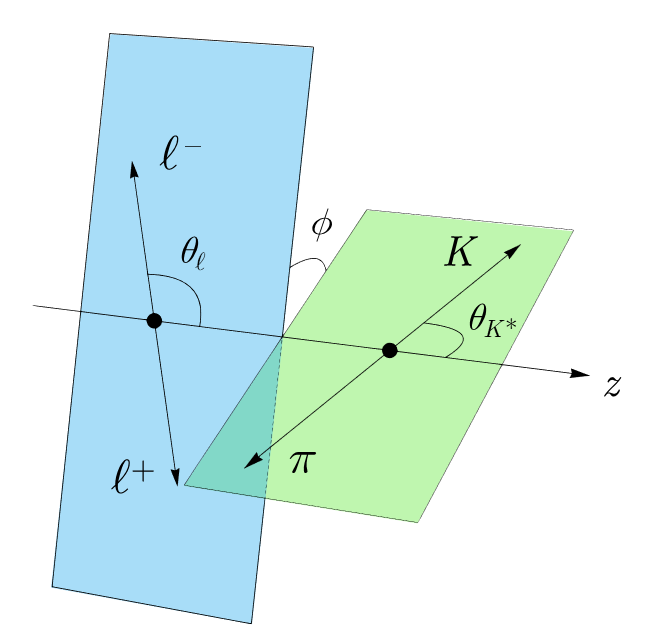}
\caption{\baselineskip 10pt  \small  Kinematics of the decay $\bar B^0 \to \bar K^{*0}(K \pi) \ell_1^- \ell_2^+$. }\label{kinematics}
\end{center}
\end{figure}
Considering the subsequent  $K^* \to K \pi$ transition and the kinematics  in Fig.~\ref{kinematics}, the fully differential decay distribution in $q^2$ and in  the angles $\theta_\ell,\,\theta_{K^*},\,\phi$ has the expression 
\begin{equation}\label{full}
  \frac{d^4\Gamma (\bar B^0 \to \bar K^{*0}(K \pi) \ell_1^- \ell_2^+)}{dq^2\, d\cos\theta_\ell\, d\cos\theta_{K^*}\, d\phi} =
   \frac{9}{32\pi} I(q^2, \theta_\ell, \theta_{K^*}, \phi)\, .
\end{equation}
The function  $I(q^2, \theta_\ell, \theta_{K^*}, \phi)$ can be written in terms  of  angular coefficient functions $I_i^{(a)}(q^2)$ using the decomposition \cite{Altmannshofer:2008dz}
\begin{align} \label{d4G}
  I(q^2,\theta_\ell, \theta_{K^*}, \phi)& = 
      I_1^s(q^2) \sin^2\theta_{K^*} + I_1^c(q^2) \cos^2\theta_{K^*}
      + \big[I_2^s(q^2) \sin^2\theta_{K^*} + I_2^c(q^2) \cos^2\theta_{K^*}\big] \cos 2\theta_\ell
\nonumber \\       
    & + I_3(q^2) \sin^2\theta_{K^*} \sin^2\theta_\ell\cos 2\phi 
      + I_4(q^2) \sin 2\theta_{K^*} \sin 2\theta_\ell \cos\phi 
\nonumber \\       
    & + I_5(q^2) \sin 2\theta_{K^*} \sin\theta_\ell\cos\phi \\      
    & + \big[I_6^s (q^2)\sin^2\theta_{K^*} +
      {I_6^c(q^2) \cos^2\theta_{K^*}}\big]  \cos\theta_\ell
      + I_7(q^2) \sin 2\theta_{K^*} \sin\theta_\ell\sin\phi
\nonumber \\ 
    & + I_8(q^2) \sin 2\theta_{K^*} \sin 2\theta_\ell \sin\phi
      + I_9(q^2) \sin^2\theta_{K^*} \sin^2\theta_\ell\sin 2\phi\,. \nonumber
\end{align}
 $I_i^{(a)}$ depend  on the Wilson coefficients in \eqref{hamil}, hence they probe the SM. They also depend on the hadronic form factors.
 Their expressions  in terms of amplitudes for the various $K^*$ polarizations and  lepton chiralities are in appendix \ref{appAngularCoeff}. 
Integration over the three angles $\phi, \theta_\ell$ and $\theta_{K^*}$ provides the $q^2$ spectrum:
\be
\frac{d \Gamma}{dq^2}=\frac{1}{4}\Big[6I_1^s(q^2)+3I_1^c(q^2)-2I_2^s(q^2)-I_2^c(q^2) \Big] \,\,,\label{dGdq2}
\ee
 further integration  over  the physical range $q^2 \in [(m_{\ell_1}+m_{\ell_2})^2, (m_B-m_{K^*})^2]$ gives the decay width.

The  hadronic uncertainties  can obscure the sensitivity to NP,  therefore it is convenient to define observables  in which such uncertainties to a large extent  cancel out.
A possibility is to  consider the  distribution  analogous to \eqref{d4G} for the CP conjugate decay mode, with  angular coefficient functions  ${\bar I}_i^{(a)}$, and define  CP-even and CP-odd quantities:
\bea
S_i^{(a)}(q^2)&=&\left(I_i^{(a)}(q^2)+{\bar I}_i^{(a)}(q^2) \right)/\left(d\Gamma/dq^2+d{\bar \Gamma}/dq^2\right) \,\, ,
\label{S} \\
A_i^{(a)}(q^2)&=&\left(I_i^{(a)}(q^2)-{\bar I}_i^{(a)} (q^2)\right)/\left(d\Gamma/dq^2+d{\bar \Gamma}/dq^2\right) \,\, .
\label{A} 
\eea
   $d{\bar \Gamma}/dq^2$ is the distribution analogous to  \eqref{dGdq2} for the CP conjugated mode.
Starting from such  expressions, several  observables  can be built.  In  SM,  $S_7,\,S_8$ and $S_9$ are tiny.

We focus on   the forward-backward asymmetry with respect to the direction of the emitted  lepton
\be 
A_{FB}(q^2)=\frac{3}{8}[2S_6^s(q^2)+S_6^c(q^2)] \,\, ,  \label{AFB}
\ee
on  the fraction of longitudinally and transversely polarized $K^*$
 \be F_L(q^2)=-S_2^c (q^2) \,\, , \qquad F_T(q^2)=1-F_L (q^2)\,\, ,\label{FLeFT}
 \ee
 and on the angular $P$-observables  \cite{Matias:2012xw}
 \bea
  P_{1}(q^2)&=&\displaystyle\frac{2}{F_T(q^2)}S_{3}(q^2)  \,\, , \quad P_{2}(q^2)=\displaystyle\frac{1}{2 F_T(q^2)}S_{6s}(q^2)  \,\, , \quad P_{3}(q^2)=-\displaystyle\frac{1}{F_T(q^2)}S_{9}(q^2) \,\, ,  \quad
  \nn \\
    P^\prime_{4,5,6}(q^2)&=&\displaystyle\frac{1}{\sqrt{F_L (q^2)\,F_T(q^2)}}S_{4,5,7}(q^2) \,\, . \label{Pobs}
 \eea

The forward-backward asymmetry  is recognised to be sensitive to NP, in particular because in SM it changes sign  in the physical $q^2$ range \cite{Ali:1991is,Ali:1994bf,Liu:1994cfa,Colangelo:1995jv,Ali:1999mm}, 
with the zero $q_0^2$  determined by 
\be
{\rm Re}[C_9]+{m_b \over
q_0^2} {\rm Re} [C_7] \left[ (m_B+m_{K^*}){T_1(q_0^2) \over V(q_0^2)}
+(m_B-m_{K^*}){T_2(q_0^2) \over A_1(q_0^2)} \right]=0
\,\,\,.\label{zeroasim} \ee
 A model independent prediction for $q_0^2$  is obtained in the large energy limit of  $K^*$,   due to  spin
symmetry relations among the form factors \cite{Charles:1998dr}
\be {T_1(E) \over V(E)}={m_B \over m_B +m_{K^*}} \,\,\, , \hskip
1cm {T_2(E) \over A_1(E)}={(m_B +m_{K^*}) \over  m_B  }
\label{symrel} \,, \ee
where $E$ is the $K^*$ energy in the $B$ rest frame.\footnote{The relations \eqref{symrel}  are broken by hard gluon corrections to the weak vertex and by hard spectator interactions \cite{Beneke:2000wa}. }
 Since $m_{K^*}^2\ll m_B^2$,  Eqs.~(\ref{zeroasim}),(\ref{symrel})   give
\be
{\rm Re}\, [C_9]q_0^2+2m_B m_b 
{\rm Re}[C_7 ]=0 \,.
\ee
 Deviations from this result would  signal  non SM operators in the effective Hamiltonian or/and modified values of the Wilson coefficients.
 Another  observable 
 \be 
 A_T^{(2)}(q^2)=\frac{2S_3(q^2)}{1-F_L(q^2)  }
 \ee
 is  sensitive to the virtual photon polarization  \cite{Kruger:2005ep,Becirevic:2011bp}, which  in SM  is  mainly left-handed, the right-handed   polarization being suppressed by $m_s/m_b$. Deviations are expected in scenarios where the photon can also be right-handed, an effect encoded in the operator  $O_7^\prime$,  which is not generated in the scenario  A of the ABCD model.
 
\subsection{LFV lepton decays}

In this section we examine the contribution of $Z^\prime$  to LFV leptonic modes. We begin with the processes $\mu^- \to e^- \gamma$, $\tau^- \to \mu^- \mu^+ \mu^-$ and  $\mu^- \to e^- e^+ e^-$. In absence of  neutrino oscillations such processes do not occur in  SM. In a minimal extension  allowing for neutrino masses and mixing, they can occur  at  rates  much smaller than the present experimental upper bounds discussed below.
We then turn to the $\mu^- \to e^- $ conversion in nuclei.

\begin{itemize}
\item $\mu^- \to e^- \gamma$

In a version of the SM extended with the see-saw mechanism to describe massive neutrinos this process can occur through a penguin diagram with internal $W$ and oscillating neutrinos. The photon is emitted by $W$. 
If there is a  $Z^\prime$ with LFV couplings,   other penguin diagrams contribute, with $Z^\prime$ and a charged lepton in the loop.  The branching fraction is \cite{Lindner:2016bgg}
\be
\mathcal{B}(\mu\to e \gamma)=\frac{3(4\pi)^3\alpha}{4G^2_F}
\left[|A^M_{e\mu}|^2+|A^E_{e\mu}|^2\right],
\ee
where
\be
A^M_{e\mu}= -\frac{1}{96\pi^2M^2_{Z^\prime}}\sum_{f=e,\mu,\tau}
\left[\Delta^{fe*}_V  \Delta^{f\mu}_V \left(1-3\frac{m_f}{m_\mu}\right)+\Delta^{fe*}_A \Delta^{f\mu}_A \left(1+3\frac{m_f}{m_\mu}\right)\right] \,\, ,   \label{AMemu}
\ee
\be
A^E_{e\mu}= \frac{i}{96\pi^2M^2_{Z^\prime}}\sum_{f=e,\mu,\tau}
\left[\Delta^{fe*}_A  \Delta^{f\mu}_V \left(1-3\frac{m_f}{m_\mu}\right)+\Delta^{fe*}_V  \Delta^{f\mu}_A \left(1+3\frac{m_f}{m_\mu}\right)\right] \,\, ,  \label{AEemu}
\ee
and   $\Delta^{ij}_V$ and  $\Delta^{ij}_A$  defined as 
\be\label{DVA}
 \Delta_V^{ij}= \Delta_R^{ij}+\Delta_L^{ij}\,\, ,\qquad
\Delta_A^{ij}= \Delta_R^{ij}-\Delta_L^{ij} \,\, .
\ee
The sum in \eqref{AMemu},\eqref{AEemu}  is over the  charged leptons in the loop. 

\item
$\ell_j\to \ell_i \bar \ell_l \ell_k  $

In the see-saw extended SM these modes are   also described by loop diagrams,  either penguins or boxes. In the ABCD model  a tree-level diagram contributes  with $Z^\prime$ as mediator.
The  generated operators and their Wilson coefficients are
\be
[O_{VXY}]^{ij}_{kl}=(\bar\psi_i\gamma^\mu P_X \psi_j)(\bar\psi_k\gamma_\mu P_Y \psi_l) \,\, ,\qquad [C_{VXY}]^{ij}_{kl}=\frac{\Delta^{ij}_X \Delta^{kl}_Y }{M^2_{Z^\prime}} \,\, , 
\ee
with $X,Y=L,R$.
 The branching fraction of  $\tau^-\to\mu^-\mu^+\mu^-$ is given by \cite{Aebischer:2019blw}:
\be \label{DecayA}
\mathcal{B}(\tau^-\to\mu^-\mu^+\mu^-)=
\frac{m_\tau^5}{1536\pi^3 \Gamma_{\tau}} \left[2 |C_{VLL}|^2
+ 2 |C_{VRR}|^2 + |C_{VLR}|^2 + |C_{VRL}|^2\right]^{\mu\tau}_{\mu\mu}\, ,
\ee
with $\Gamma_\tau$  the  $\tau$ full decay width and a  $1/2$  factor due to the  identical leptons in the final state.  
Analogous expressions hold for $\tau^-\to e^-e^+e^-$ and $\mu^-\to e^-e^+e^-$.

\item   $\mu^--e^-$ conversion in nuclei

The $\mu^- - e^-$ conversion in a nucleus of atomic number $Z$ and mass number $A$ is the process
\be
\mu^- +(A,Z) \rightarrow e^- + (A,Z)\,. 
\ee
The  transition rate  $\Gamma(\mu\rightarrow e)$ in the ABCD model is obtained 
  considering the dominant tree-level $Z^\prime$ contributions inducing  four-fermion operators,  and assuming that the conversion leaves the nucleus in its ground state \cite{Hisano:1995cp,Kitano:2002mt}:
\begin{eqnarray}\label{eq:Gmue}
\Gamma(\mu^- \rightarrow e^-)
&=& \frac{\alpha^3}{16\pi^2}\frac{Z_{eff}^4}{Z}|F(q)|^2~\frac{m_{\mu}^5}{M^4_{Z^\prime}}
\left [|\Delta_L^{\mu e}(Z^\prime)|^2+|\Delta_R^{\mu e}(Z^\prime)|^2\right] \nonumber \\
&&\times
\left |(2Z+N)\Delta_V^{uu}(Z^\prime)+(Z+2N)\Delta_V^{dd}(Z^\prime) \right|^2 .
\end{eqnarray}
 $\Delta^{ij}_V(Z^\prime)$ are  in (\ref{DVA}). $N=A-Z$ is the number of neutrons in the nucleus, $Z_{eff}$ is  an effective parameter and $F(q^2)$  the nuclear form factor. 
The  observable to investigate is the branching fraction  defined as 
 \begin{equation}
   \mathcal{B}(\mu^- \to e^-) = \frac{\Gamma(\mu^- \to e^-)}{\Gamma_{\rm capt}} \,\, ,  \label{Brcapture}
 \end{equation}
which involves   the muon capture rate $\Gamma_{\rm capt}$ in the nucleus.
The upper bound for   $ \mathcal{B}(\mu\to e)$ in $^{48}_{22}{\rm{Ti}}$  established by the SINDRUM II experiment at PSI \cite{Kaulard:1998rb} is
\be
 \mathcal{B}(\mu^- \to e^-)\le 4.3 \times10^{-12} \quad ({\rm at}\,\,\, 90\% \,\, {\rm C.L.}) \,\, .
 \ee
 For $^{48}_{22}{\rm{Ti}}$ the values of the relevant parameters are $Z_{eff}=17.6$ and
$F(q^2\simeq -m_{\mu}^2) \simeq 0.54$ \cite{Bernabeu:1993ta},  with  the muon capture rate  $\Gamma_{\rm capt}\simeq(2.590 \pm 0.012) \times 10^6$ s$^{-1}$ \cite{Suzuki:1987jf}. 
This is the case considered in our analysis. An upper bound has  been established by the SINDRUM II Collaboration also for the gold target:  $\mathcal{B}(\mu^- \, {\rm Au} \to e^- \, {\rm Au} )\le 7 \times10^{-13}$\cite{SINDRUMII:2006dvw}. Its analysis requires a different set of nuclear input parameters, that
can be inferred from \cite{Kitano:2002mt,Suzuki:1987jf}. The results of our study do not change  if  the experimental bound  and the parameters for the gold target are used.

\end{itemize}

\subsection{$(g-2)_{e, \mu}$}
The $Z^\prime$ contribution to   the muon anomalous magnetic moment  $\dd a_\mu=\frac{(g-2)_\mu}{2}$ has been obtained in \cite{Aebischer:2019blw}
using the general formulae  in \cite{Lindner:2016bgg}:
\be
\Delta a_\mu  \simeq  -\frac{1}{16\pi^2}\frac{m_\mu^2}{M^2_{Z^\prime}}
\sum_{f=e,\mu,\tau}
\left[\left|\Delta^{f\mu}_V\right|^2\left(\frac{2}{3}-\frac{m_f}{m_\mu}\right)+\left|\Delta^{f\mu}_A \right|^2\left(\frac{2}{3}+\frac{m_f}{m_\mu}\right)\right] \,\, . \label{damu}
\ee
 $\Delta^{ij}_{V (A)}$ are  defined in (\ref{DVA}), and the sum is over internal charged leptons in the loop.   
 For the electron, $\Delta a_e $ is obtained  replacing $\mu$ by $e$ in \eqref{damu}.

\section{Numerical analysis, results and correlations } \label{numerics}
As mentioned in Sec.~\ref{abcd}, in the ABCD model various scenarios have been considered for the  $Z^\prime $ couplings to fermions  \cite{Aebischer:2019blw}.  One of them, the scenario A, consists in assuming  flavour violating couplings of $Z^\prime$  only with left-handed fermions,   obtained setting to zero  the parameters in \eqref{quarks}-\eqref{leptons}. 
 The present study is based on this minimal scenario.

The list of  the parameters consists of the four entries in   \eqref{4par}  and  of  the parameters of the CKM and PMNS matrices.
The coupling $g_{Z^\prime}$ is varied in the range $g_{Z^\prime} \in [0.01,\,0.1]$,  and two cases are considered for the $Z^\prime$ mass, $M_{Z^\prime}=1$ TeV and $M_{Z^\prime}=3$ TeV.
The four independent parameters of the CKM matrix are taken as $V_{us},\,\gamma,\,|V_{cb}|$ and $|V_{ub}|$, see appendix  \ref{AppCKM}.  The strategy adopted in the present study consists in
 varying $|V_{cb}|$ and $|V_{ub}|$ in the ranges determined by  the exclusive and inclusive measurements,
 \be
 |V_{cb}| \in \big[ |V_{cb}|_{exc},\,|V_{cb}|_{inc} \big] \,\, , \qquad |V_{ub}| \in \big[ |V_{ub}|_{exc},\,|V_{ub}|_{inc} \big] \,\,,
 \label{VcbVubRanges}
 \ee
 instead of choosing some preferred value.
All values are  quoted in the appendix \ref{AppCKM}, together with  the  parameters of the PMNS matrix.
\begin{figure}[b!]
\begin{center}
\includegraphics[width = 0.40\textwidth]{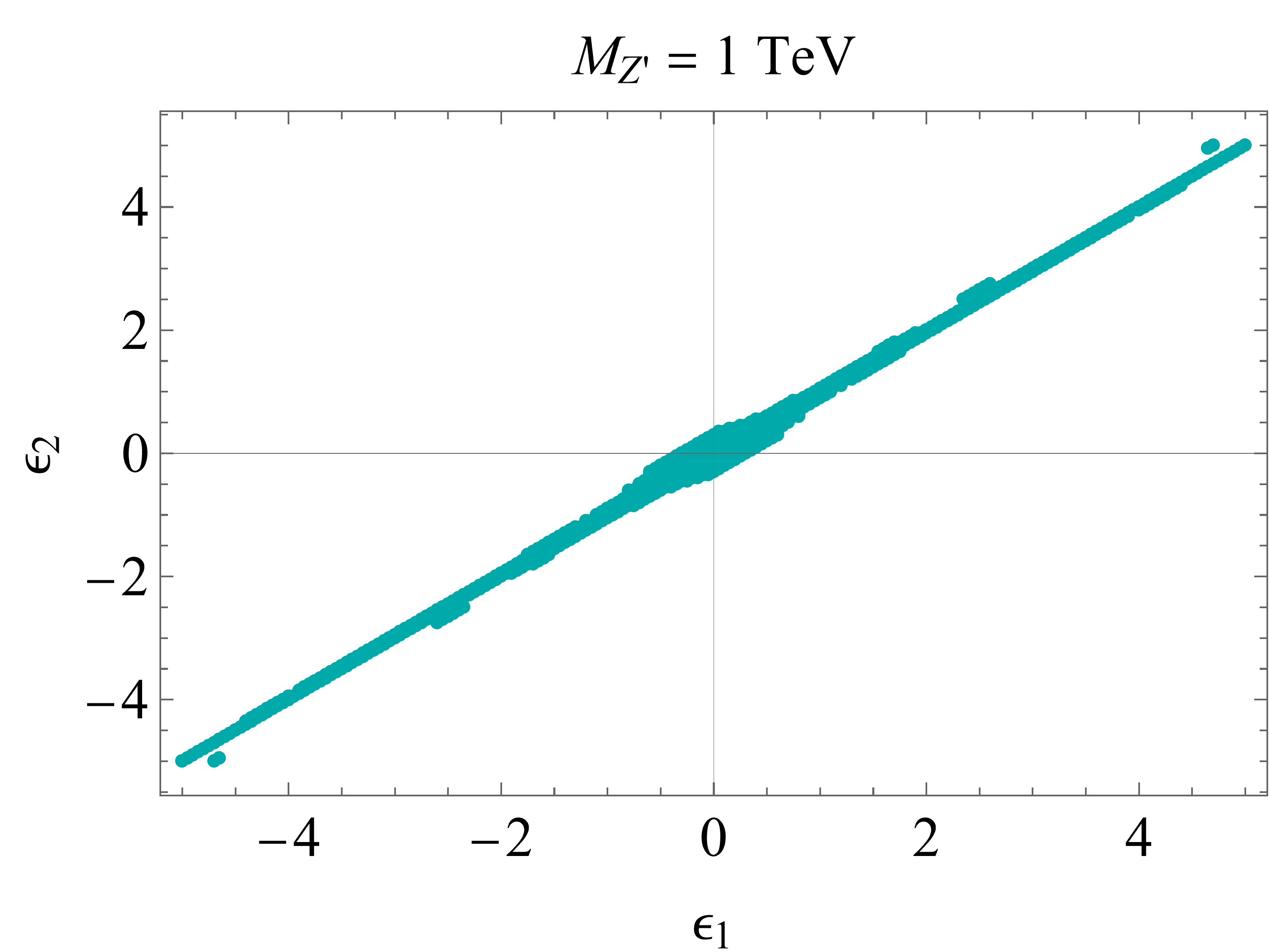} 	\hskip 0.5cm
\includegraphics[width = 0.40\textwidth]{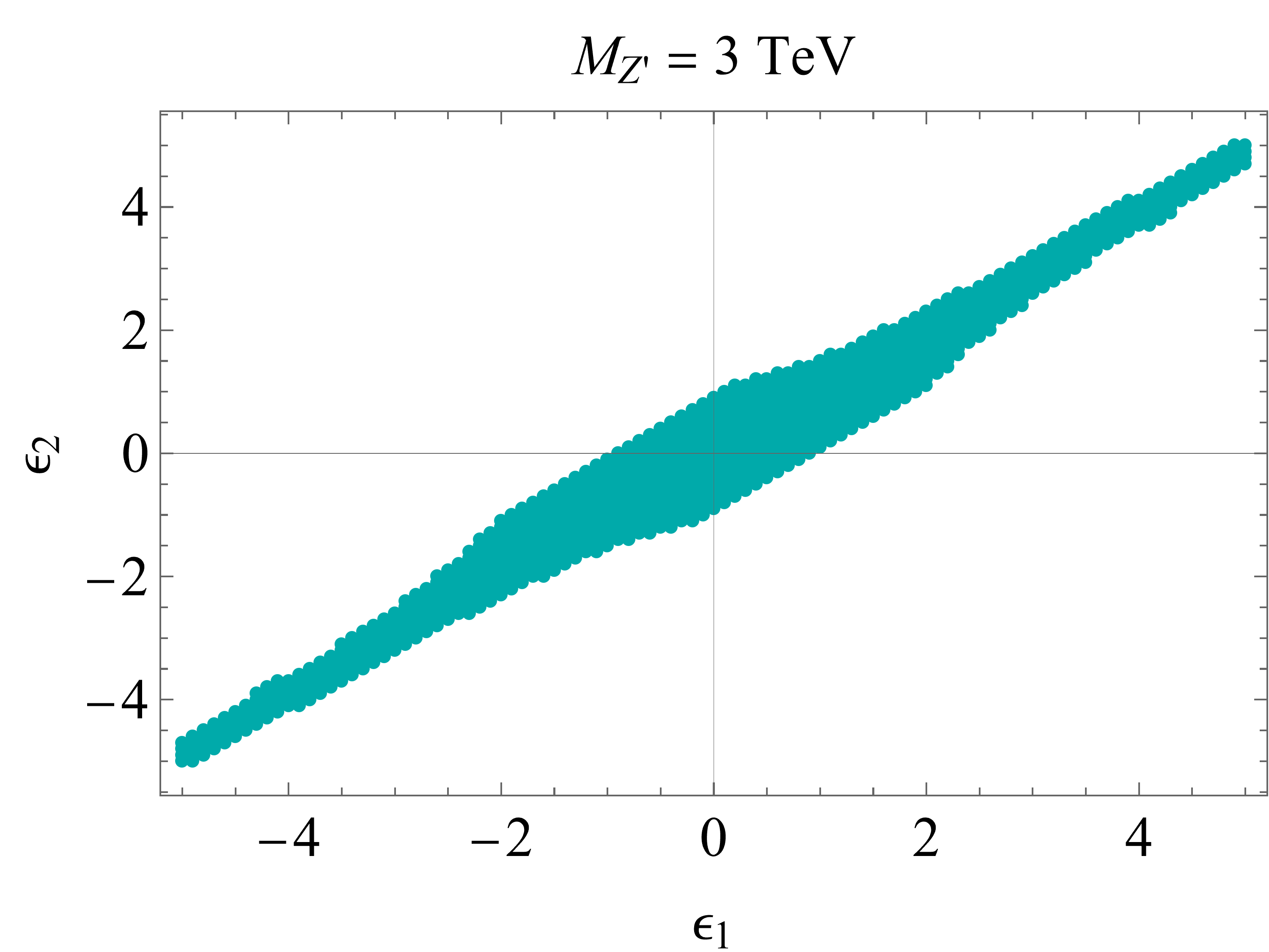}
\caption{\baselineskip 10pt  \small  Allowed region  in the $(\epsilon_1,\,\epsilon_2)$ plane after imposing $\Delta F=2$ constraints,
for   $M_{Z^\prime}=1$ TeV (left panel) and $M_{Z^\prime}=3$ TeV (right panel).}\label{fig:epspar}
\end{center}
\end{figure}
\begin{figure}[bh!]
\begin{center}
\includegraphics[width =0.60 \textwidth]{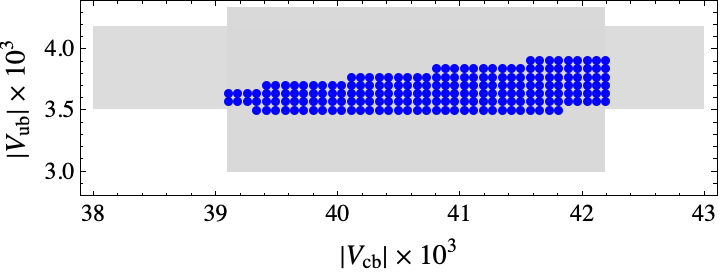}
\caption{\baselineskip 10pt  \small  Allowed region in the $|V_{cb}|,\,|V_{ub}|$ plane after imposing the $\Delta F=2$ constraints  (dark blue points)  for $M_{Z^\prime}=1$ TeV and $3$ TeV.  The light shaded area corresponds to the ranges  \eqref{VcbVubRanges}.  }\label{fig:ckm}
\end{center}
\end{figure}
The parameter space $(\epsilon_1, \epsilon_2)$ is constrained  using  $\Delta F=2$ observables,  modifying the SM expressions  to  account for  the $Z^\prime$  contribution.
Such observables are the mass differences $\Delta M_d$, $\Delta M_s$, $\Delta M_K$ between the neutral mesons $B_d$, $B_s$ and kaons, the CP asymmetries $S_{\psi K_s}$, $S_{\psi \phi}$ and the CP violating parameter in the kaon sector $\epsilon_K$.
All  quantities are related to the neutral meson  $M-{\bar M}$  mixing ($M = K^0,\,B_{d,s}$), which in  SM  only appears at
the one-loop level due to box diagrams mediated by internal up-type quark exchanges \cite{Buras:2020xsm}.
In the $B_d$ and $B_s$ cases the dominant contribution is  by the intermediate top quark,  for kaon the charm exchange must also be taken into account. The SM result depends on the Inami-Lim function $S_0(x_t)$ for $B_{d,s}$, and on $S_0(x_c,\,x_t)$ for kaon. Moreover, the GIM mechanism allows to get rid of mass independent terms. 
In the ABCD model, neutral meson mixings receive a tree-level contribution
from the $Z^\prime$ exchange. This  leads to a modification of the  Inami-Lim function,  $ S(M)=S_0+\Delta S(M)$ for $M = K^0,\,B_{d,s}$,
with $\Delta S(M)$ given in  \cite{Buras:2012jb}.
The  ABCD parameters  enter in $\Delta S(M)$ through the $Z^\prime$ couplings to fermions, which allows to effectively  constrain  the parameter space  by the precise experimental values for such observables. 
We proceed as in \cite{Aebischer:2019blw}:
 $\Delta M_d$, $\Delta M_s$ and the CP asymmetries $S_{J\psi K_s}$, $S_{\phi \psi}$ are required to stay  in a range around $5\%$ of the central value of the experimental datum,   $\Delta M_K$ is required to lie within 25$\%$ of the SM value,  accounting for possible long-distance contributions,  and   $\epsilon_K$  in the range $\epsilon_K\in[2.0,\,2.5]\times 10^{-3}$.

The allowed regions in the  $(\epsilon_1,\,\epsilon_2)$ plane  are displayed in Fig.~\ref{fig:epspar} for  the two values of $M_{Z^\prime}$.  The regions are  broader than those found in \cite{Aebischer:2019blw}, a consequence of   scanning  $|V_{cb}|$ and $|V_{ub}|$ in the exclusive-inclusive ranges instead of choosing a preferred value. 

A  remarkable result, shown in Fig.~\ref{fig:ckm}, is that the scan over the parameter space subject to the $\Delta F=2$ constraints does not yield any configuration with $|V_{ub}|$ equal to the inclusive determination. The largest obtained value  is $ |V_{ub}|=3.92 \times 10^{-3}$ (within the resolution set by the scan step), which indicates a preference for smaller values. This agrees with the conclusions of the  discussion in  \cite{Blanke:2018cya} on the sensitivity of $\Delta F=2$ observables to the precise values of the CKM parameters.

 After having imposed the $\Delta F=2$ constraints, we first look for deviations in lepton-flavour-conserving (LFC) $B$ and $B_s$
decays and for the correlations with the corresponding lepton-flavour-violating (LFV) modes. The LFC processes provide non-trivial constraints on the LFV ones. 

We then study  the correlations with purely leptonic LFV processes, in particular with  $\tau^- \to \mu^- \mu^+ \mu^-$, $\mu^- \to e^- \gamma$, $\mu^- \to e^- e^+ e^-$ and $ \mu^- \to e^-$ conversion in nuclei, obtaining that the experimental upper bounds on such purely leptonic channels play a hierarchical role in constraining the branching ratios of LFV $B$ and $B_s$ decays.  The bound on $\tau^- \to \mu^- \mu^+ \mu^-$ does not impose significant restrictions, whereas the other three channels set progressively more stringent limits.
 The correlations with $\tau^- \to \mu^- \mu^+ \mu^-$, $\mu^- \to e^- \gamma$  allow branching ratios for the LFV beauty meson decays of ${\cal O}(10^{-9})$.  Including constraints from $\mu^- \to e^- e^+ e^-$ and $ \mu^- \to e^-$ conversion in nuclei, the branching fractions are pushed 
down to  ${\cal O}(10^{-11})$ and ${\cal O}(10^{-13})$, as discussed below.

\subsection{ $b \to s \ell_1^+ \ell_2^-$}
In  SM where  only  $\ell_1= \ell_2$  is allowed, the values of the Wilson coefficients of the operators $O_7$ and $O_{9,10}$ in the low-energy Hamiltonian \eqref{hamil} read \cite{EOSAuthors:2021xpv}: 
\be
C_7^{\rm eff}(m_b)=-0.450 \,\, , \hskip 1cm C_9(m_b) =4.273 \,\, , \hskip 1 cm C_{10}(m_b) =-4.166 \,\,, \label{C7910}
\ee
with $C_7^{\rm eff}$  defined in Sec.~\ref{section-hamil}.
Efforts have been  spent in global analyses to identify  the coefficients deviating from  the SM value, and to which extent,  to reproduce the experimental data \cite{Altmannshofer:2014rta,Capdevila:2017bsm,Alguero:2021anc,Alguero:2023jeh}.
\begin{figure}[th!]
\begin{center}
\includegraphics[width =0.35 \textwidth]{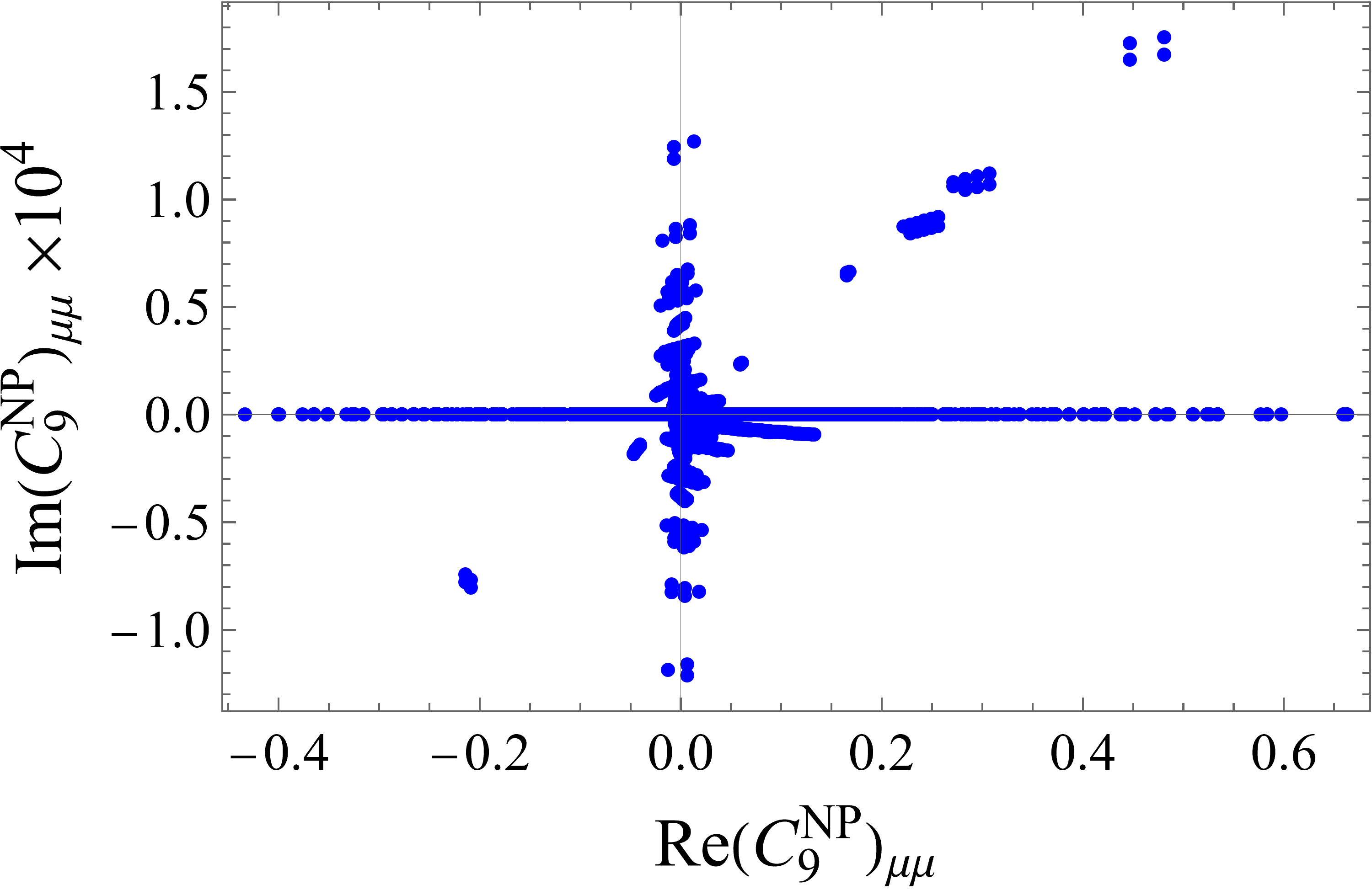}\hskip 0.4cm
\includegraphics[width =0.35 \textwidth]{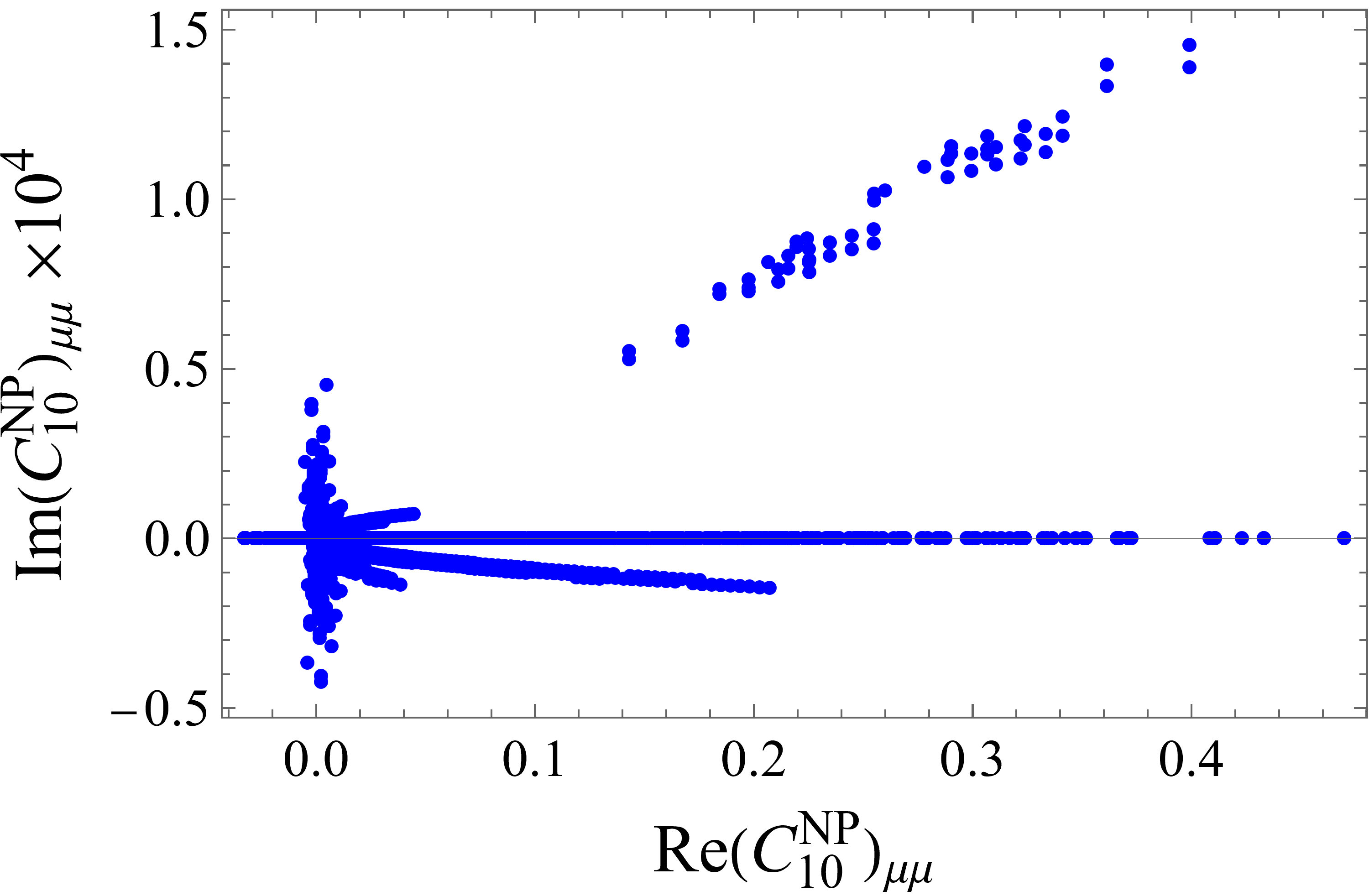}\\
\vskip 0.2cm
\includegraphics[width =0.35 \textwidth]{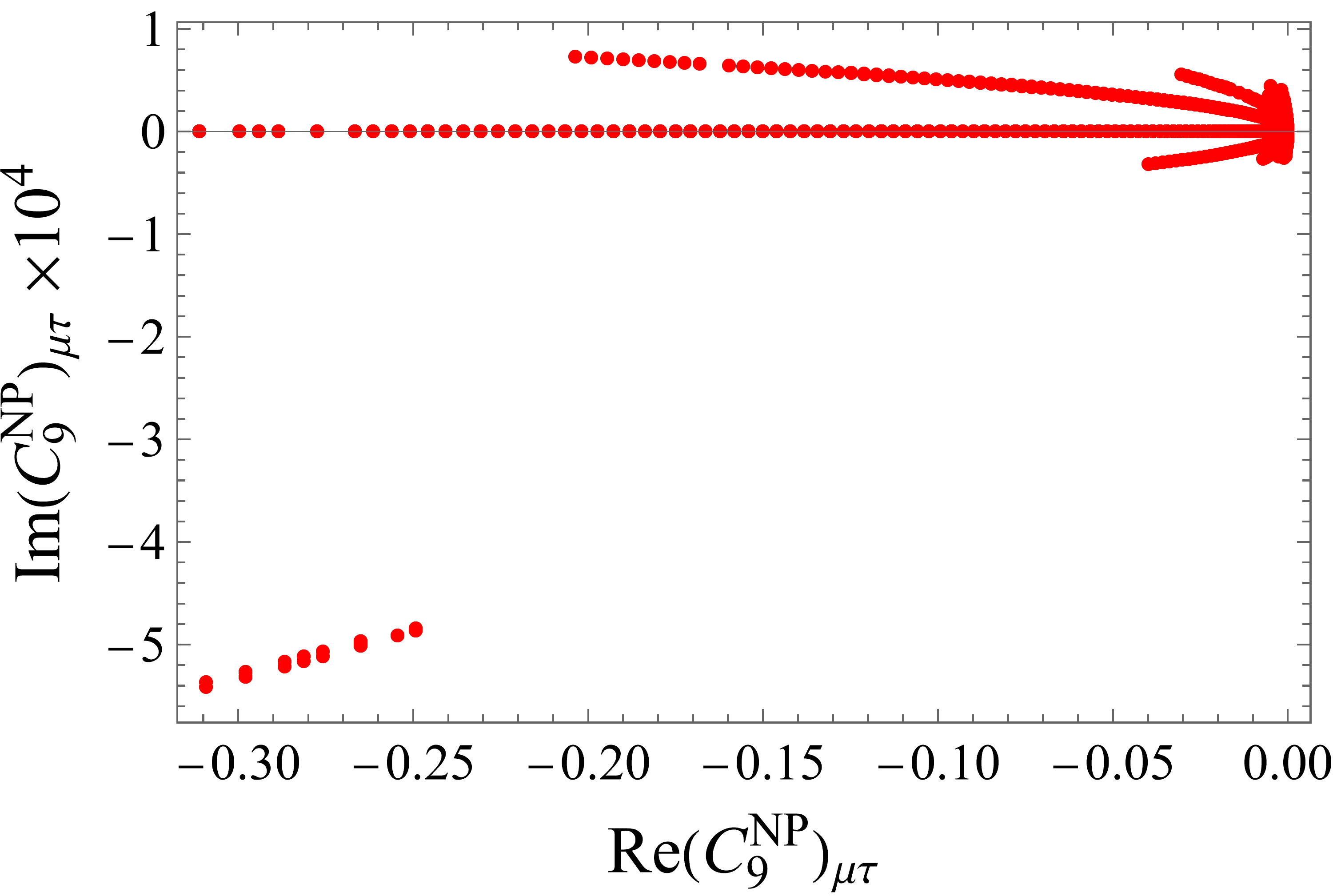}\hskip 0.4cm
\includegraphics[width =0.35 \textwidth]{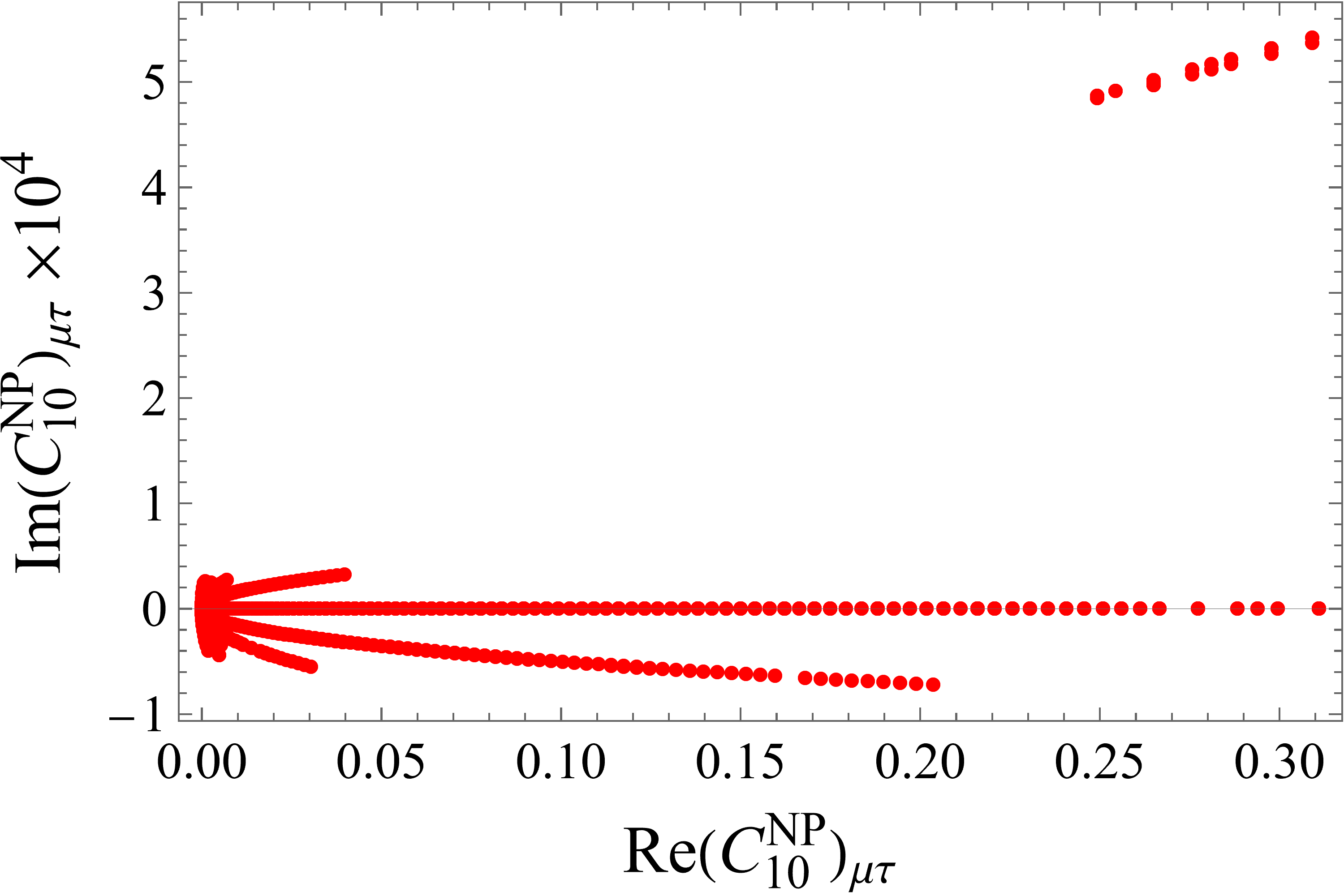}
\caption{\baselineskip 10pt  \small  Correlation between real and  imaginary parts of the $Z^\prime$ contributions  to the Wilson coefficients $C_9$ (left panels) and $C_{10}$ (right panels) for the  LFC  $b \to s \mu^- \mu^+$  (top panels) and  LFV   $b \to s \mu^- \tau^+$ transition (bottom panels). The $Z^\prime$ mass is set to $M_{Z^\prime}=1$  TeV. }\label{Wilson1000}
\end{center}
\end{figure}
\begin{figure}[hb!]
\begin{center}
\includegraphics[width =0.35 \textwidth]{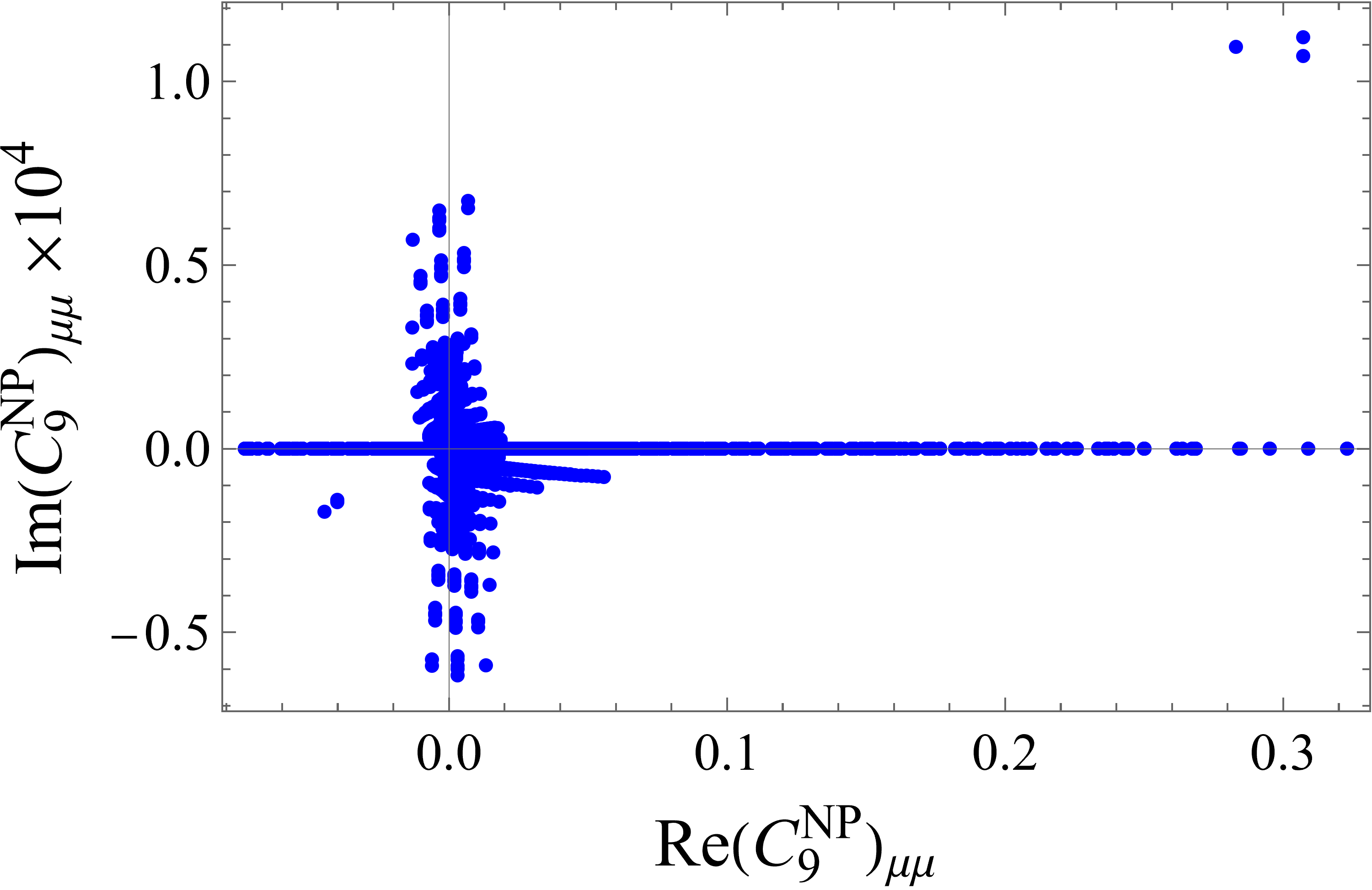}\hskip 0.4cm
\includegraphics[width =0.35 \textwidth]{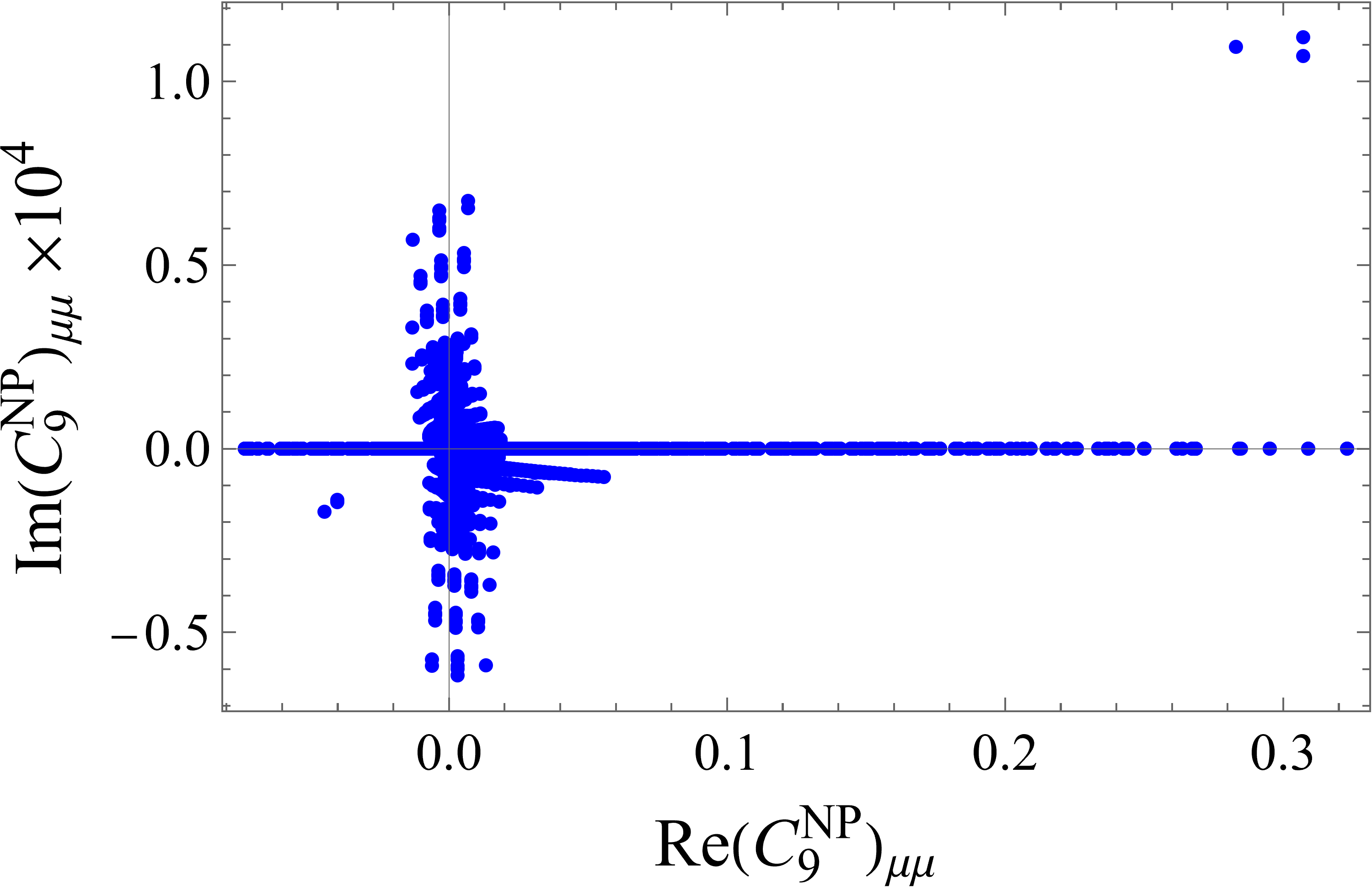}\\
\vskip 0.2cm
\includegraphics[width =0.35 \textwidth]{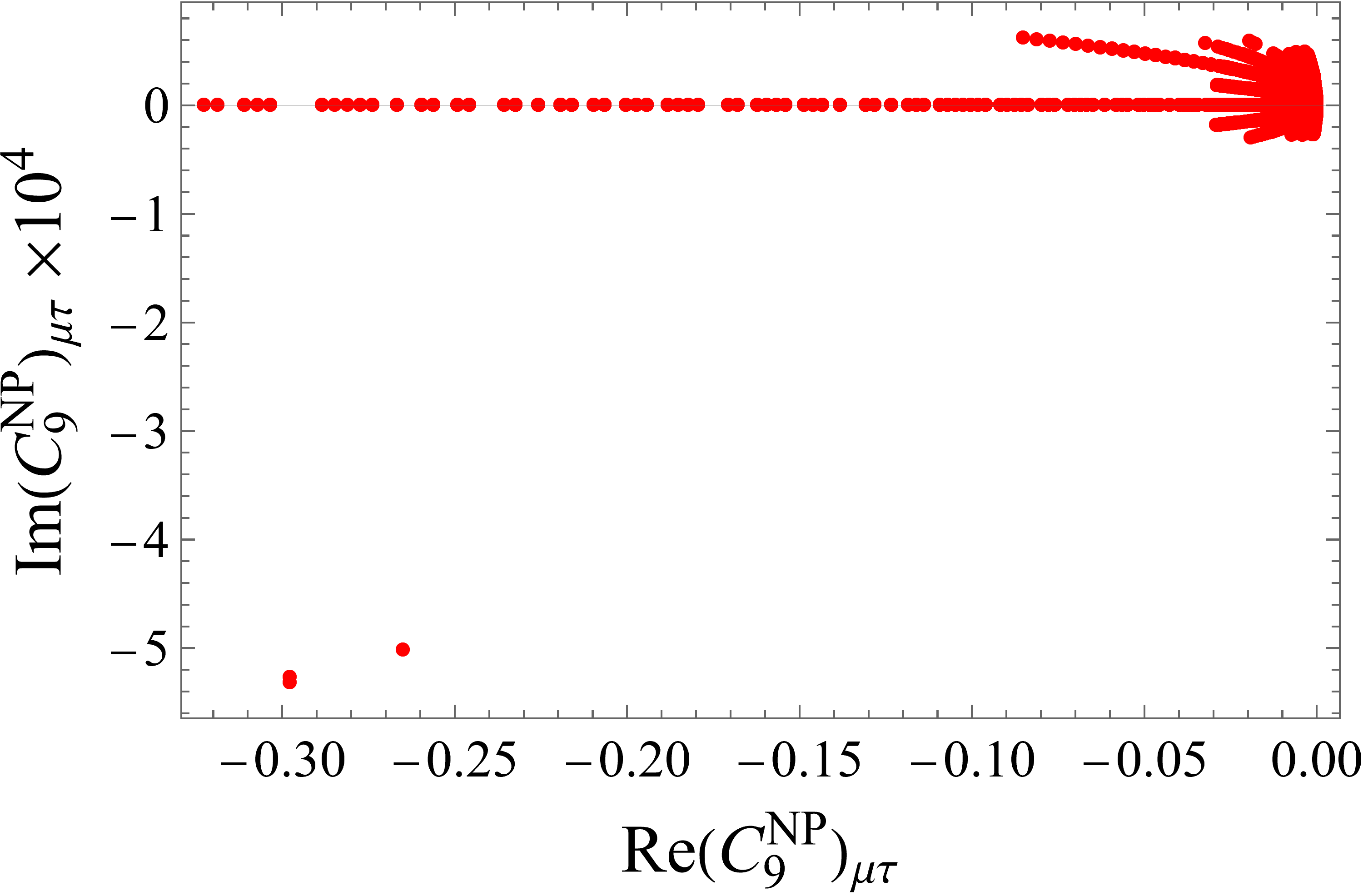}\hskip 0.4cm
\includegraphics[width =0.35 \textwidth]{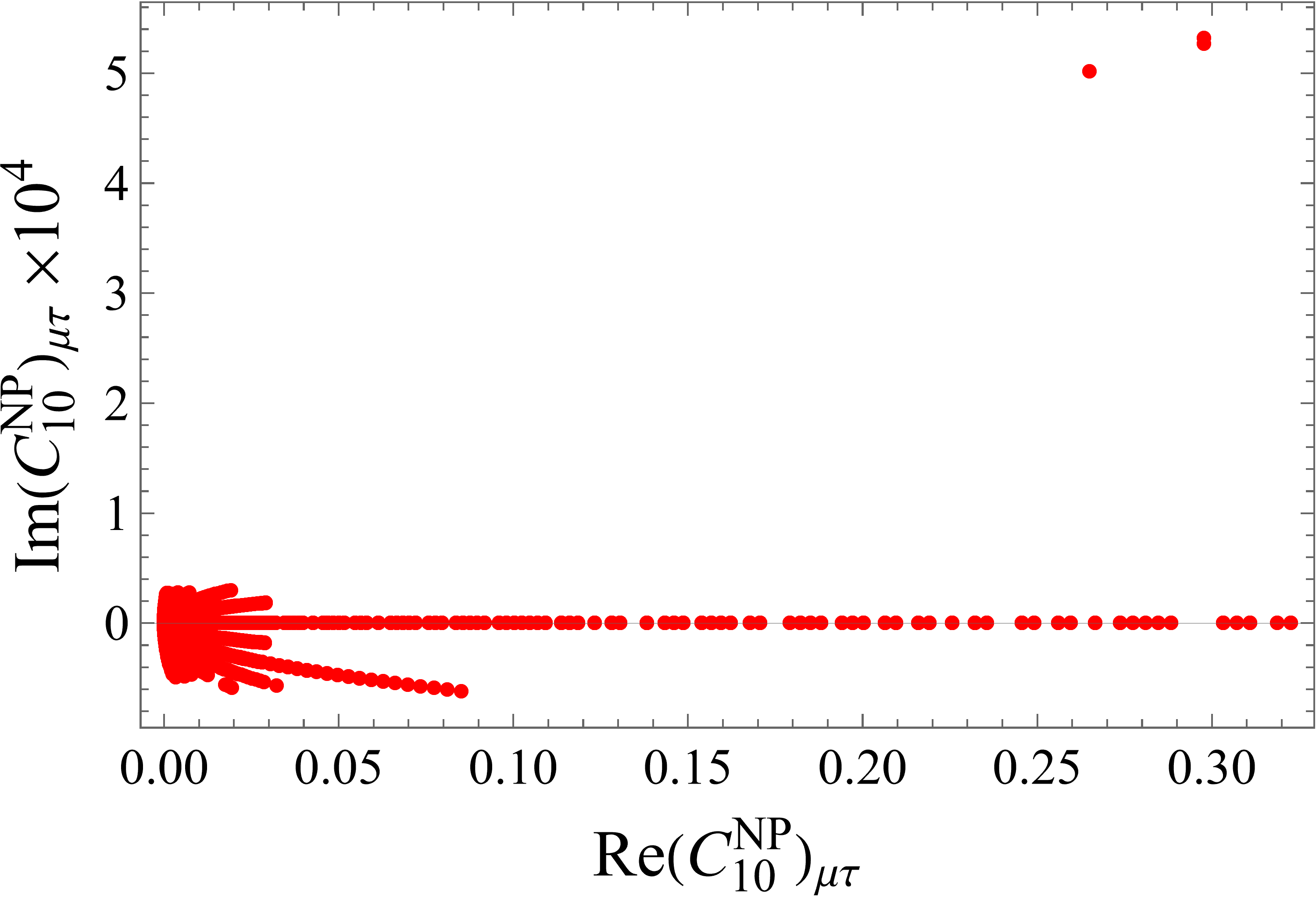}
\caption{\baselineskip 10pt  \small  Correlation between real and imaginary parts of the $Z^\prime$ contributions to the Wilson coefficients $C_9$ and $C_{10}$ as in Fig.~\ref{Wilson1000}, for  $M_{Z^\prime}=3$ TeV.  }\label{Wilson3000}
\end{center}
\end{figure}
In the  scenario A of the ABCD model  there are no contributions to the coefficients of the primed operators in \eqref{hamil}. The  $Z^\prime$ contribution to $C_7$ is negligible with respect to  SM    \cite{Buras:2012jb}.    The contributions to $C_9$ and $C_{10}$  for $M_{Z^\prime}=1$ TeV  and $3$ TeV are displayed in Figs.~\ref{Wilson1000} and \ref{Wilson3000}, where  each point corresponds to a pair real/imaginary part of the coefficients computed for the same values of the parameters.   The contributions to  the real parts can be sizeable, up to $\pm 10 \%$ of the SM values.
 Imaginary parts of ${\cal O} (10^{-4})$ are generated,  originating from the phases of the CKM and PMNS matrices. The correlations between the real  parts of the coefficients are displayed in the plots in Fig.~\ref{Wilsonmix}  for $M_{Z^\prime}=1$  and $3$ TeV.

The values of $C_9^{NP}$ and $C_{10}^{NP}$ obtained for the lepton flavour violating modes $b \to s \mu^- \tau^+$, together with their correlation, are displayed in the same figures \ref{Wilson1000},\ref{Wilson3000} and \ref{Wilsonmix}.  
The anticorrelation between  $(C_9^{NP})_{\mu \tau}$ and $(C_{10}^{NP})_{\mu \tau}$  is a feature of the scenario A of the ABCD model. Indeed, the expressions of the coefficients are
\be
(C_9^{NP})_{\ell_1 \ell_2} = - \frac{1}{g_{SM}^2 s_W^2 M_{Z^\prime}^2\lambda_{ts}^*}  \big(\Delta_L^{bs}\big)^* \Delta_V^{\ell_1 \ell_2} \,\, , \qquad
(C_{10}^{NP})_{\ell_1 \ell_2} = - \frac{1}{g_{SM}^2 s_W^2 M_{Z^\prime}^2\lambda_{ts}^*}  \big(\Delta_L^{bs}\big)^* \Delta_A^{\ell_1 \ell_2} \,\, , \label{C9eC10NP}
\ee
with $\dd g_{SM}^2= \frac {4 G_F}{\sqrt 2} \frac{e^2}{8 \pi^2 s_W^2}$, $s_W$ the sine of the Weinberg angle,  and $\Delta_{V(A)}^{\ell_1 \ell_2}$ defined in \eqref{DVA}. In the scenario A where the $Z^\prime$ flavour violating  couplings to right-handed fermions vanish,  we have  $\Delta_{V}^{\ell_1 \ell_2}=-\Delta_{A}^{\ell_1 \ell_2}$ for $\ell_1 \neq \ell_2$.
\begin{figure}[h]
\begin{center}
\includegraphics[width =0.35 \textwidth]{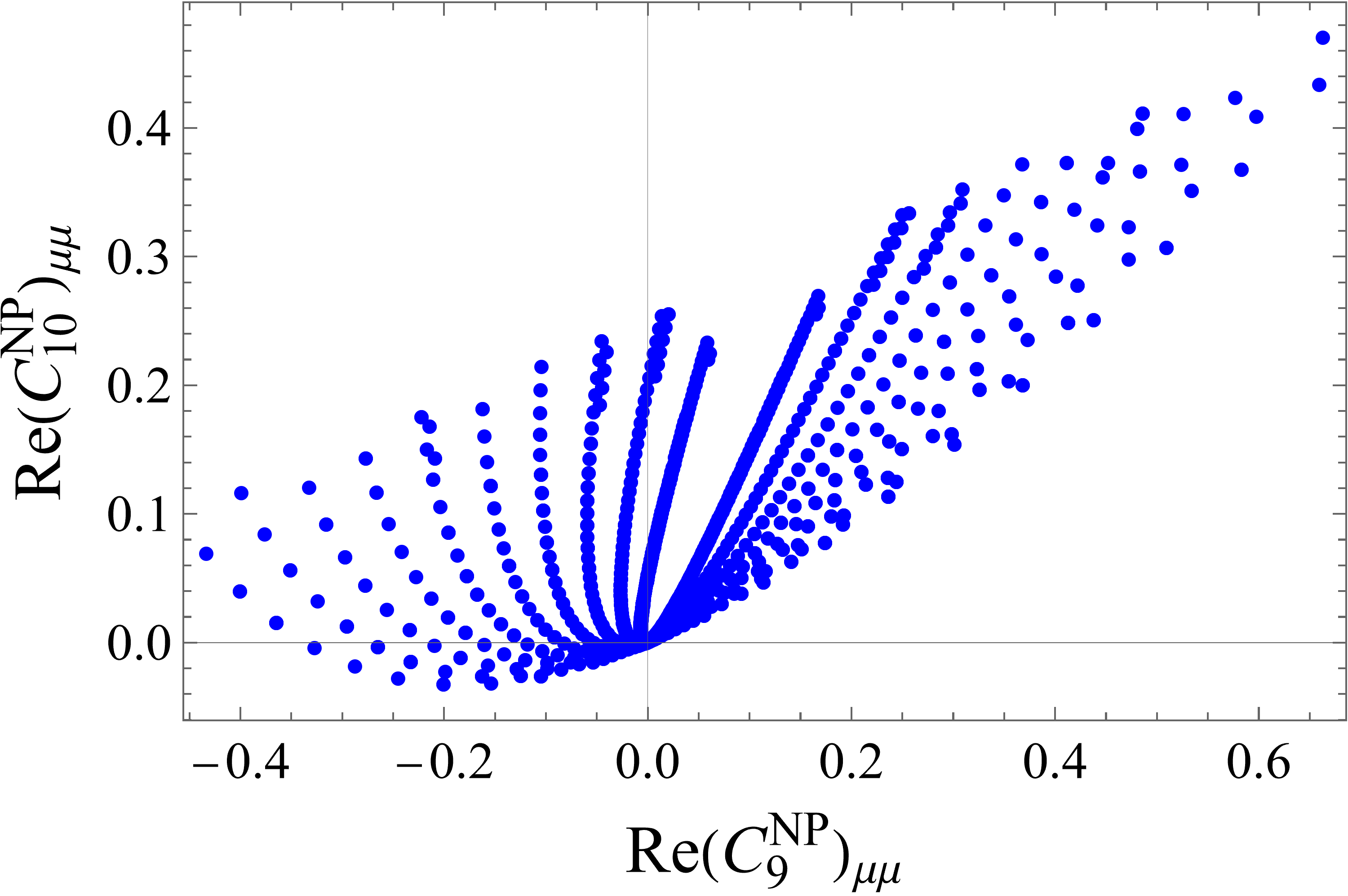}\hskip 0.4cm
\includegraphics[width =0.35 \textwidth]{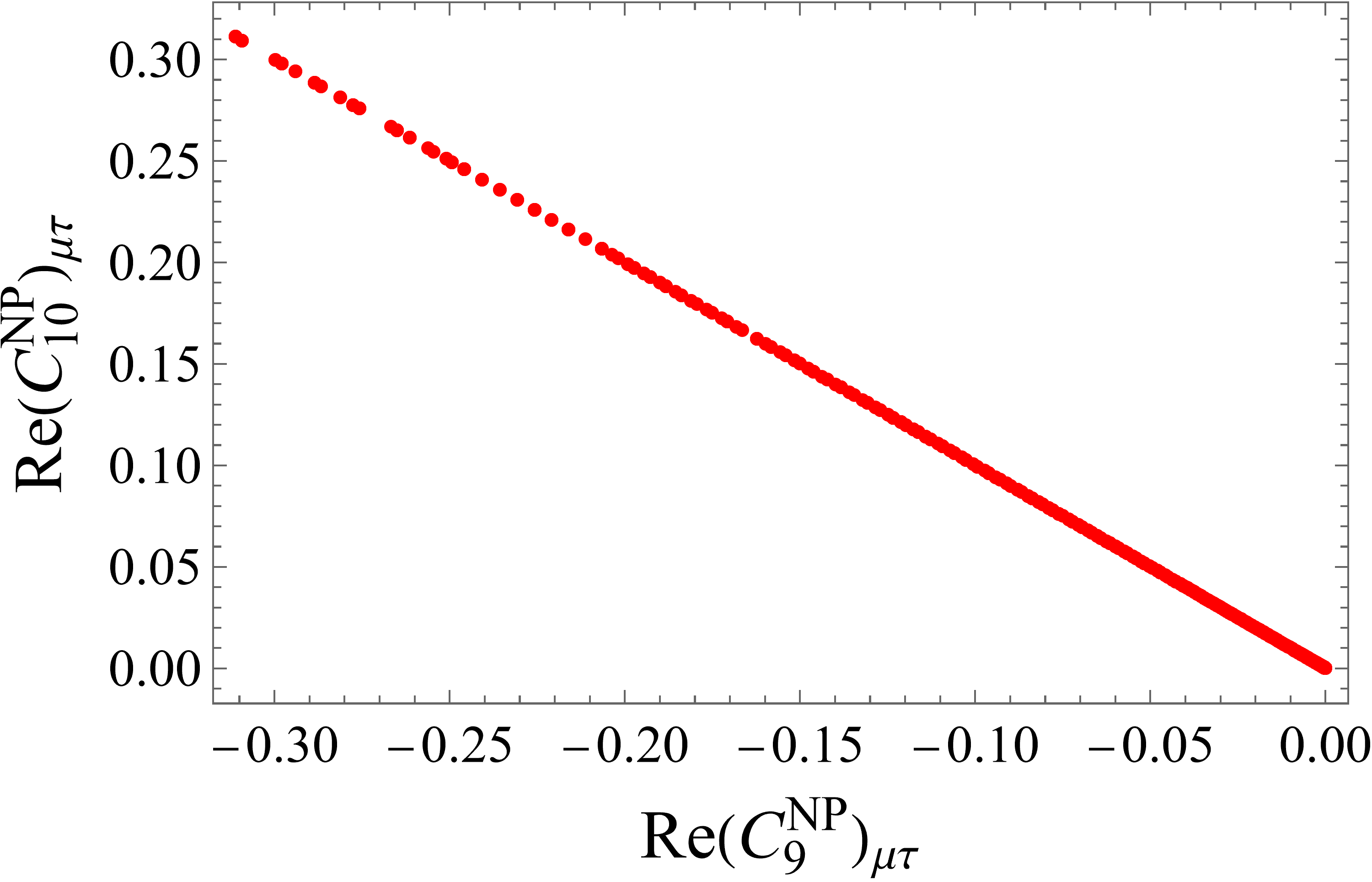}
\\
\vskip 0.2cm
\includegraphics[width =0.35 \textwidth]{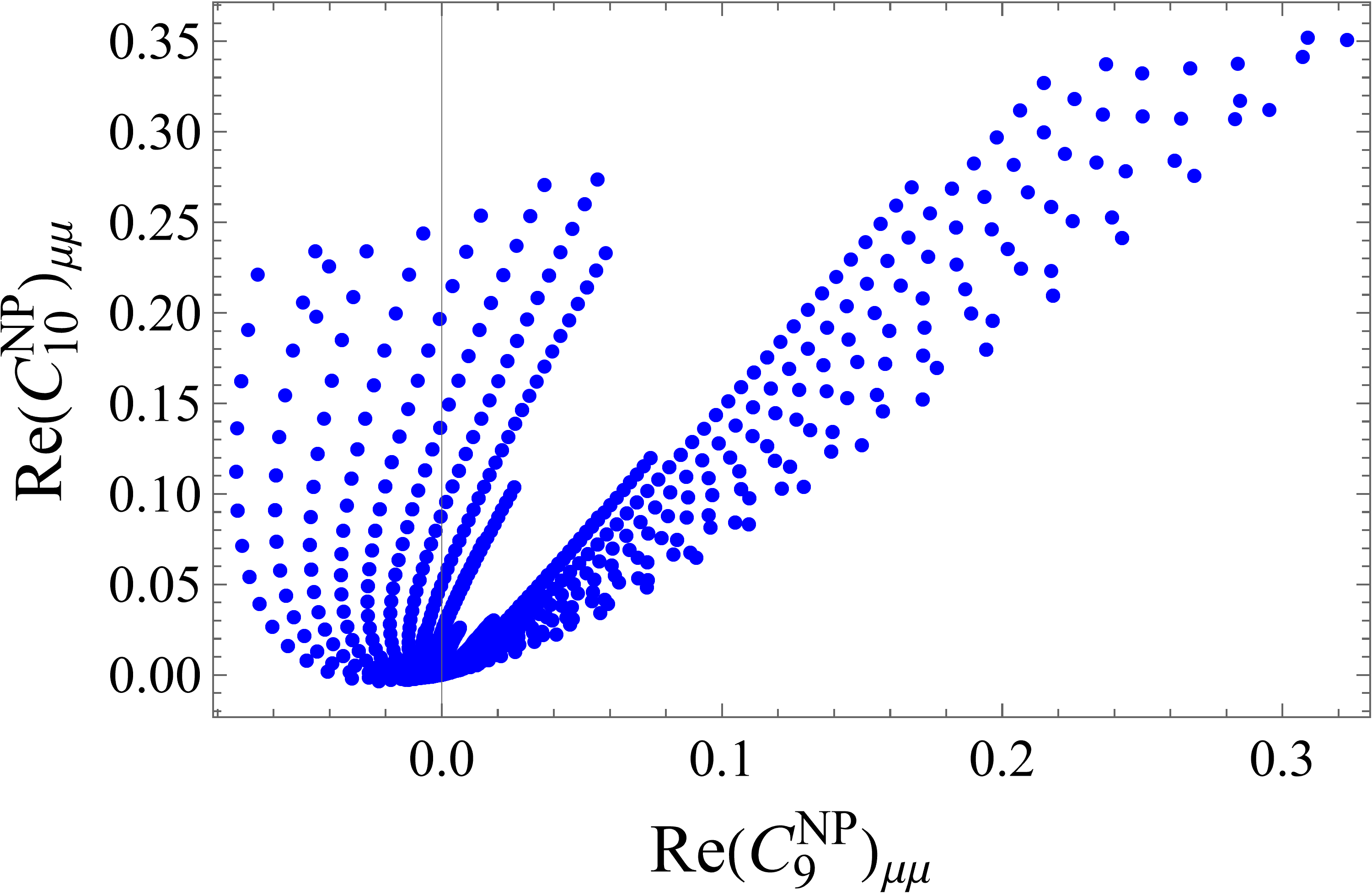}\hskip 0.4cm
\includegraphics[width =0.35 \textwidth]{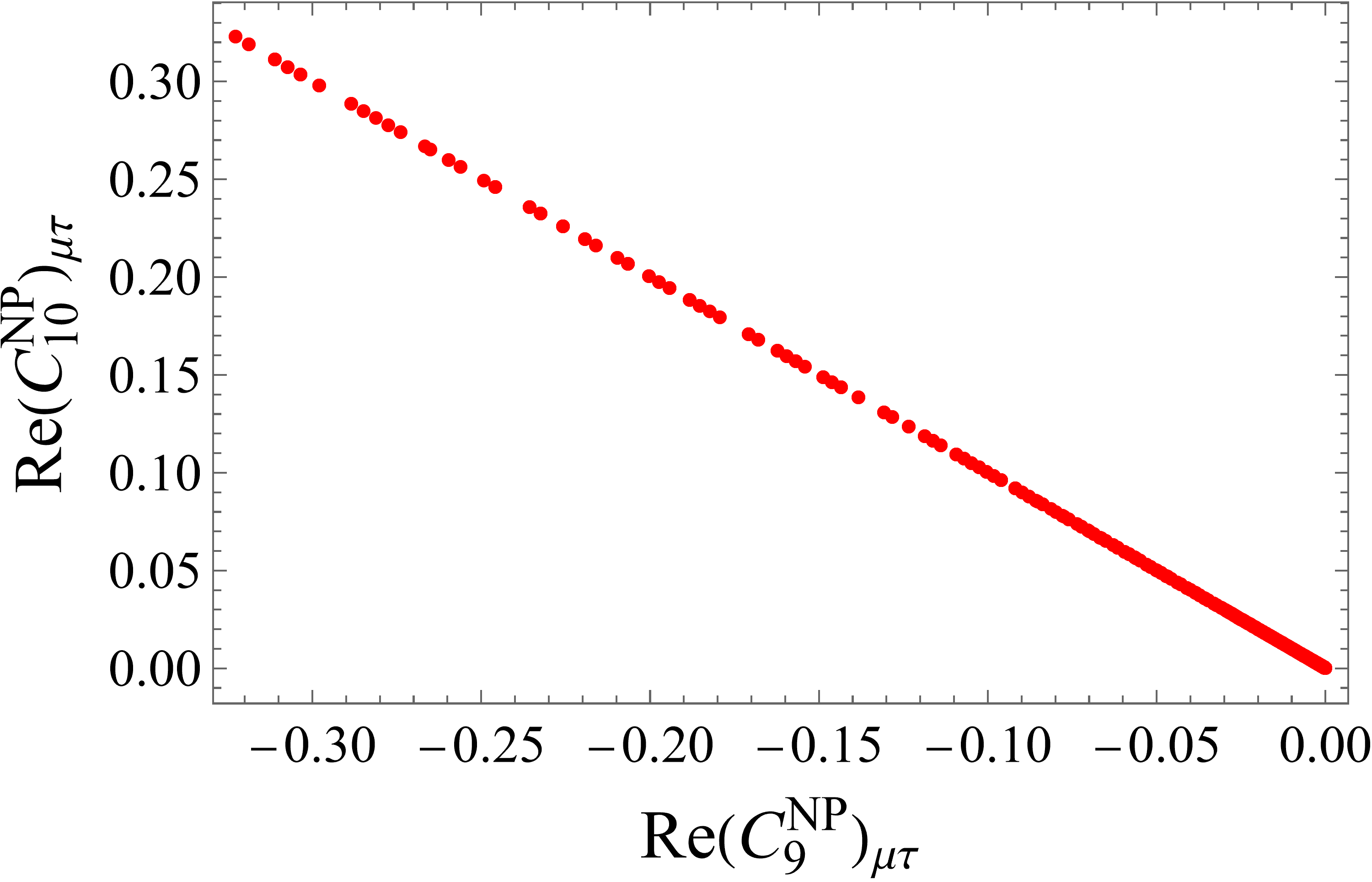}
\caption{\baselineskip 10pt  \small  Correlation between  the real  parts   of the $Z^\prime$  contributions to Wilson coefficients $(C_9)_{\mu \mu}$ and $(C_{10})_{\mu \mu}$ (left panels)  and $(C_9)_{\mu \tau}$ and $(C_{10})_{\mu \tau}$ (right panels) for $M_{Z^\prime}=1$ TeV (top panels) and for $M_{Z^\prime}=3$ TeV(bottom panels).  }\label{Wilsonmix}
\end{center}
\end{figure}
\subsubsection{Exclusive lepton flavour conserving  $b \to s \ell^+ \ell^-$ modes}
\begin{figure}[t]
\begin{center}
\includegraphics[width =0.4 \textwidth]{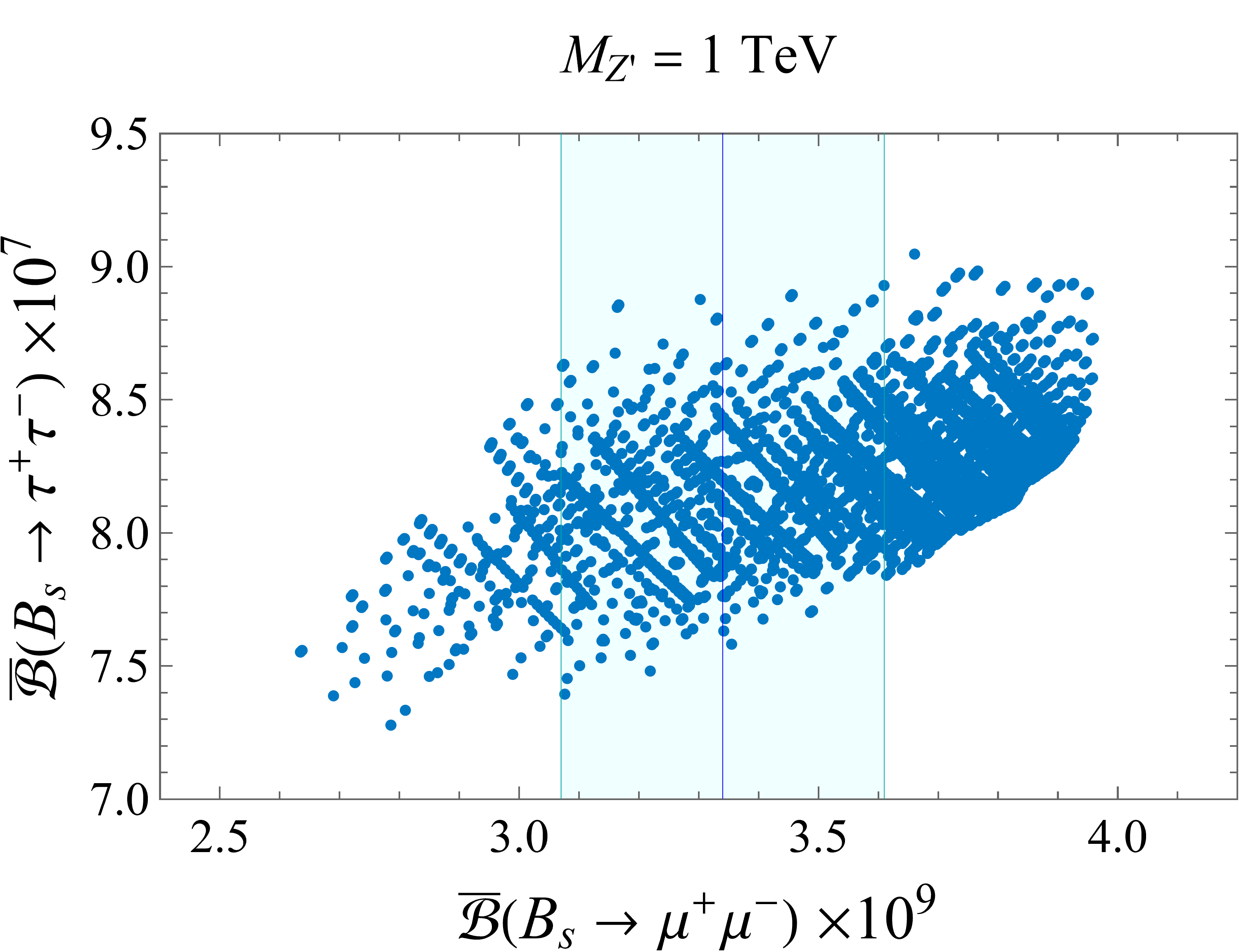}
\hskip .4cm 
\includegraphics[width =0.4 \textwidth]{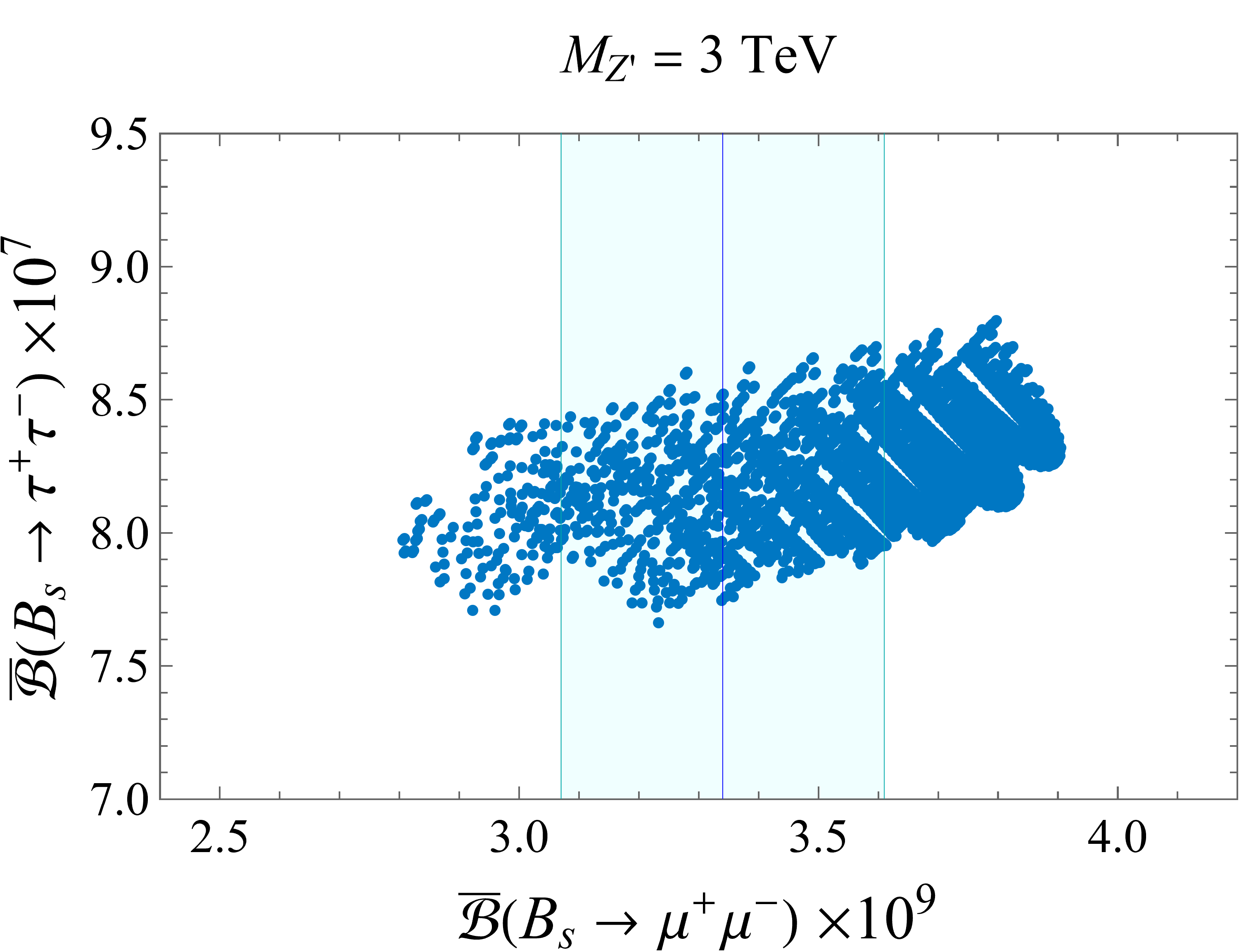}
\caption{\baselineskip 10pt  \small   Correlation between ${\bar {\cal B}}(B_s \to \mu^+ \mu^-)$ and ${\bar {\cal B}}(B_s \to \tau^+ \tau^-)$ for $M_{Z^\prime}=1$ TeV (left) and  $M_{Z^\prime}=3$ TeV (right panel).  The shaded  region  corresponds to the experimental world average for ${\bar {\cal B}}(B_s \to \mu^+ \mu^-)$\cite{Navas:PDG}.}\label{BsLFC}
\end{center}
\end{figure}

The simplest process  induced by  $b \to s \ell^+ \ell^-$  is the purely leptonic $B_s$ decay, that we analyse using $f_{B_s}=230.3 \pm 1.3$ MeV \cite{HeavyFlavorAveragingGroupHFLAV:2024ctg}.
Fig.~\ref{BsLFC}  shows the branching fractions ${\bar {\cal B}}(B_s \to \mu^+ \mu^-)$ and ${\bar {\cal B}}(B_s \to \tau^+ \tau^-)$  for $M_{Z^\prime}=1$  TeV and $3$ TeV.
 In SM we find  ${\bar {\cal B}}(B_s \to \mu^+ \mu^-)=(3.65 \pm 0.10)\times 10^{-9}\left(|\lambda_{ts}^*|/0.04 \right)^2$ and ${\bar {\cal B}}(B_s \to \tau^+ \tau^-)=(7.70 \pm 0.10)\times 10^{-7}\left(|\lambda_{ts}^*|/0.04 \right)^2$. The experimental world average  ${\bar {\cal B}}(B_s \to \mu^+ \mu^-)_{\rm exp}=(3.34 \pm 0.27) \times 10^{-9}$ \cite{Navas:PDG} is indicated in  
 Fig.~\ref{BsLFC}.  The results in the ABCD model encompass the experimental range. Moreover,  requiring that  ${\bar {\cal B}}(B_s \to \mu^+ \mu^-)$ agrees with experiment, ${\bar {\cal B}}(B_s \to \tau^+ \tau^-)=(8.2 \pm 0.5) \times 10^{-7}$ is predicted for $M_{Z^\prime}=1$  TeV, and  ${\bar {\cal B}}(B_s \to \tau^+ \tau^-)=(8.2 \pm 0.8) \times 10^{-7}$ for $M_{Z^\prime}=3$ TeV.   The  correlation between the two purely leptonic modes is shown in the figure, where each point corresponds to the branching fractions of $B_s \to \mu^+ \mu^-$ and $B_s \to \tau^+ \tau^-$ computed with the same set of parameters.
 Fig.~7  shows that when ${\bar {\cal B}}(B_s \to \mu^+ \mu^-)$  is predicted within its  experimental range, the branching ratio ${\bar {\cal B}}(B_s \to \tau^+ \tau^-)$
is also constrained. This feature emerges from the numerical scan, and  can be  understood from the analytic structure of the amplitudes in scenario A. Indeed, 
Eq.~\eqref{Bslept}  shows that both branching ratios are proportional to $|C_{10}|^2$,
 \be
{\bar {\cal B}}(B_s \to \ell^+ \ell^-)=K_\ell \times |C_{10}|^2=K_\ell \times \left|C_{10}^{SM}+(C_{10}^{NP})_{\ell \ell} \right|^2 \,\,\, ,
 \ee
 with $K_\ell$  a  numerical factor  different for $\ell=\mu$ and $\ell=\tau$.
 From Eq.~\eqref{C9eC10NP} one  observes that $(C_{10}^{NP})_{\mu \mu}$ and $(C_{10}^{NP})_{\tau \tau}$ differ for the term $\Delta_A^{\ell \ell}$ that depends on the scanned  parameters. Using Eq.~\eqref{LHQD1} one finds 
 \be
 (C_{10}^{NP})_{\tau \tau} \simeq (C_{10}^{NP})_{\mu \mu}+ f(g_Z,\,\epsilon_1,\,\epsilon_2) \,\, , \label{C10NPnew}
 \ee
 where $f$ depends on the scanned parameters. Hence, one writes
 \be
 {\bar {\cal B}}(B_s \to \tau^+ \tau^-) \simeq \frac{\lambda^{1/2}(m_{B_s}^2,m_\tau^2,m_\tau^2)}{\lambda^{1/2}(m_{B_s}^2,m_\mu^2,m_\mu^2)}\left[{\bar {\cal B}}(B_s \to \mu^+ \mu^-)+{\tilde f}(g_Z,\,\epsilon_1,\,\epsilon_2) \right] \,\, ,  \label{corr}
 \ee
 where ${\tilde f}(g_Z,\,\epsilon_1,\,\epsilon_2)$ denotes the terms  taking into account  the function $f$ in Eq.~\eqref{C10NPnew}. This  relation shows that the experimental bound on ${\bar {\cal B}}(B_s \to \mu^+ \mu^-)$ also  limits ${\bar {\cal B}}(B_s \to \tau^+ \tau^-)$.

For the  lepton flavour conserving  modes $B \to K^{*} \ell^+ \ell^-$, with $\ell=\mu,\,\tau$,  we use the local form factors  determined in \cite{Bharucha:2015bzk} and  described in appendix \ref{appA}.
\begin{figure}[b!]
\begin{center}
\includegraphics[width =0.4 \textwidth]{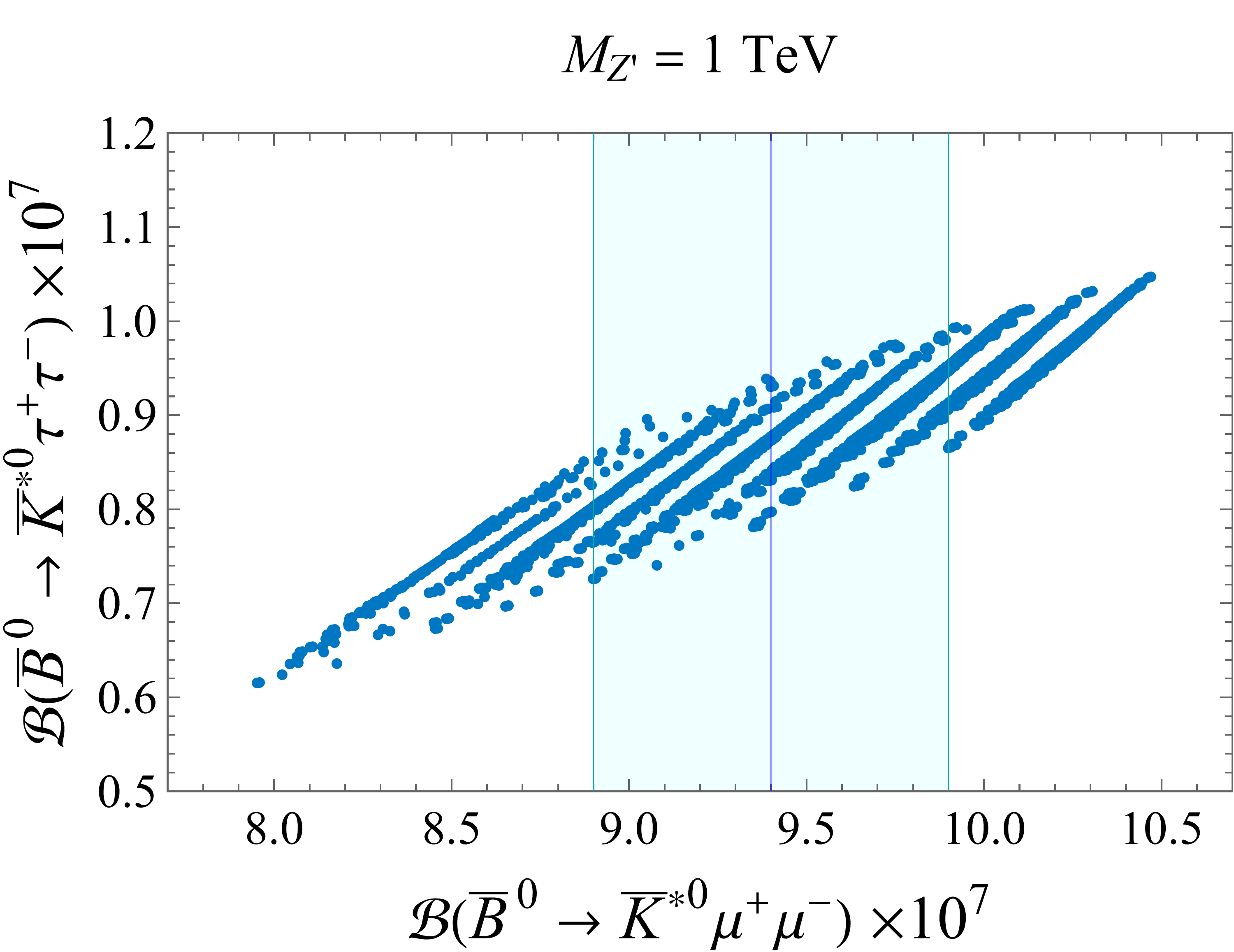}
\hskip .4cm
\includegraphics[width =0.4 \textwidth]{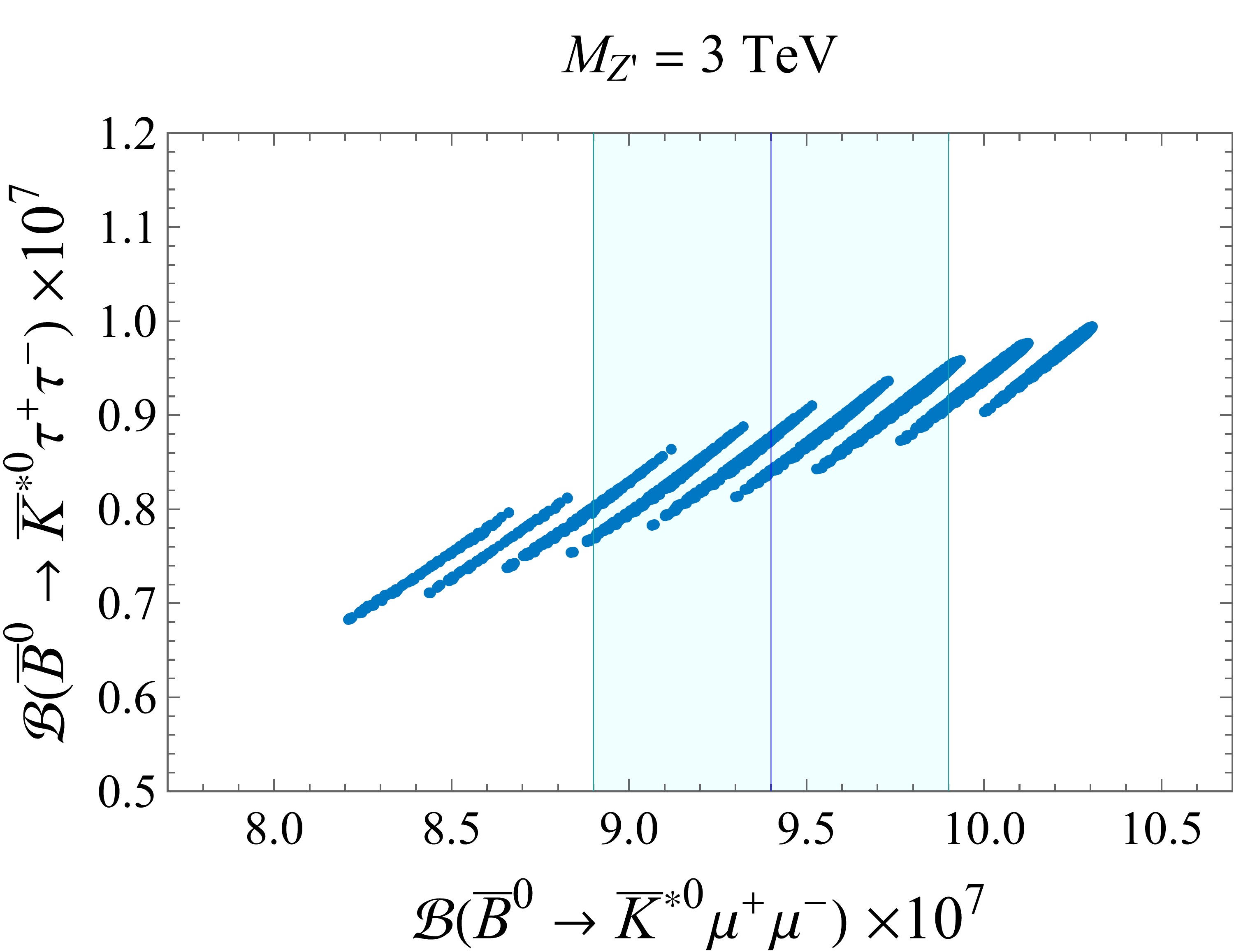}
\caption{\baselineskip 10pt  \small  Correlation between  ${\cal B}(\bar B^0 \to \bar K^{*0} \mu^+ \mu^-)$ and ${\cal B}(\bar B^0 \to \bar K^{*0} \tau^+ \tau^-)$ for $M_{Z^\prime}=1$ TeV (left panel) and $M_{Z^\prime}=3$ TeV (right panel).  The shaded  band corresponds to the experimental result  ${\cal B}(\bar B^0 \to \bar K^{*0} \mu^+ \mu^-)=(9.4 \pm 0.5) \times 10^{-7}$  \cite{Navas:PDG}.}\label{KstarLFC}
\end{center}
\end{figure}
Figure \ref{KstarLFC} shows the correlation between the branching fractions ${\cal B}(\bar B^0 \to \bar K^{*0} \mu^+ \mu^-)$ and ${\cal B}(\bar B^0 \to \bar K^{*0} \tau^+ \tau^-)$. The vertical band corresponds to the measurement ${\cal B}(\bar B^0 \to \bar K^{*0} \mu^+ \mu^-)_{\rm exp}=(9.4 \pm 0.5) \times 10^{-7}$ \cite{Navas:PDG}.
In the case of $B \to K^{*} \tau^+ \tau^-$ only an upper limit is known: ${\cal B}(\bar B^0 \to \bar K^{*0} \tau^+ \tau^-)<3.1 \times 10^{-3}$.
In the ABCD model  the result for  ${\cal B}(\bar B^0 \to K^{*0} \mu^+ \mu^-)$ encompasses the experimental interval. Requiring  that  it lies in the measured range, the predictions ${\cal B}(\bar B^0 \to \bar K^{*0} \tau^+ \tau^-)=( 8.6\pm 1.3)\times 10^{-8}$ and ${\cal B}(\bar B^0 \to \bar K^{*0} \tau^+ \tau^-)=( 8.6\pm 0.9)\times 10^{-8}$ are obtained for  $M_{Z^\prime}=1$ and $3$ TeV, respectively.
For the   SM we have: ${\cal B}(\bar B^0 \to \bar K^{*0} \mu^+ \mu^-)=(9.6 \pm 2.0) \times 10^{-7}\left(|\lambda_{ts}^*|/0.04 \right)^2$ and ${\cal B}(\bar B^0 \to \bar K^{*0} \tau^+ \tau^-)=(9.2 \pm 2.0)\times 10^{-8}\left(|\lambda_{ts}^*|/0.04 \right)^2$.

\begin{figure}[t]
\begin{center}
\includegraphics[width = \textwidth]{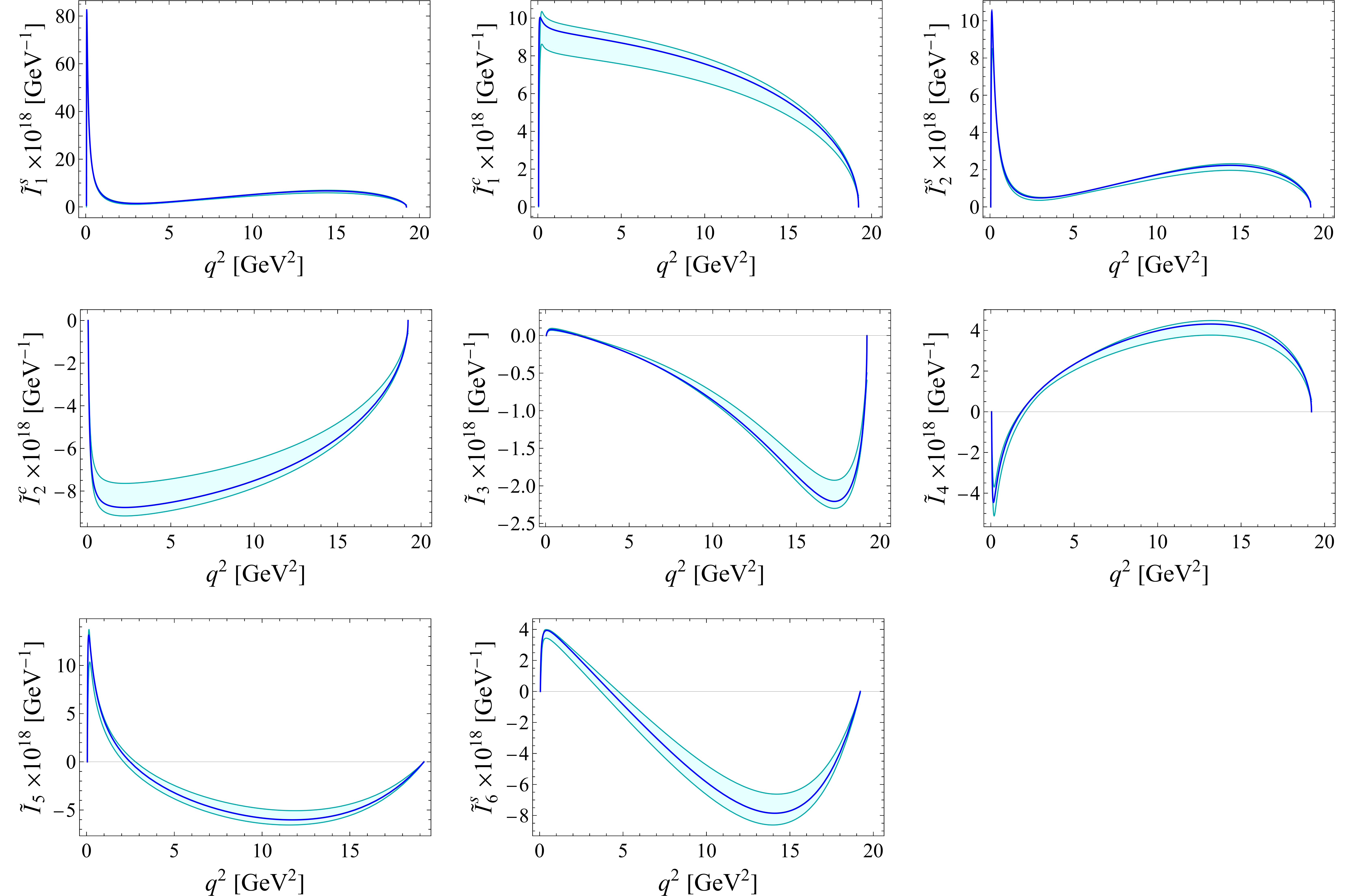}
\caption{\baselineskip 10pt  \small  Angular coefficient functions $\tilde I_i^{(a)}(q^2)$ 
for $\bar B^0 \to \bar K^{*0} (K \pi) \mu^+ \mu^-$. The continuous line is the SM result for the central value of the form factor parameters,
the light shaded region  the result in the ABCD model varying the parameters in the allowed ranges, with   $M_{Z^\prime}=1$ TeV.}\label{AngMu1000}
\end{center}
\end{figure}

\begin{figure}[t]
\begin{center}
\includegraphics[width = \textwidth]{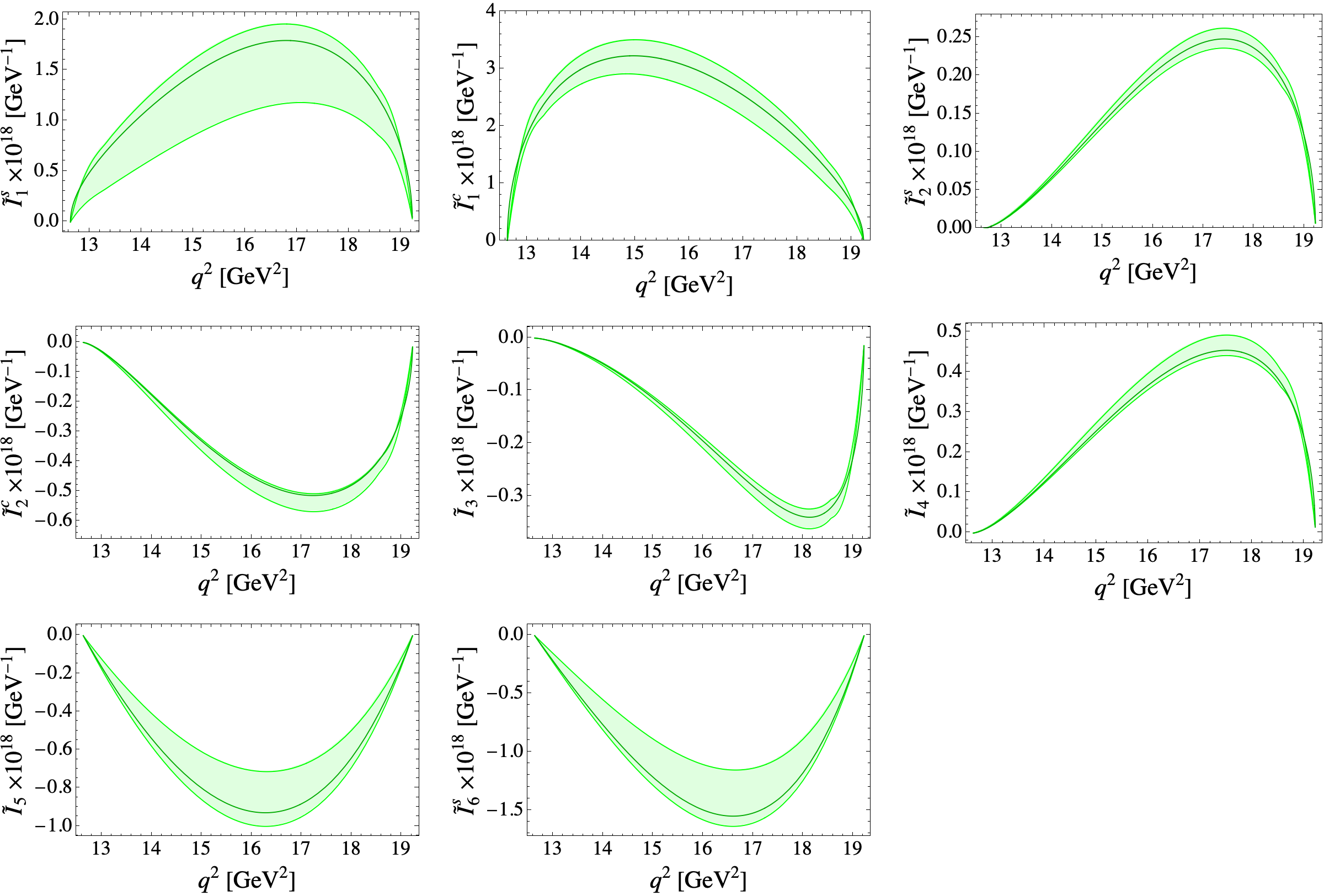}
\caption{\baselineskip 10pt  \small  Angular coefficient functions for   $\bar B^0 \to \bar K^{*0} (K \pi) \tau^+ \tau^-$. The continuous line is the SM result for the central value of the form factor parameters,
the light shaded region  the result in the ABCD model varying the parameters in the allowed ranges, with   $M_{Z^\prime}=1$ TeV.}\label{AngTau1000}
\end{center}
\end{figure}

The  $B \to K^*(K \pi)  \mu^+ \mu^-$ angular distribution  has been  experimentally  studied  in details \cite{BaBar:2006tnv,BaBar:2015wkg,Belle:2009zue,Belle:2016fev,CDF:2011tds,CMS:2015bcy,CMS:2017rzx,ATLAS:2018gqc}, finding a deviation  from SM predictions in some angular $P$-observables defined in \eqref{Pobs}  \cite{LHCb:2013zuf,LHCb:2013ghj,LHCb:2015svh,LHCb:2020lmf,LHCb:2020gog,LHCb:2023gpo,LHCb:2024onj}.
In the analysis of  the ABCD model predictions,  to make the comparison with  SM more transparent, we define the rescaled angular coefficient functions  ${\tilde I}_i=I_i/|\lambda_{ts}^*|^2$,  with $i=1s, \dots 6s$, and display them in Figs.~\ref{AngMu1000}, \ref{AngTau1000}   for  $\ell=\mu$  and $\ell=\tau$, setting $M_{Z^\prime}=1$ TeV (for $M_{Z^\prime}=3$ TeV the results do not  change sensibly).
The  functions $I_{6c,\,7,\,8,\,9}$ vanish in  SM;   in the ABCD model they are   about $10^{-6}$ times smaller than all others,  therefore we do not display them.
In the figures, the dark continuous line is the SM result for the central values of the form factor parameters,  the light shaded band  the ABCD result.   Modest deviations with respect to SM are possible, they are more pronounced in the case of  $\tau$.

\begin{figure}[t]
\begin{center}
\includegraphics[width =0.3 \textwidth]{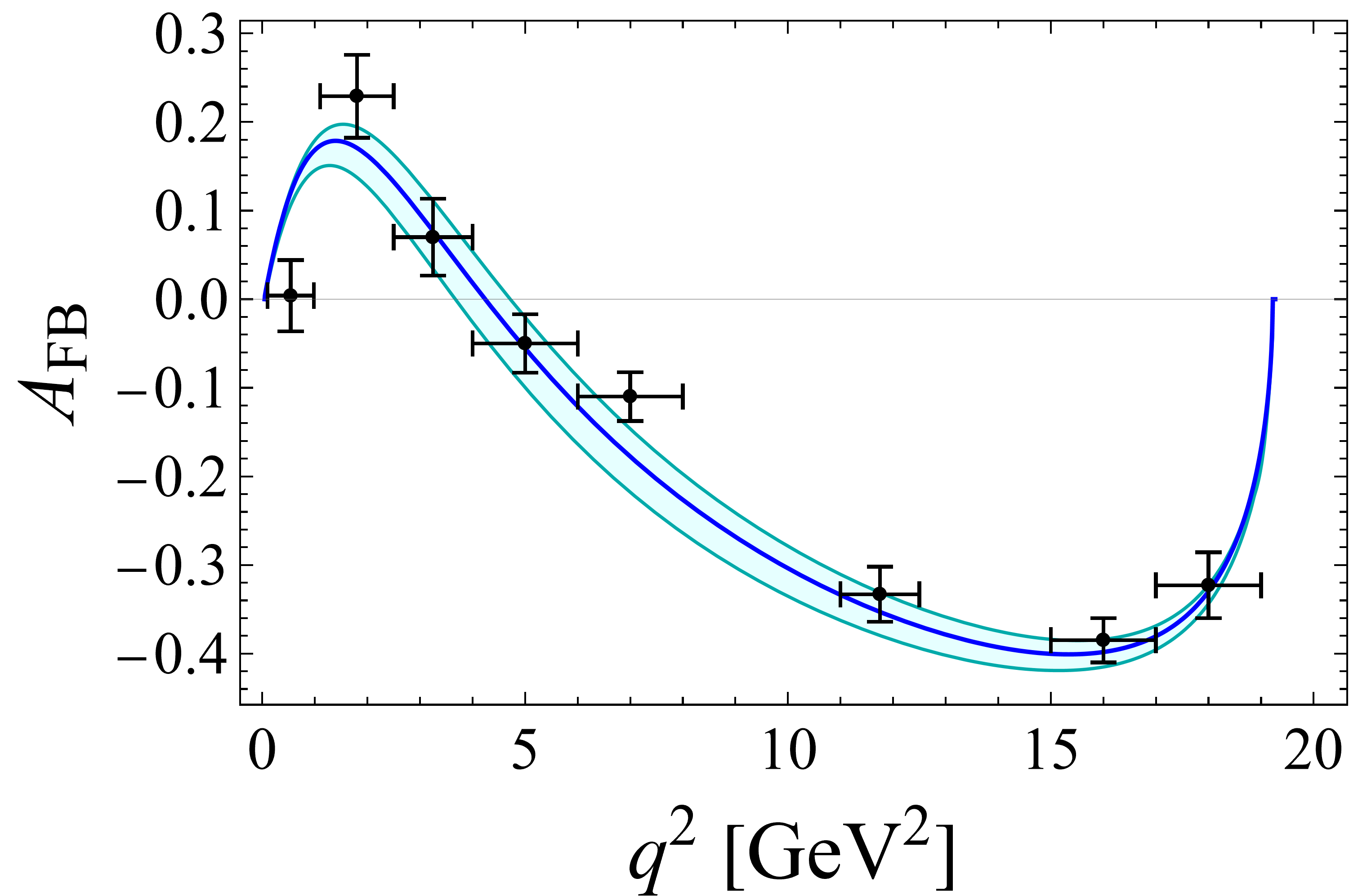}
\hskip 0.5cm \includegraphics[width =0.3 \textwidth]{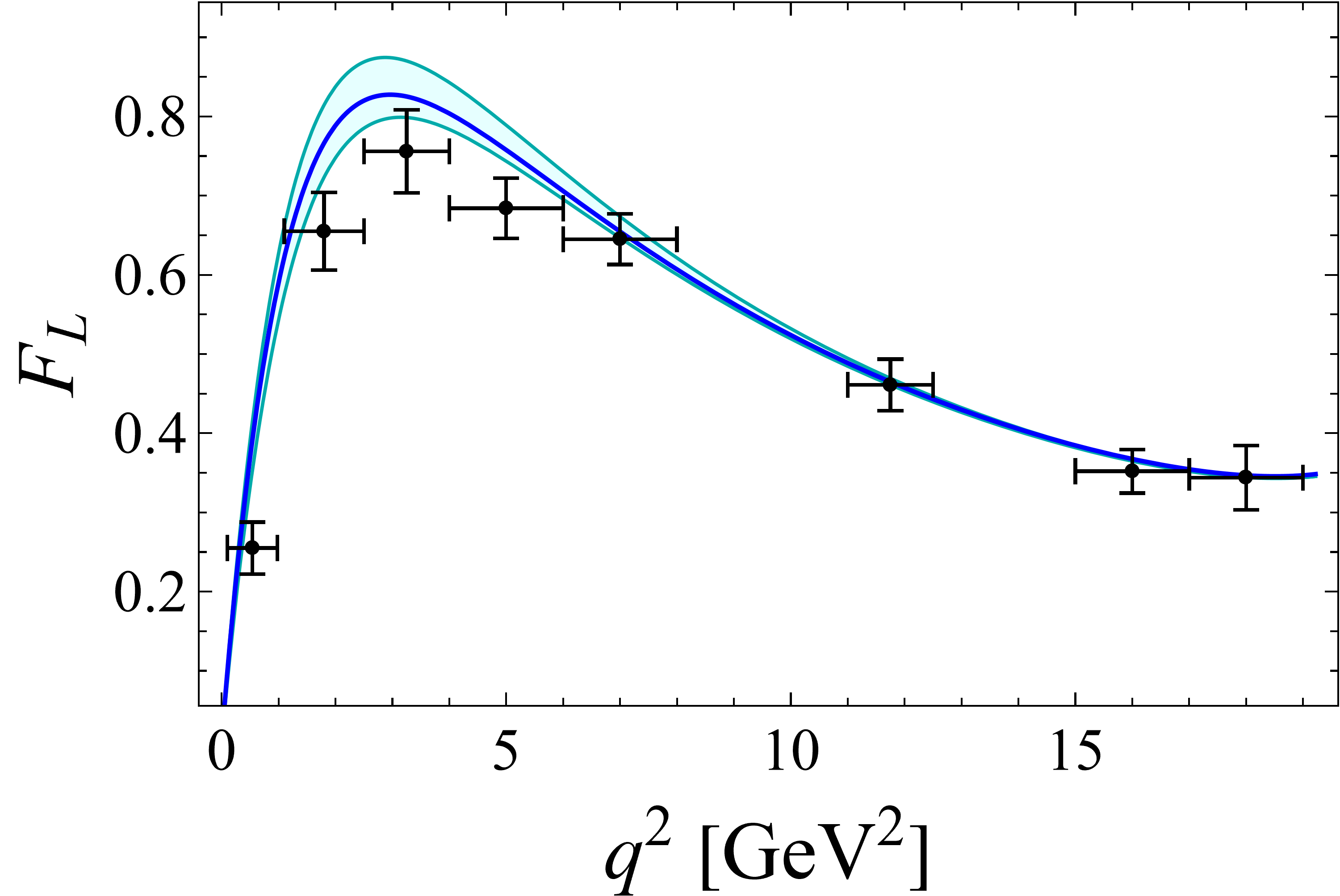}\\ \vskip 0.5cm
\includegraphics[width =0.31 \textwidth]{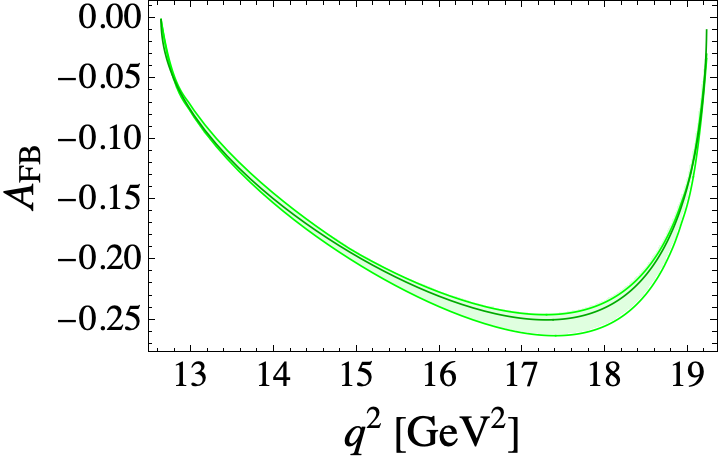}\hskip 0.5cm 
\includegraphics[width =0.31 \textwidth]{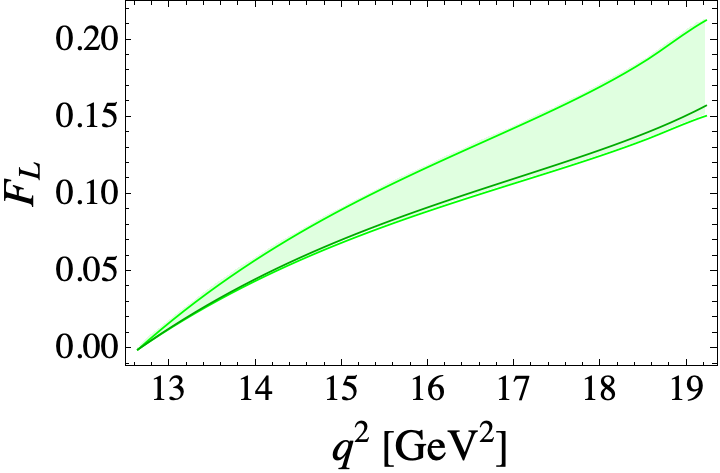}\hskip 0.5 cm
\caption{\baselineskip 10pt  \small Forward-backward asymmetry $A_{FB}(q^2)$ (left) and $K^*$ polarization fraction $F_L(q^2)$  (right plots) for  $\bar B^0 \to \bar K^{*0} (K \pi) \mu^+ \mu^-$ (top panels) and 
$\bar B^0 \to \bar K^{*0} (K \pi) \tau^+ \tau^-$ (bottom panels).   The dark continuous line is the SM result, the shaded region  the ABCD result for  $M_{Z^\prime}=1$ TeV.
The black dots in the top panels correspond to the LHCb measurement  \cite{LHCb:2020lmf}.}\label{obs-mu}
\end{center}
\end{figure}

\begin{figure}[t]
\begin{center}
\includegraphics[width =0.3 \textwidth]{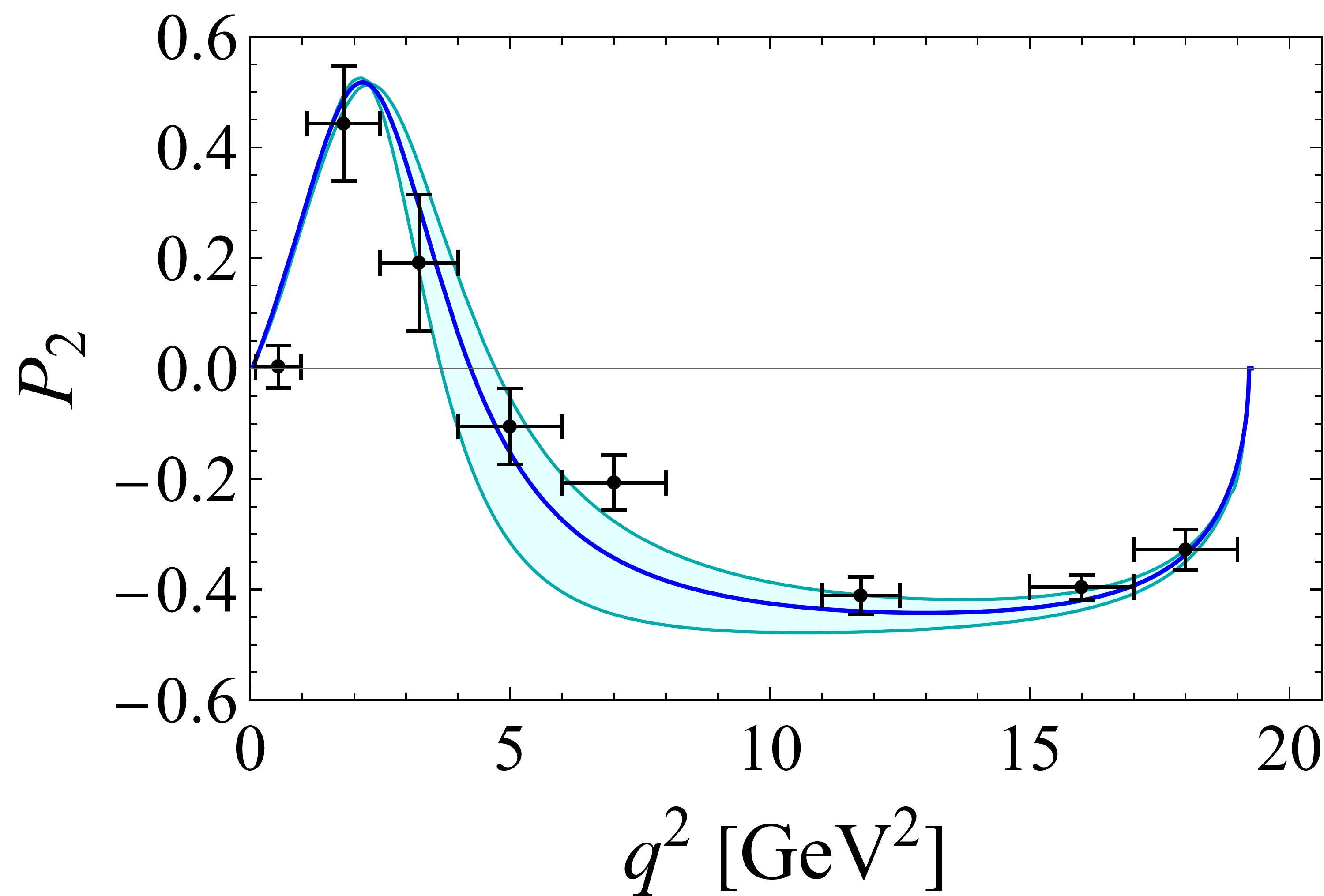}
\hskip 0.5cm \includegraphics[width =0.3 \textwidth]{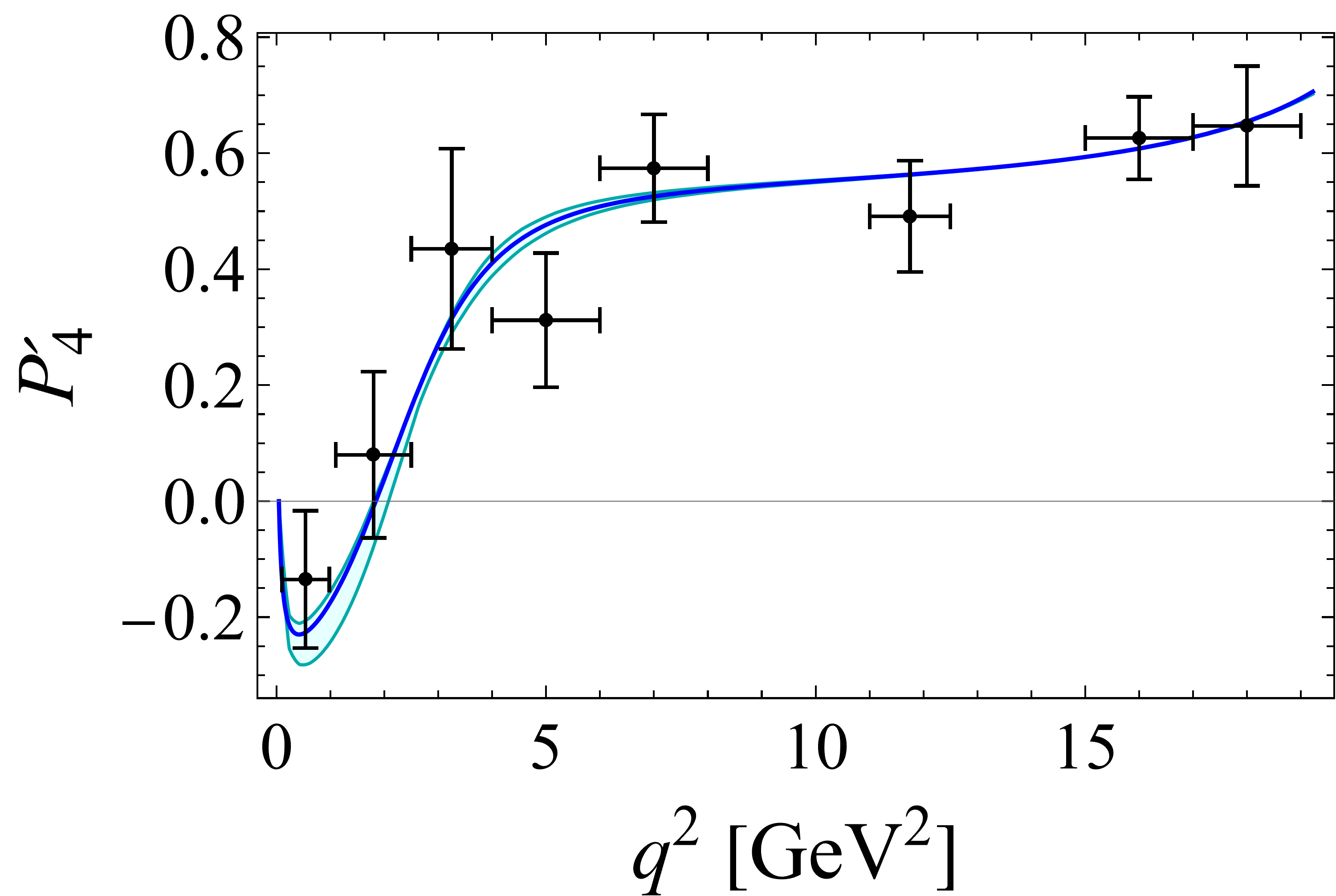}
\hskip 0.5cm \includegraphics[width =0.3 \textwidth]{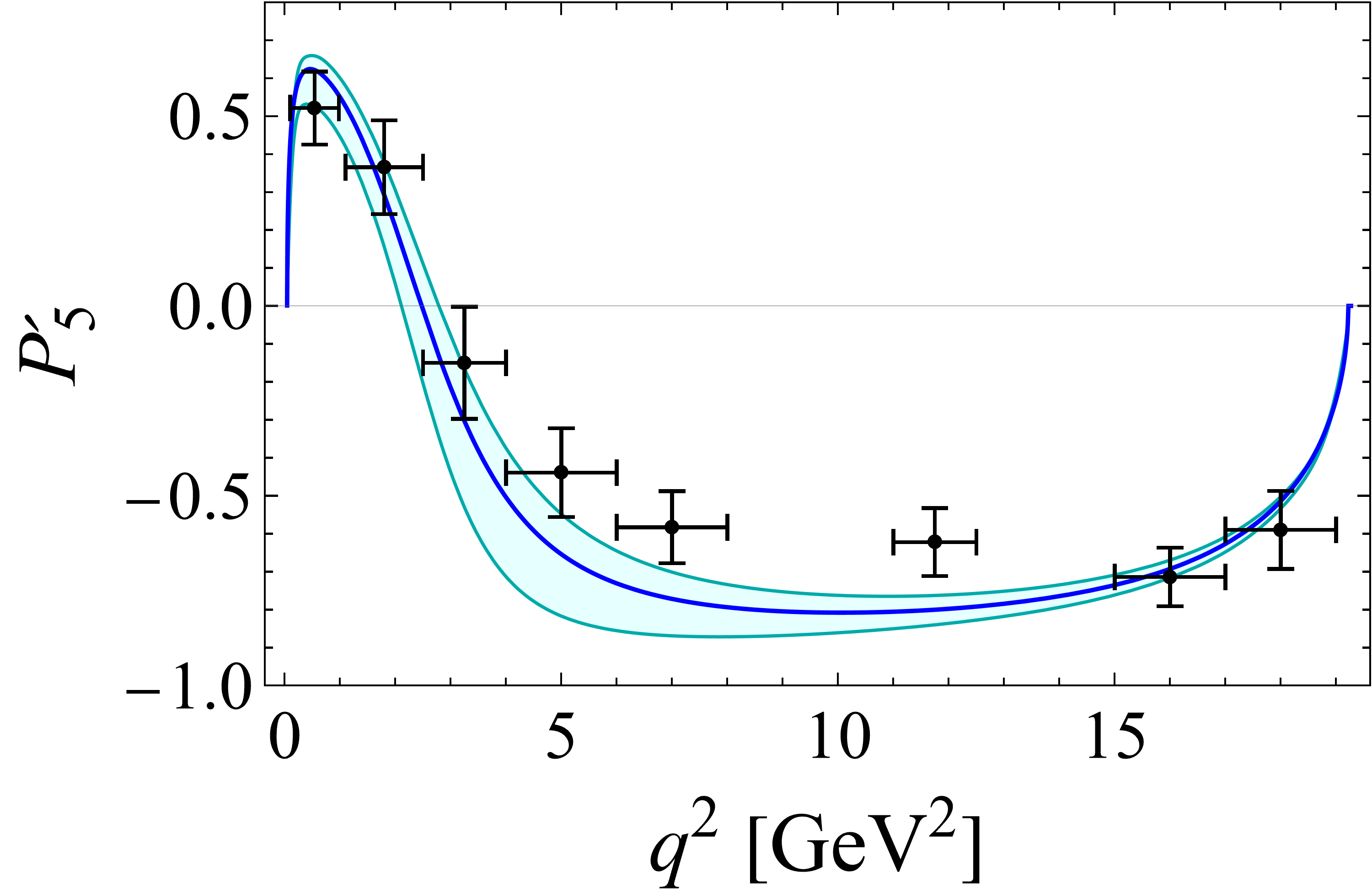}\\ \vskip 0.5cm
\includegraphics[width =0.31 \textwidth]{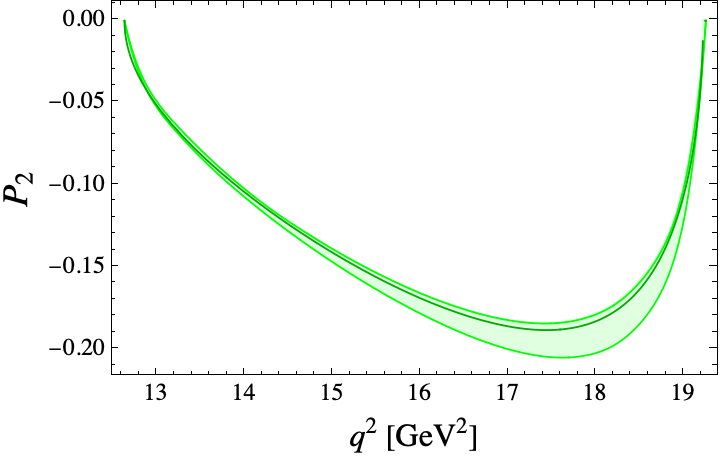}\hskip 0.5cm 
\includegraphics[width =0.31 \textwidth]{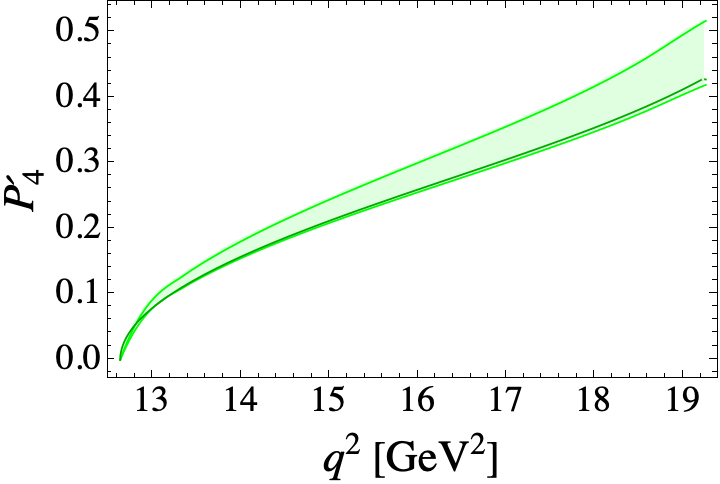}\hskip 0.5 cm
\includegraphics[width =0.31 \textwidth]{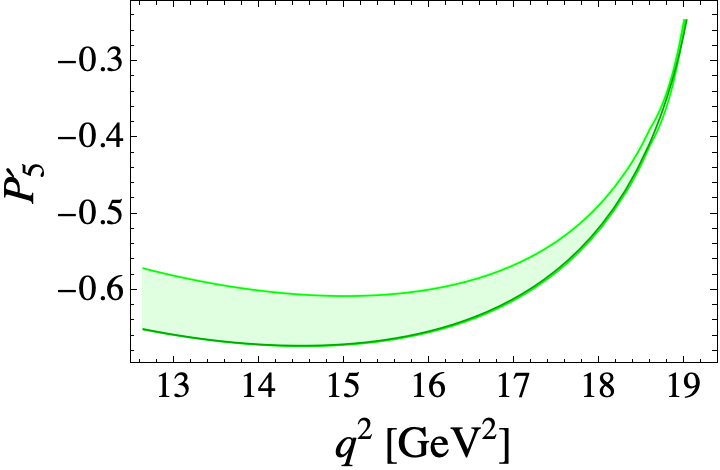}
\caption{\baselineskip 10pt  \small Angular $P$-observables in Eq.~\eqref{Pobs} for  $\bar B^0 \to \bar K^{*0} (K \pi) \mu^+ \mu^-$ (top panels) and $\bar B^0 \to \bar K^{*0} (K \pi) \tau^+ \tau^-$ (bottom panels). The black dots in the top panels correspond to LHCb measurement \cite{LHCb:2020lmf}. Same color code as in Fig.~\ref{obs-mu}.}\label{Pobs-mu}
\end{center}
\end{figure}

For  the forward-backward asymmetry and the fraction $F_L$ of longitudinally polarized $K^*$,   Eqs.~\eqref{AFB},\eqref{FLeFT}, the results are displayed in Fig.~\ref{obs-mu} for $\mu$ and $\tau$.
No measurement  is available for  $\tau$.  The largest deviation  from SM is found in  $F_L$  at high $q^2$ in  the $\tau$ mode.
For the $P$-observables  in \eqref{Pobs}  where tensions   emerged \cite{LHCb:2013zuf,LHCb:2013ghj,LHCb:2015svh,LHCb:2020lmf,LHCb:2020gog,LHCb:2023gpo,LHCb:2024onj},  we display  the  most significant  results in Fig.~\ref{Pobs-mu}.
 The effects obtained in the ABCD model  are  more pronounced for $\tau$, namely   for low $q^2$ the observable $P^\prime_5$  can differ from  SM. 

\begin{figure}[h]
\begin{center}
\includegraphics[width =0.4 \textwidth]{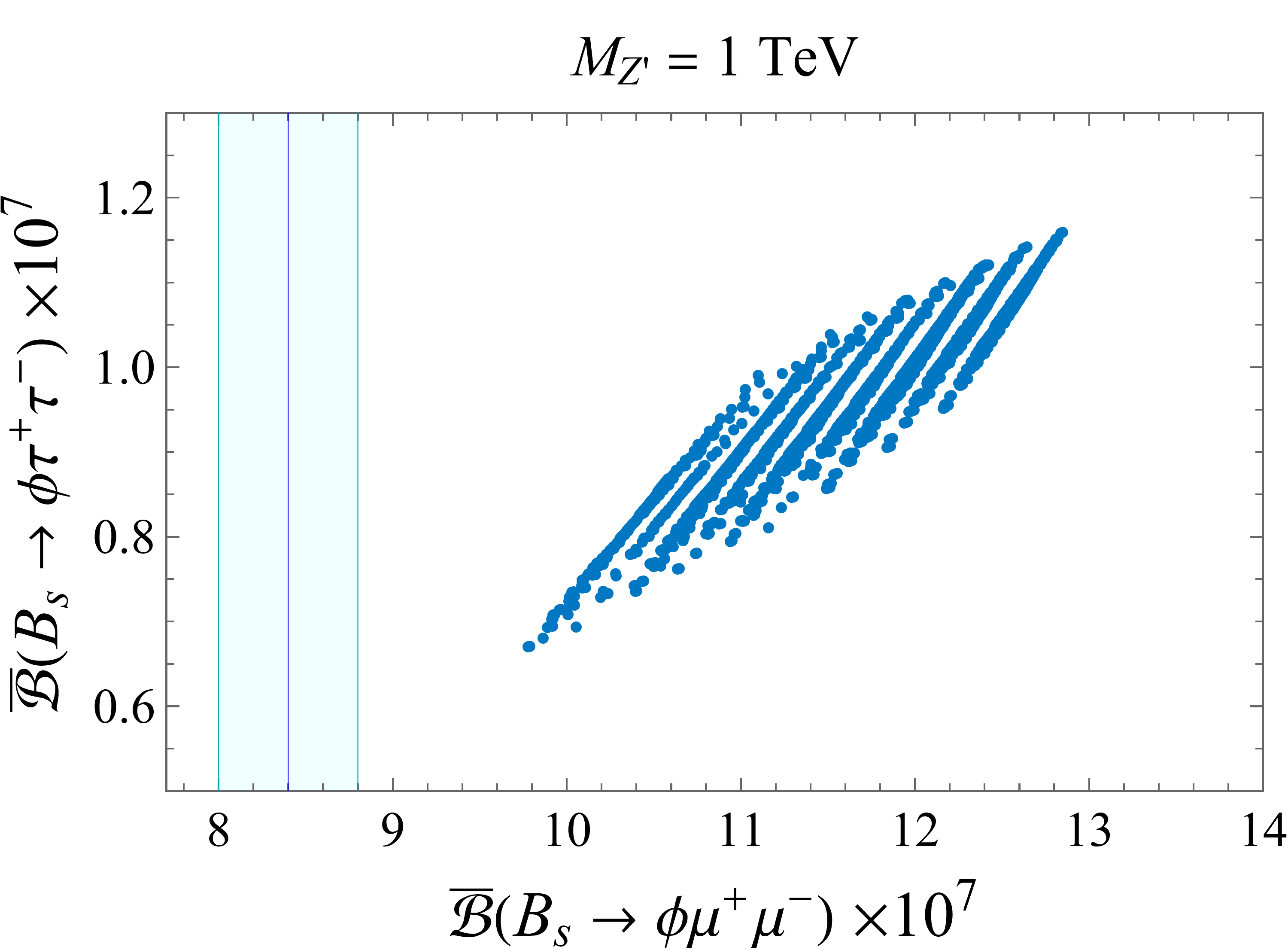}\hskip 0.4cm 
\includegraphics[width =0.4 \textwidth]{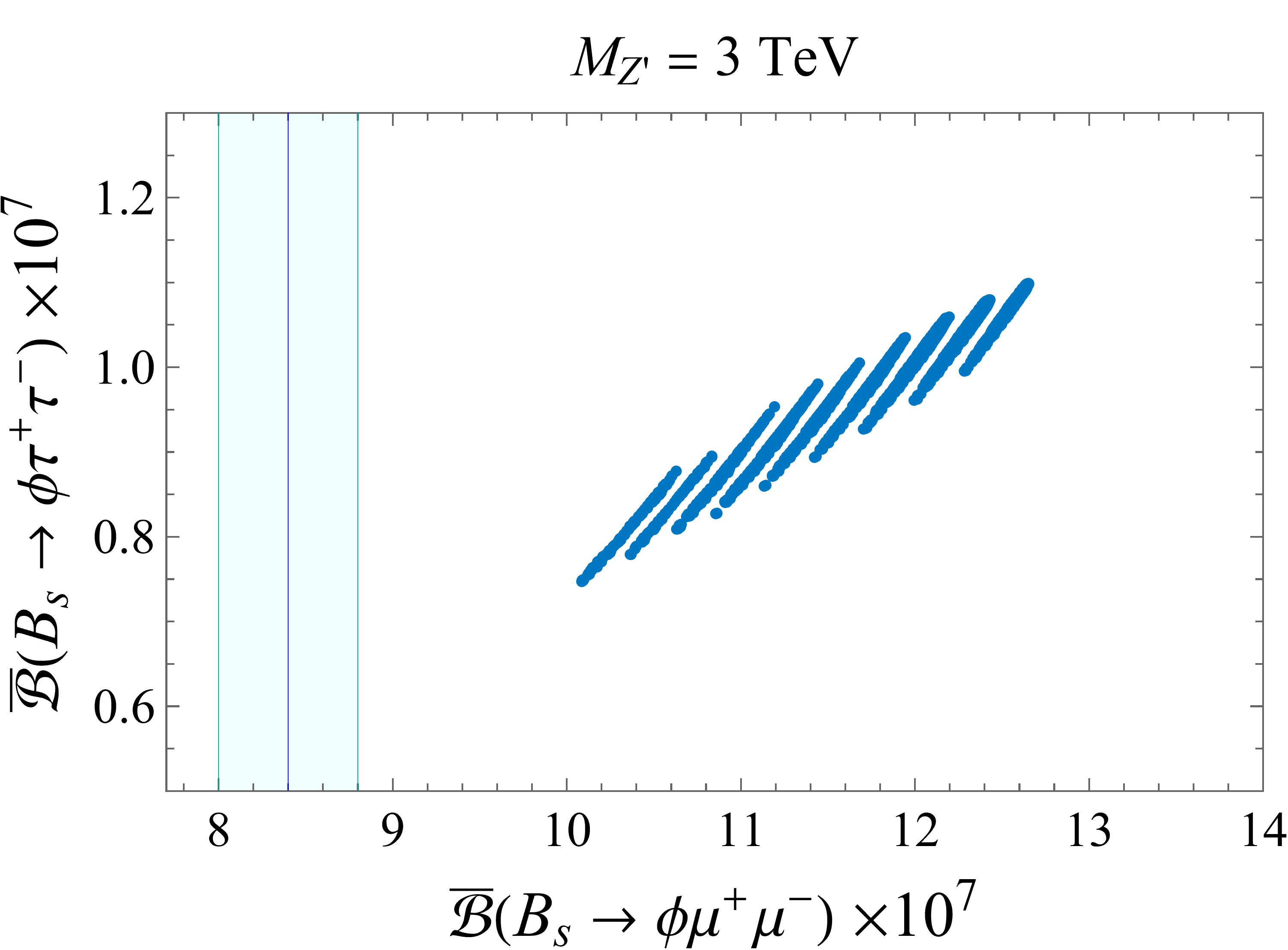}
\caption{\baselineskip 10pt  \small  Correlation between  ${\bar {\cal {B}}}(B_s \to \phi \mu^+ \mu^-)$ and ${\bar {\cal {B}}}(B_s \to \phi \tau^+ \tau^-)$ for $M_{Z^\prime}=1$ TeV (left panel) and $M_{Z^\prime}=3$ TeV (right panel).    The shaded vertical band corresponds to the  result  $ {\bar {\cal B}}(B_s \to \phi \mu^+ \mu^-)|_{exp}=(8.4\pm0.4)\times 10^{-7}$ quoted in \cite{Navas:PDG}.  }\label{phiLFC}
\end{center}
\end{figure}

For  $B_s \to \phi \mu^- \mu^+$,
using the form factors in \cite{Bharucha:2015bzk} with  parameters   in Table \ref{tabFF},  we find in SM:   ${\bar {\cal {B}}}(B_s \to \phi \mu^- \mu^+)=(11.8 \pm 2.3)\times 10^{-7}\left(|\lambda_{ts}^*|/0.04 \right)^2$, confirming that a higher value is foreseen for this rate with respect to the measured    ${\bar {\cal {B}}}(B_s \to \phi \mu^- \mu^+)_{\rm exp}=(8.4 \pm 0.4)\times 10^{-7}$    \cite{Navas:PDG}.  Measurements are not available for the  $\tau$ mode:  we find  ${\bar {\cal {B}}}(B_s \to \phi \tau^- \tau^+)=(1.02 \pm 0.20)\times 10^{-7}\left(|\lambda_{ts}^*|/0.04 \right)^2$ in  SM.
  Fig.~\ref{phiLFC} shows the correlation between the two LFC  decays in the ABCD  model: 
the   plot  for  $M_{Z^\prime}=1$ TeV   shows that the range for the $\tau$ mode is  ${\bar {\cal {B}}}(B_s \to \phi \tau^- \tau^+)=(0.9 \pm 0.25) \times 10^{-7}$,
for $M_{Z^\prime}=3$ TeV the  range is ${\bar {\cal B}}(B_s \to \phi \tau^- \tau^+)=(0.9 \pm 0.2) \times 10^{-7}$.  
 However, since  ${\bar {\cal B}}(B_s \to \phi \mu^- \mu^+)_{\rm exp}$ is not reproduced,   a puzzle remains unsolved for this mode.  If the measurement is confirmed, the first issue to scrutinise is  the uncertainty of the $B_s \to \phi$ form factors.
%
%%%%%%%%%%%%%%%
\begin{table}[h]
\center{\begin{tabular}{|l|c|c|c|c|c|}
\hline
\quad {\rm LFV mode} & ${\cal B}^{(1)}\times10^{9}$  & ${\cal B}^{(2)}\times10^{9}$ & ${\cal B}^{(3)}\times10^{11}$ & ${\cal B}^{(4)}\times10^{13}$ & experiment 
\\
\hline
$B_s \to \tau^+ \mu^-$ &  $0.00 \div  2.10$     &  $0.00 \div  1.60$ &    $0.00 \div  1.20$   &   $0.00 \div  9.20$    & $ <4.2 \times 10^{-5}$ \cite{Navas:PDG}\\
$\bar B^0 \to \bar K^{*0} \tau^+ \mu^-$ &  $0.00 \div 2.90$ &    $0.00 \div 1.15$ &    $0.00 \div  1.60$    & $0.00 \div  5.10$    & $<1.0 \times 10^{-5}$ \cite{Navas:PDG}
\\
$B_s \to \phi \tau^+ \mu^-$ &  $0.00 \div  3.43$ &      $0.00 \div  2.60$  &  $0.00 \div  1.95$& $0.00 \div  6.10$&  $<1.0 \times 10^{-5}$\cite{LHCb:2024wve} 
\\
\hline
$B_s \to \tau^+ \mu^-$ &   $0.00 \div 2.30$     & $0.00 \div 0.14$ &  $0.00 \div  1.10$  & $0.00 \div  1.05$ &$ <4.2 \times 10^{-5}$ \cite{Navas:PDG}
\\
$\bar B^0 \to \bar K^{*0} \tau^+ \mu^-$ &  $0.00 \div 3.10$ &   $0.00 \div 0.20$   &  $0.00 \div  1.53$   &  $0.00 \div  1.42$  &$<1.0 \times 10^{-5}$ \cite{Navas:PDG}
\\
$B_s \to \phi \tau^+ \mu^-$ &  $0.00 \div 3.70$ &     $0.00 \div 0.25$  & $0.00 \div  1.80$    &  $0.00 \div  1.70$   & $<1.0 \times 10^{-5}$\cite{LHCb:2024wve} 
\\
\hline
\end{tabular}  }
\caption {\small  Ranges of branching fractions of LFV $B$, $B_s$ decays in  the scenario A of the ABCD model. The first three rows correspond to  $M_{Z^\prime}=1$ TeV, the last three rows to $M_{Z^\prime}=3$ TeV. 
The four quoted ranges are obtained after imposing the constraints from the upper bounds on LFV charged  lepton decay modes  specified in Table~\ref{tabLFV-leptons}:
${\cal B}^{(1)}$    from $\tau^- \to \mu^- \mu^+ \mu^-$,  ${\cal B}^{(2)}$   from  ${\cal B}(\mu^- \to e^- \gamma)$,  ${\cal B}^{(3)}$    from  $\mu^- \to e^- e^+ e^-$,  ${\cal B}^{(4)}$    from  $\mu^- \to e^-$ conversion in titanium.   The  experimental upper bounds   (at $90\%$ C.L.) are quoted in the last column. }
\label{tabLFV1000}
\end{table}
%%%%%%%%%%%%%%%%%

%%%%%%%%%%%%%%%%%%%%%%
%
%
\begin{figure}[t]
\begin{center}
\includegraphics[width =0.4 \textwidth]{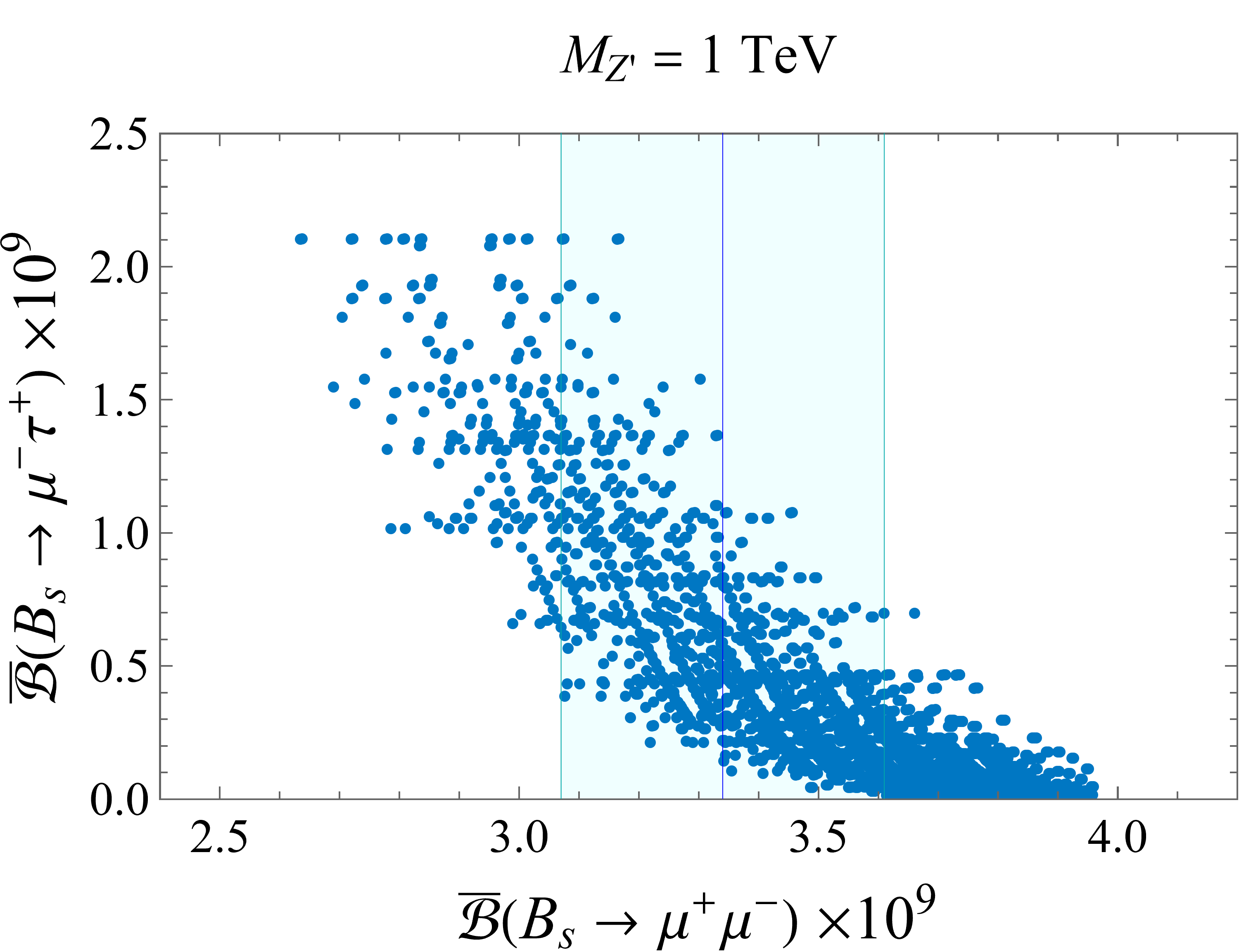} \hskip .4cm
\includegraphics[width =0.4 \textwidth]{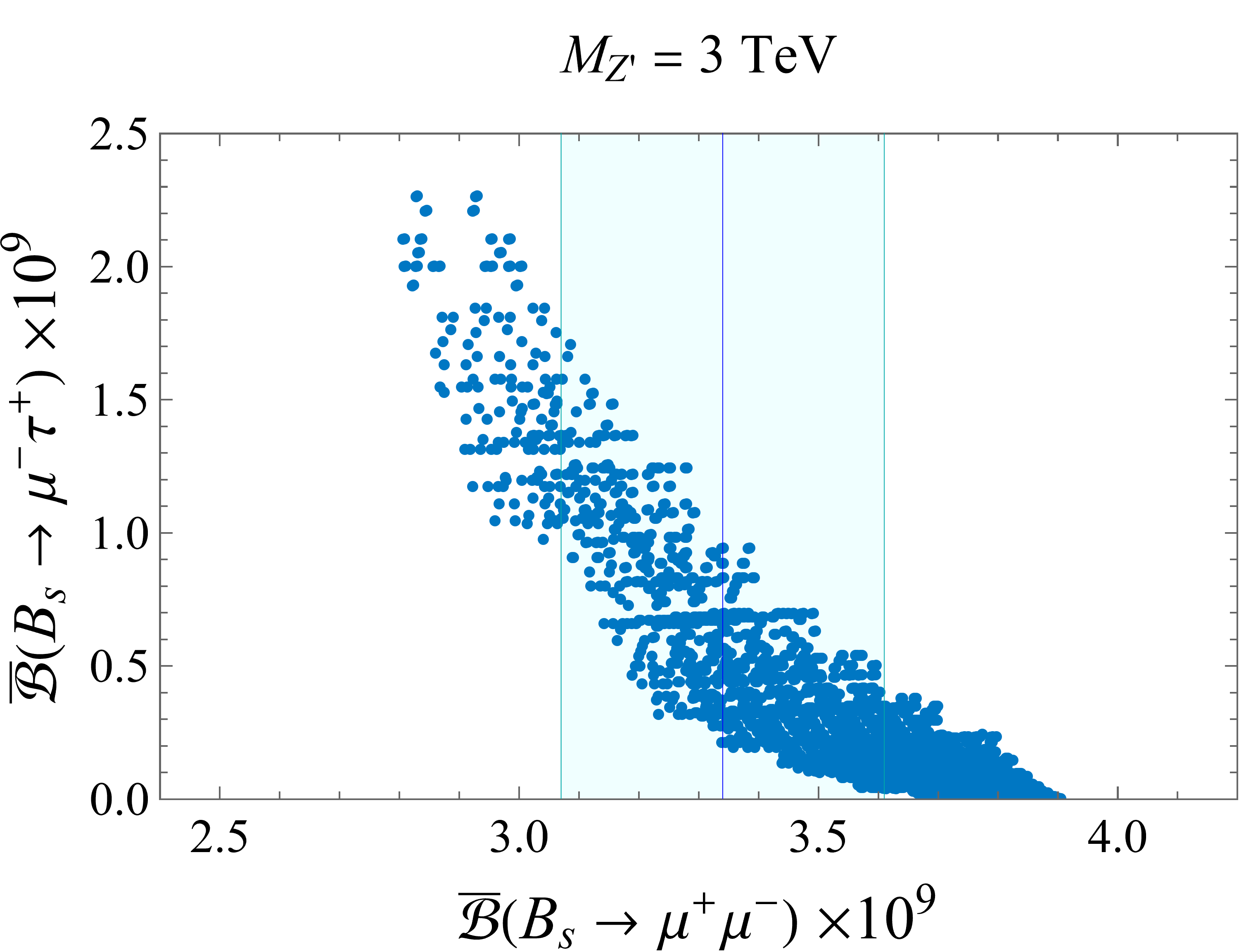}
\caption{\baselineskip 10pt  \small  Correlation between the branching fractions ${\bar {\cal B}}(B_s \to \mu^+ \mu^-)$ and ${\bar {\cal B}}(B_s \to \tau^+ \mu^-)$ for $M_{Z^\prime}=1$ TeV (left panel) and  $M_{Z^\prime}=3$ TeV (right panel).  The shaded vertical band corresponds to the experimental world average  \cite{Navas:PDG}. }\label{BsLFV}
\end{center}
\end{figure}
\begin{figure}[t]
\begin{center}
\includegraphics[width =0.4 \textwidth]{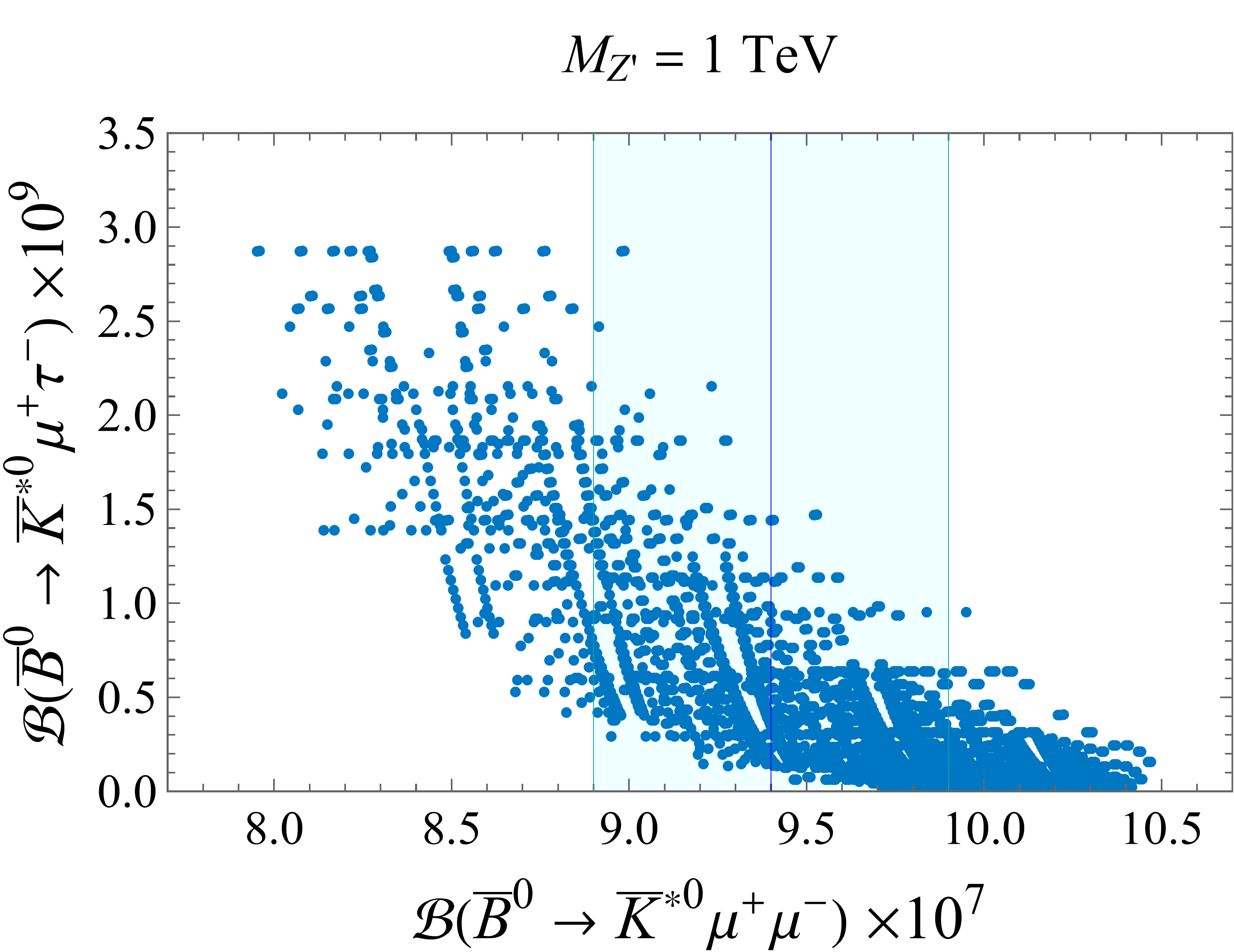} \hskip .4cm
\includegraphics[width =0.4 \textwidth]{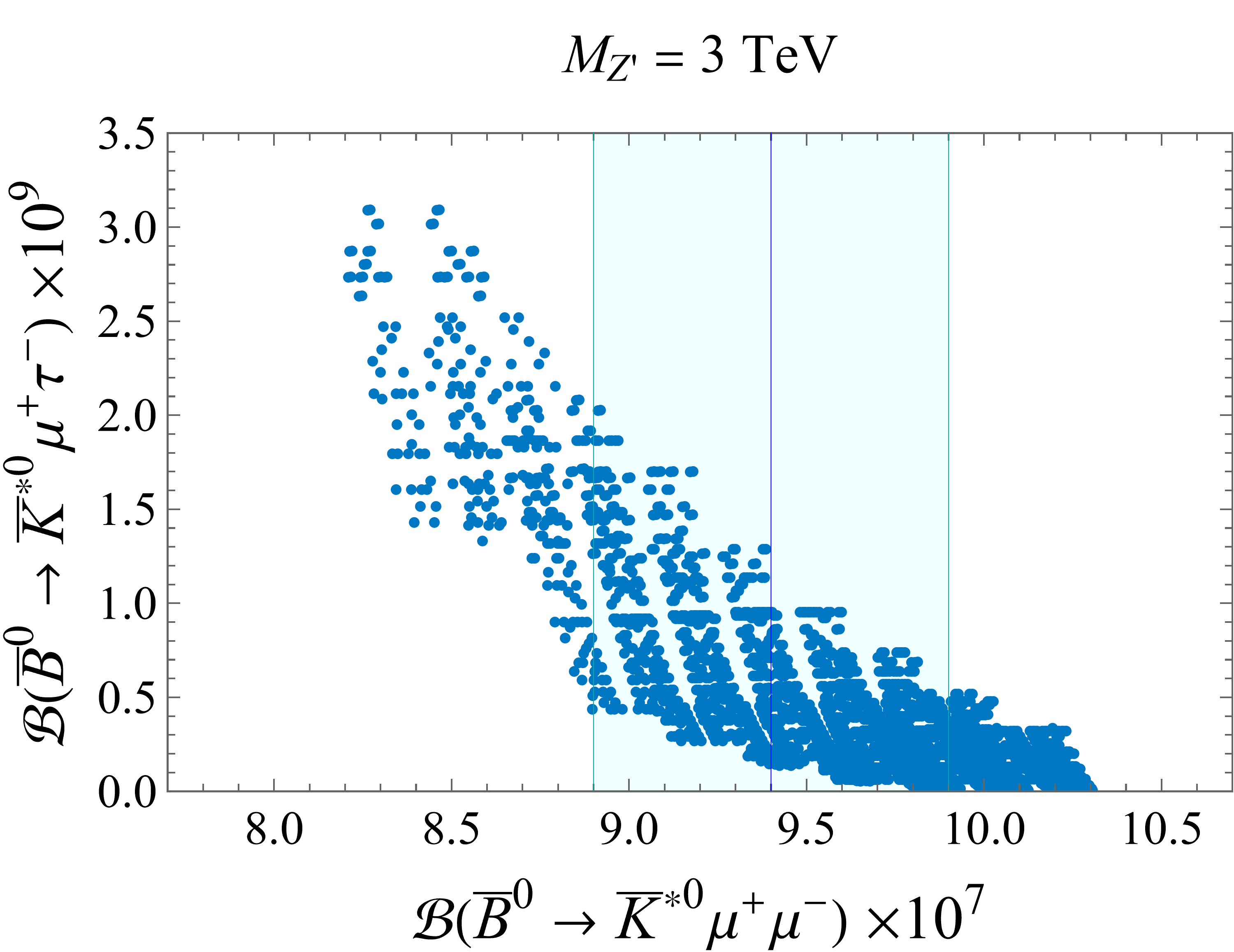}
\caption{\baselineskip 10pt  \small  Correlation between  ${\cal B}(\bar B^0 \to \bar K^{*0} \mu^+ \mu^-)$ and ${\cal B}(\bar B^0 \to \bar K^{*0} \mu^+ \tau^-)$ for $M_{Z^\prime}=1$ TeV (left panel) and  $M_{Z^\prime}=3$ TeV (right panel). Same color code  as in Fig.~\ref{KstarLFC}. }\label{KstarLFV}
\end{center}
\end{figure}
\subsubsection{Exclusive lepton flavour violating  modes $b \to s \ell_1^- \ell_2^+$ ($\ell_1 \neq \ell_2$)}
A feature of the ABCD model is that  correlations are generated between lepton flavour conserving  and   lepton flavour violating modes. In Table \ref{tabLFV1000} we quote  the   predictions  for a set of LFV modes, for  $M_{Z^\prime}=1$ TeV and $3$ TeV. The predictions can be further constrained  requiring agreement of the rates of corresponding LFC modes with experiment. In the Table we quote the unconstrained result and  the one constrained by the corresponding LFC process. 
The  current experimental upper bounds for the LFV modes are  also included in the Table, they are expected to be  sensibly improved in the near future.\footnote{Recent experimental investigations of  $\bar B^0 \to \bar K^{*0} \tau^+ \mu^-$ are described in \cite{LHCb:2022wrs,Belle-II:2025hpl}.}  

Figure \ref{BsLFV} shows the correlation between the branching ratios of the  LFC  $B_s \to \mu^+ \mu^-$ and LFV  $B_s \to \tau^+ \mu^-$ modes for the two  values of $M_{Z^\prime}$. The experimental upper bound for  ${\bar {\cal B}}(B_s \to \tau^+ \mu^-)$ is quoted in Table \ref{tabLFV1000}.
 In the ABCD model the branching ratio of the LFV $ \tau^+ \mu^- $ mode can be comparable to that of the LFC $\mu^+ \mu^- $ mode. Indeed,  the decay to same flavour leptons is determined  only by the $C_{10}$ term, while for $\ell_1 \neq \ell_2$ also $C_9$ contributes. Moreover, the helicity suppression in the  $\ell_1=\ell_2=\mu$ mode is lifted in the $ \tau^+ \mu^- $ mode producing
a sizable  enhancement.

\begin{figure}[t]
\begin{center}
\includegraphics[width =0.4 \textwidth]{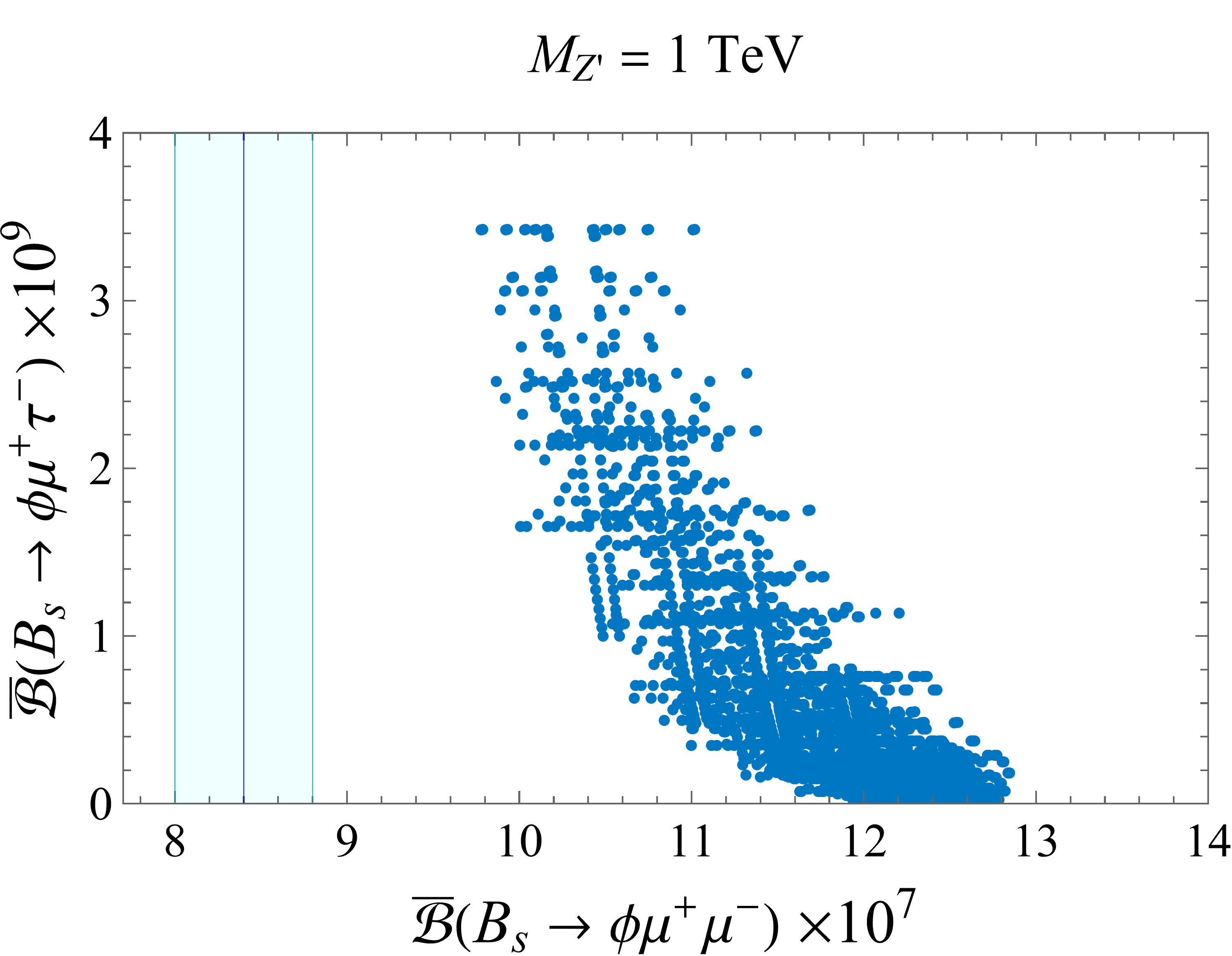} \hskip .4cm
\includegraphics[width =0.4 \textwidth]{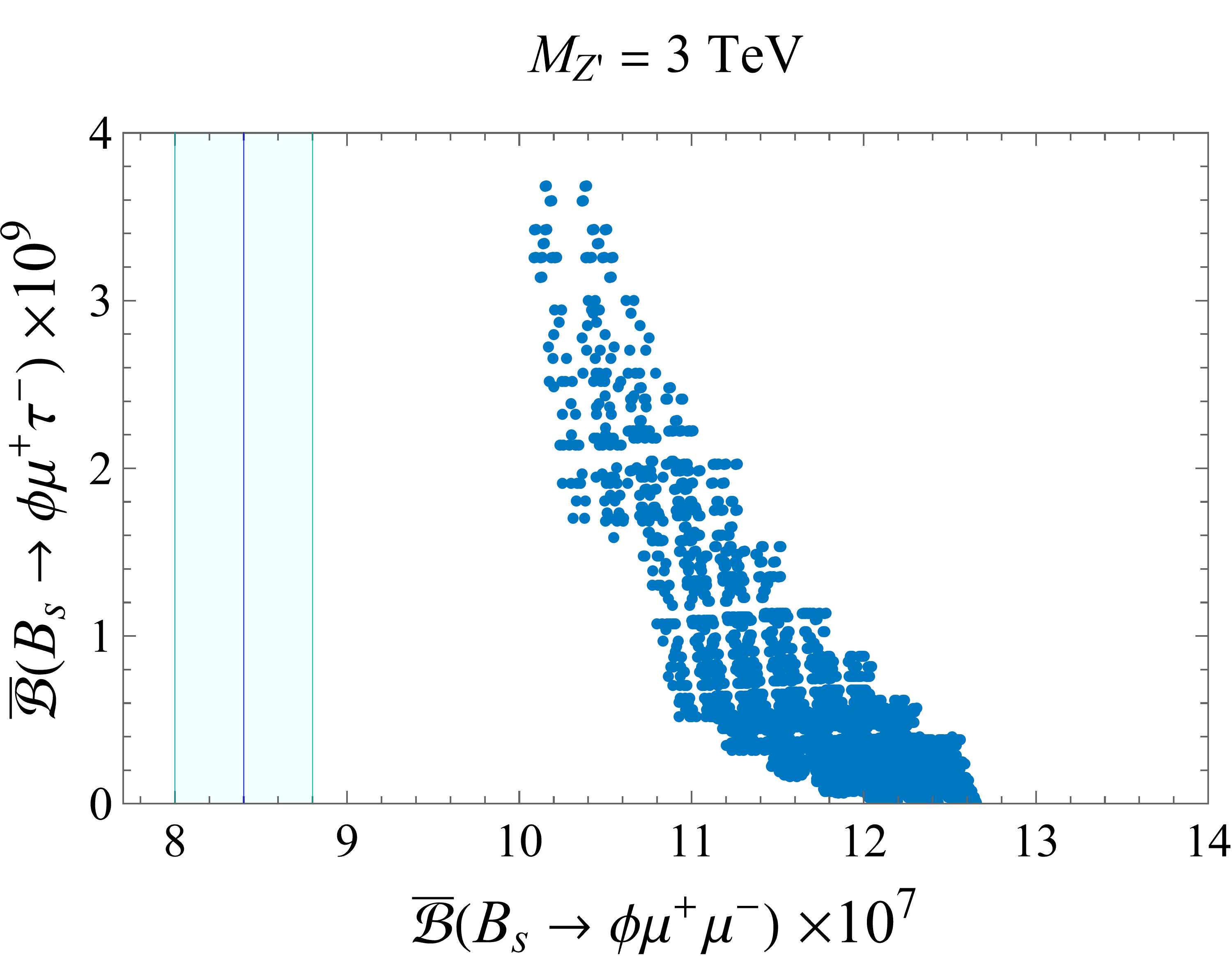}
\caption{\baselineskip 10pt  \small  Correlation between  $ {\bar {\cal B}}(B_s \to \phi \mu^+ \mu^-)$ and $ {\bar {\cal B}}(B_s \to \phi \mu^+ \tau^-)$ for $M_{Z^\prime}=1$ TeV (left panel) and  $M_{Z^\prime}=3$ TeV (right panel).  Same color code  as in Fig.~\ref{phiLFC}. }\label{BsphiLFV}
\end{center}
\end{figure}

While for the LFC modes the results do not sensibly  change  varying $M_{Z^\prime}$ from 1 TeV to 3  TeV, this is not true for the LFV modes, since 
  ABCD does not predict  large deviations from SM: increasing the $Z^\prime $ mass, the   SM  results are approached. For modes  allowed in  SM, the deviations are small even for small values of  $M_{Z^\prime}$. On the other hand, for modes  forbidden in SM,  increasing  $M_{Z^\prime}$ results in  a sensible decrease of the branching fractions.
The same observation holds for  ${\bar B}^0 \to {\bar K}^{*0} \tau^+ \mu^-$:  the branching fraction predicted in the ABCD model  is quoted  in Table \ref{tabLFV1000} both without and with the constraint given  by the measured ${\cal B}({\bar B}^0 \to {\bar K}^{*0} \mu^+ \mu^-)$. 
As we discuss below, constraints provided by  LFV leptonic decays further reduce the range for this rate.
Predictions for   $B_s \to \phi \tau^+ \mu^-$ are  presented in Table \ref{tabLFV1000} and in Fig.\ref{BsphiLFV}.

\subsection{Correlations among  lepton flavour violating processes in the quark and lepton sectors}
 Since the $Z^\prime$ couplings to leptons and quarks involve the same pair of parameters, $\epsilon_1$ and $\epsilon_2$,  correlations can be established among observables in the different sectors. Particularly relevant are the correlations between the LFV $b \to s \ell_1^+ \ell_2^-$ modes and the LFV leptonic decays $\mu^- \to e^- \gamma$, $\tau^- \to \mu^- \mu^+ \mu^-$, $\mu^- \to e^- e^+ e^-$, and the $\mu^- \to e^-$ conversion in nuclei. As we shall see, in the scenario A of the ABCD model
the   upper bounds on the last four channels play a hierarchical role in constraining the branching ratios of LFV $B$ and $B_s$ decays.   Their combined effect  prevents the branching fractions of LFV $B$ and $B_s$ decays from exceeding ${\cal O}(10^{-12})$.

It is  useful to analyse the individual constraints from the four leptonic processes, which in scenarios different from scenario A depend on distinct parameters. This provides us with a guidance on possible modifications of the scenario. For instance, since $\tau^- \to \mu^- \mu^+ \mu^-$ provides the weakest constraint, one may consider assigning a different role to the third lepton family compared to the other two: we defer such an  investigation to future work.
In the following  we analyze the impact of  $\tau^- \to \mu^- \mu^+ \mu^-$ and $\mu^- \to e^- \gamma$, then we consider  the  most stringent channels  $\mu^- \to e^- e^+ e^-$ and $\mu^- \to e^-$ conversion in nuclei.

\subsubsection 
{Rare $B_{(s)}$ decays versus $\tau^- \to \mu^- \mu^+ \mu^-$ and  $\mu^- \to e^-  \gamma$.}
%%%%%%%%%%%%%%%
%
\begin{table}[t]
\center{\begin{tabular}{|l|c|}
\hline
\quad {\rm LFV mode} &  ${\cal B}$ experiment \\
\hline
$\tau^- \to \mu^- \mu^+ \mu^-$ &  $<2.1 \times 10^{-8}$ \cite{Navas:PDG}\\
$\mu^- \to e^- \gamma$ &  $<1.5 \times 10^{-13}$ \cite{MEGnew}\\
$\mu^- \to e^- e^+ e^-$ & \,\,\,\, \,\,$<1.0 \times 10^{-12}$\cite{SINDRUM:1987nra,Navas:PDG} \\
$\mu^- \to e^-$ conversion in $^{48}_{22}{\rm{Ti}}$ &  \quad \,\,  $<4.3 \times 10^{-12}$ \cite{Wintz:1998rp,Kaulard:1998rb}\\
\hline
\end{tabular}  }
\caption {\small  Experimental upper bounds (at $90 \, \%$ C.L.)  for  LFV charged lepton modes. The branching fraction for the $\mu^- \to e^-$ conversion  is defined in Eq.~\eqref{Brcapture}. }
\label{tabLFV-leptons}
\end{table}
\begin{figure}[bh!]
\begin{center}
\includegraphics[width = 0.4\textwidth]{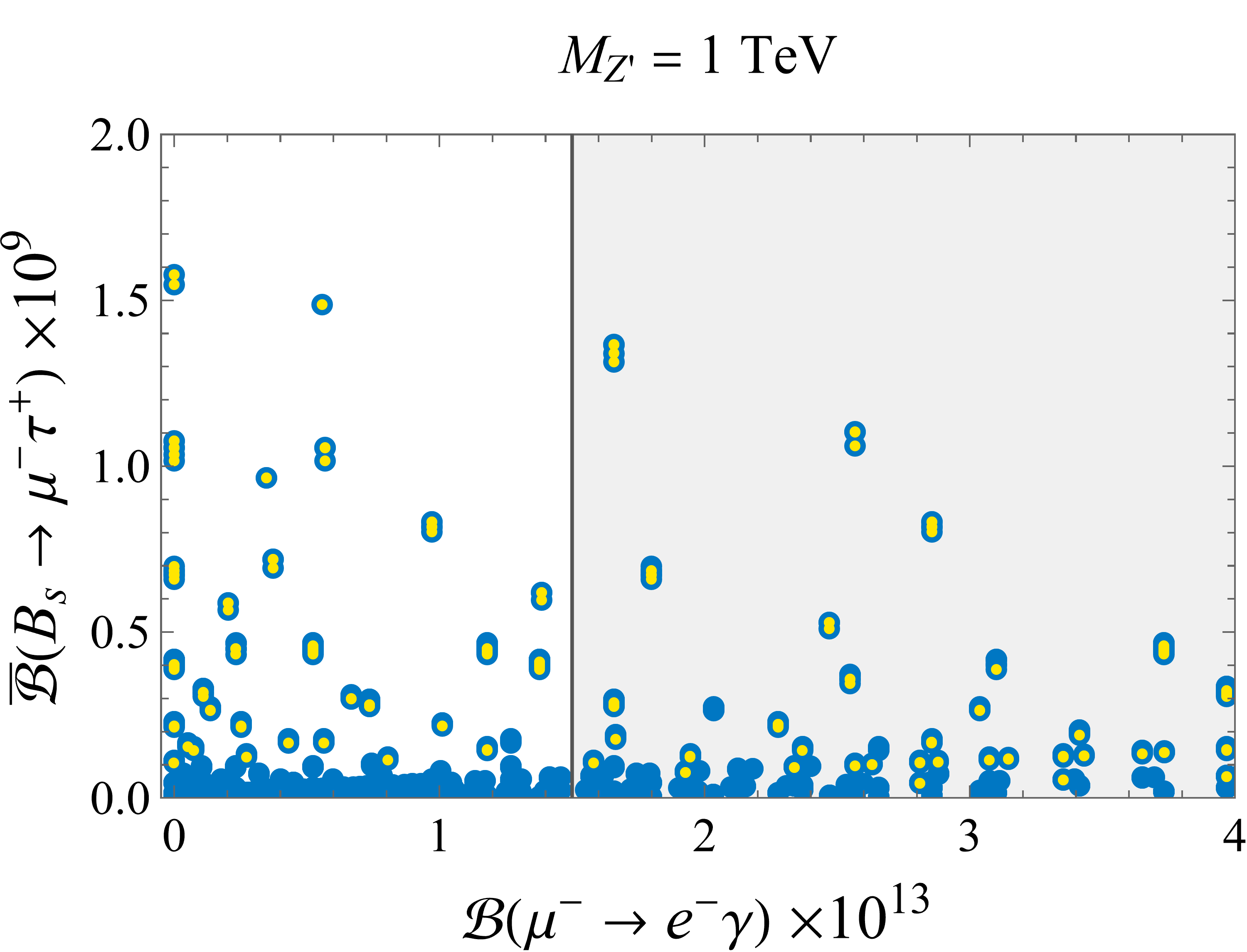}  \hskip 0.4cm
\includegraphics[width =0.4 \textwidth]{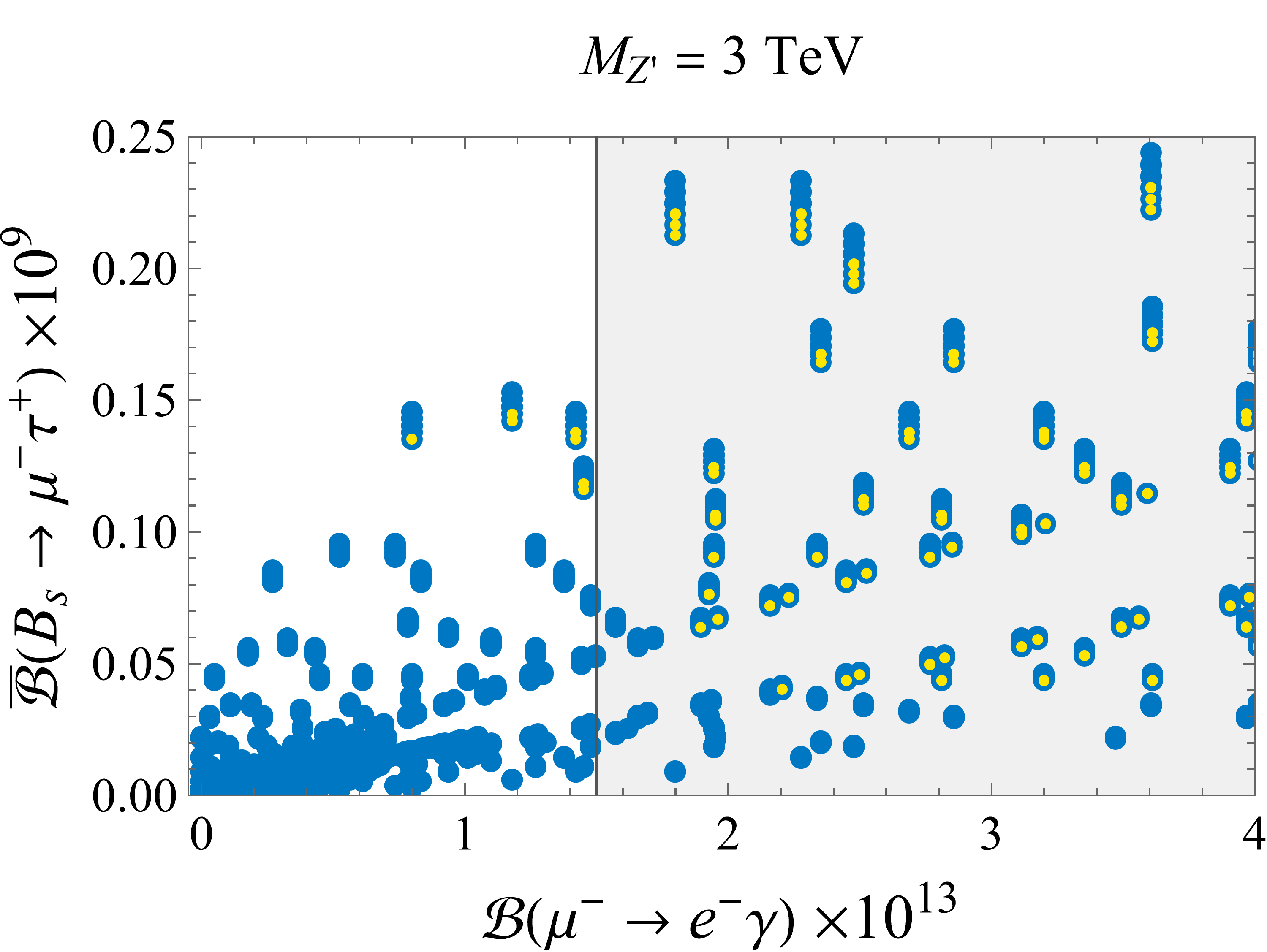}
\caption{\baselineskip 10pt  \small  Correlation between ${\bar {\cal B}}(B_s \to \tau^+ \mu^-)$ and ${\cal B}(\mu^- \to e^- \gamma)$ for $M_{Z^\prime}=1$ TeV (left panel) and $M_{Z^\prime}=3$ TeV (right panel).  The vertical line indicates  the experimental upper bound ${\cal B}(\mu^- \to e^- \gamma)<1.5 \times 10^{-13} \,\, (\rm{at}\,  90\% \, {\rm CL})$   \cite{MEGnew}. 
The light yellow points  are selected requiring that  the LFC   ${\bar {\cal B}}( B_s \to  \mu^+ \mu^-)$ agrees within $1\sigma$ with experiment.
}\label{Bsvsmueg1000}
\end{center}
\end{figure}
\begin{figure}[th!]
\begin{center}
\includegraphics[width = 0.4\textwidth]{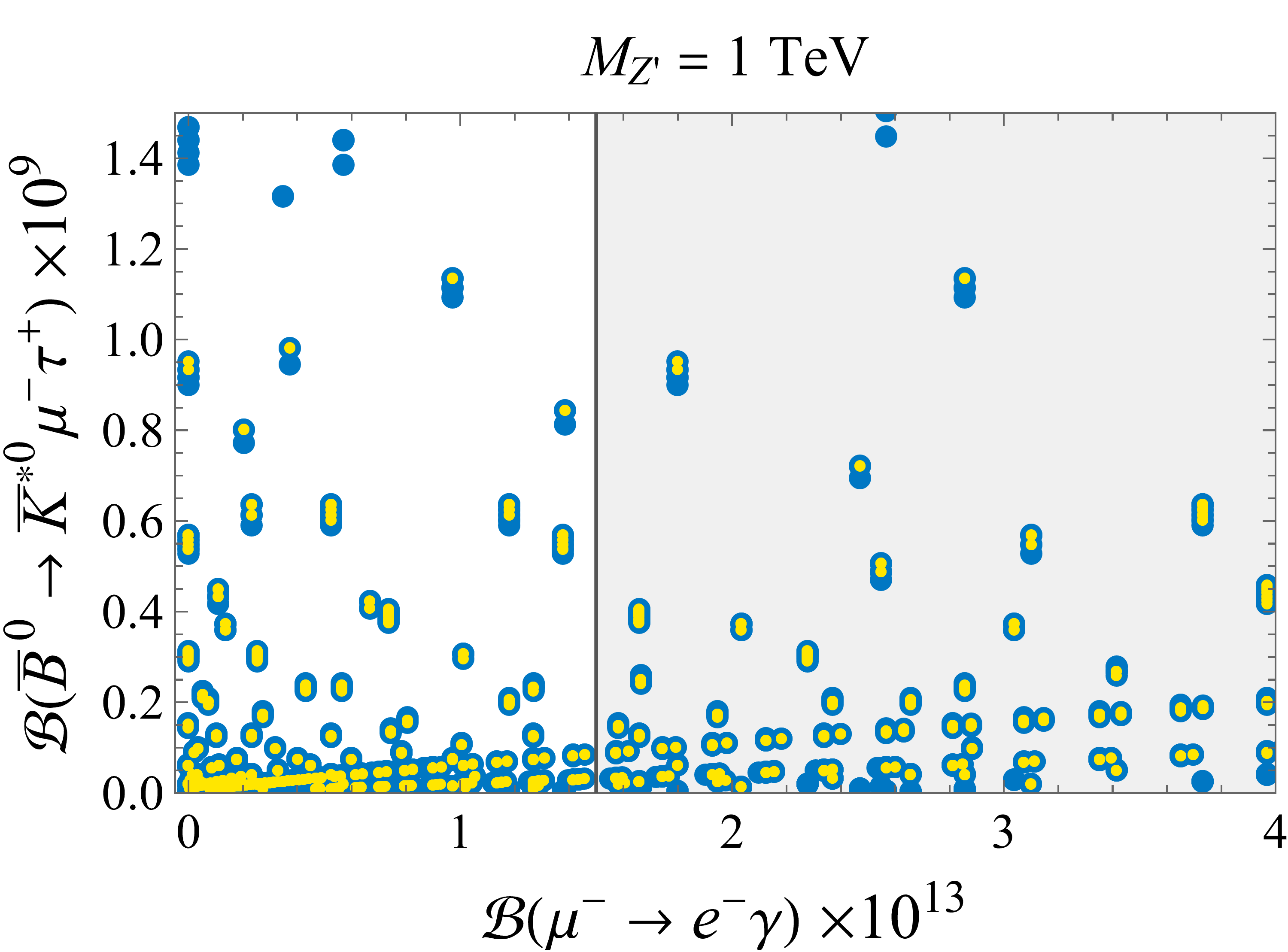} \hskip 0.4 cm
\includegraphics[width =0.4 \textwidth]{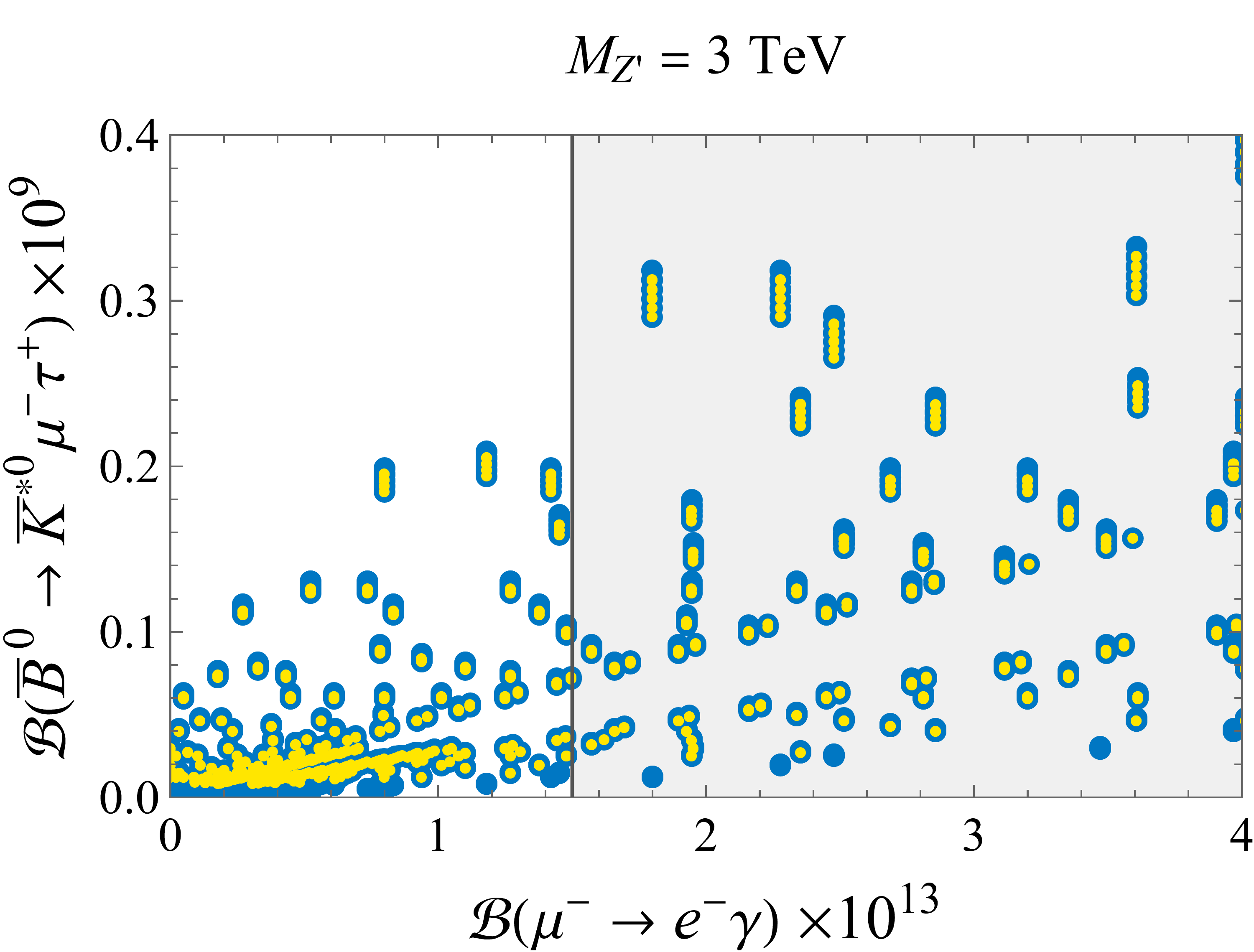}
\caption{\baselineskip 10pt  \small  Correlation between  ${\cal B}(\bar B^0 \to \bar K^{*0} \tau^+ \mu^-)$ and ${\cal B}(\mu^- \to e^- \gamma)$ for $M_{Z^\prime}=1$ TeV (left panel) and $M_{Z^\prime}=3$ TeV (right panel).  The vertical line corresponds to the experimental upper bound for ${\cal B}(\mu^- \to e^- \gamma)$   \cite{MEGnew}.
 The light yellow points  are selected requiring that  the LFC   ${\cal B}(\bar B^0 \to \bar K^{*0} \mu^+ \mu^-)$ agrees within $1\sigma$ with experiment.}   \label{Kstarvsmueg1000}
\end{center}
\end{figure}

The correlations involving $\tau^- \to \mu^- \mu^+ \mu^-$,  with the experimental upper bound ${\cal B}(\tau^- \to \mu^- \mu^+ \mu^-)<2.1 \times 10^{-8}$  \cite{Navas:PDG}
 do not provide significant constraints for  $B_s$ and $B$ modes.
Accounting for the measurement of the width of $B_s \to \mu^+ \mu^-$, the largest value for ${\cal B}(\tau^- \to \mu^- \mu^+ \mu^-)$ is about $1.2 \times 10^{-8}$. 

 $\mu^- \to e^- \gamma$ puts more stringent limits.
Figure \ref{Bsvsmueg1000}  shows the correlation between ${\bar {\cal B}}(B_s \to \tau^+ \mu^-)$ and ${\cal B}(\mu^- \to e^- \gamma)$   for $M_{Z^\prime}=1$ TeV and $3$ TeV. 
The dark points show the results without imposing constraints from LFC mode, the light points the results obtained  imposing that the measured ${\bar {\cal B}}(B_s \to \mu^+ \mu^-)$ is recovered.   The  upper bound is displayed ${\cal B}(\mu^- \to e^- \gamma)<1.5 \times 10^{-13} \,\, (\rm{at}\,  90\% \, {\rm CL})$ recently established by the MEG II Collaboration\cite{MEGnew}   which has improved the previous limit ${\cal B}(\mu^- \to e^- \gamma)<4.2 \times 10^{-13} \,\, (\rm{at}\,  90\% \, {\rm CL})$ \cite{MEG:2016leq}. The bound  reduces  the room  for ${\bar {\cal B}}(B_s \to \tau^+ \mu^-)$.
This provides us with a clear example of  the most peculiar feature of the ABCD model, the interplay between quark and lepton observables. Within the model, quite large values of ${\cal B}(\mu^- \to e^- \gamma)$  could be obtained, however, the requirement that the rate of $B_s \to \mu^+ \mu^-$ agrees with experiment  suppresses ${\cal B}(\mu^- \to e^- \gamma)$. 
The correlation between ${\cal B}(\bar B^0 \to \bar K^{*0} \tau^+ \mu^-)$ and ${\cal B}(\mu^- \to e^- \gamma)$ is displayed in Fig.~\ref{Kstarvsmueg1000}, and similar remarks hold:
the  bound on ${\cal B}(\mu^- \to e^- \gamma)$ \cite{MEGnew}    reduces the allowed range for ${\cal B}(B \to K^{*} \tau^+ \mu^-)$, see Table \ref{tabLFV1000}.
The same occurs for $B_s \to \phi \tau^+ \mu^-$.

\subsubsection
{Rare $B_{(s)}$ decays versus $\mu^- \to e^- e^+ e^-$ and $\mu^- \to e^-$ conversion in nuclei}
\begin{figure}[h]
\begin{center}
\includegraphics[width =0.4 \textwidth]{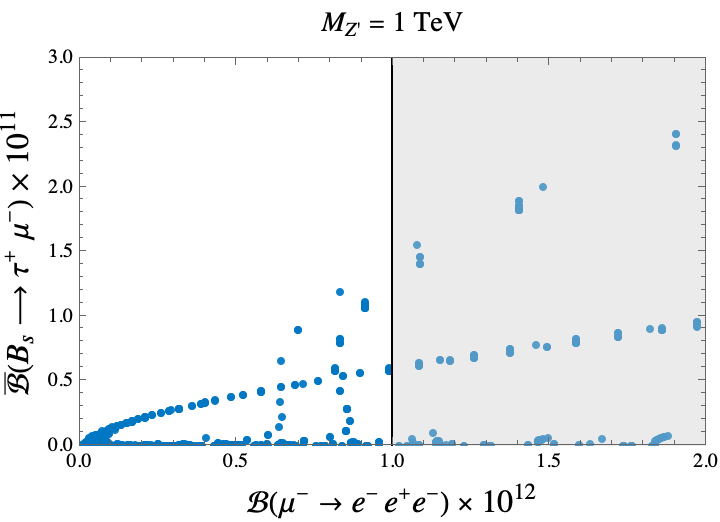} \hskip 0.4cm 
\includegraphics[width =0.4 \textwidth]{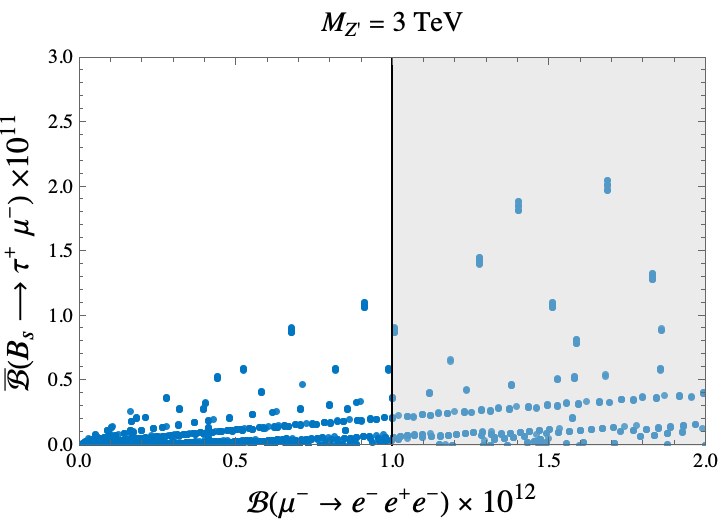} \\ \vskip 0.4cm 
\includegraphics[width = 0.4\textwidth]{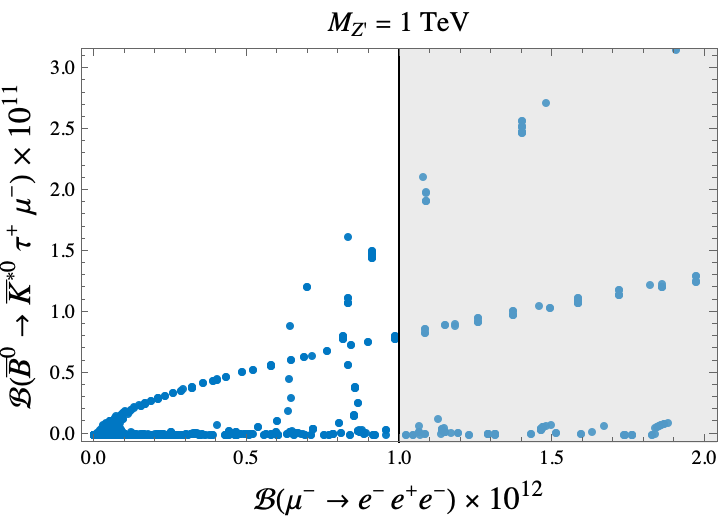} \hskip 0.4cm 
\includegraphics[width = 0.4\textwidth]{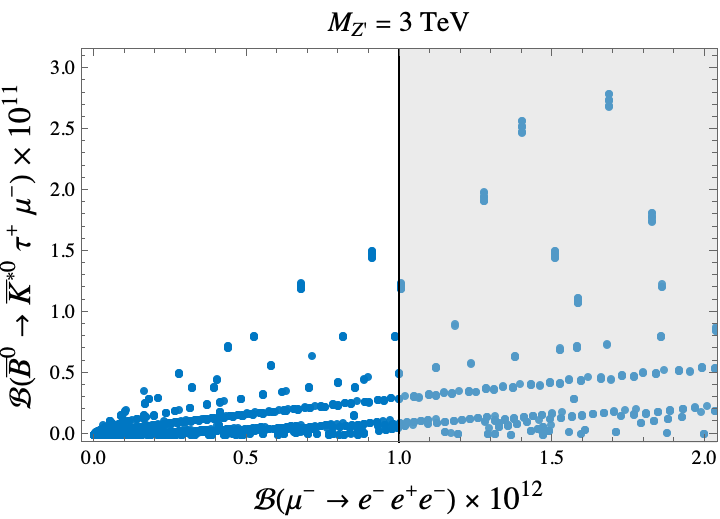}
\caption{\baselineskip 10pt  \small  Correlation between  ${\bar {\cal B}}(B_s \to \tau^+ \mu^-)$ and ${\cal B}(\mu^- \to e^- e^+ e^-)$ (top panels) and 
 between ${\cal B}(\bar B^0 \to \bar K^{*0} \tau^+ \mu^-)$ and ${\cal B}(\mu^- \to e^- e^+ e^-)$ (bottom panels), for
$M_{Z^\prime}=1$ TeV (left panel) and $M_{Z^\prime}=3$ TeV (right panel).  The gray band corresponds to the upper bound ${\cal B}(\mu^- \to e^- e^+ e^-)<1 \times 10^{-12}\,\, (\rm{at}\,  90\% \, {\rm CL})$ \cite{SINDRUM1988,Navas:PDG}.
}\label{BsvsMu3e}
\end{center}
\end{figure}

\begin{figure}[h]
\begin{center}
\includegraphics[width =0.4 \textwidth]{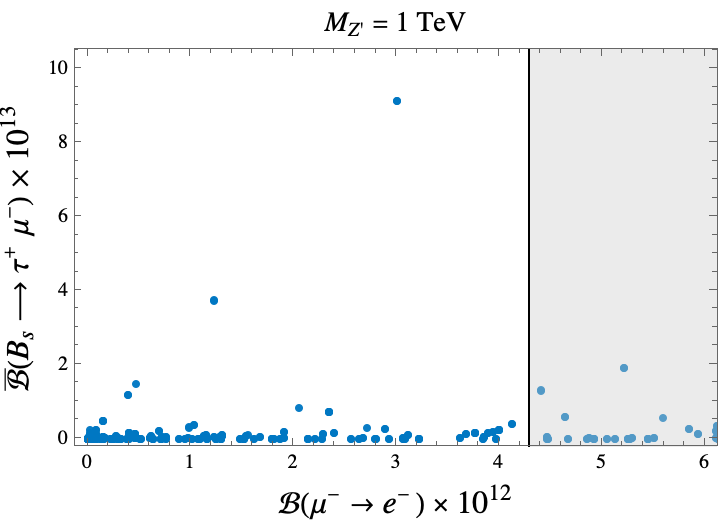} \hskip 0.4cm 
\includegraphics[width =0.4 \textwidth]{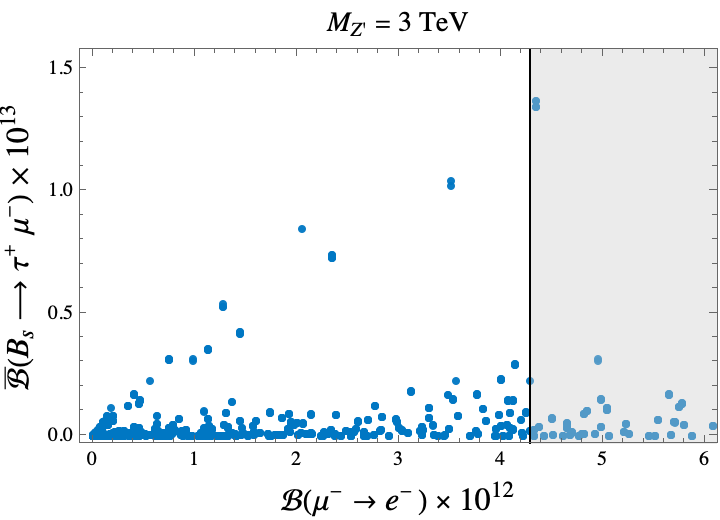} \\ \vskip 0.4cm 
\includegraphics[width = 0.4\textwidth]{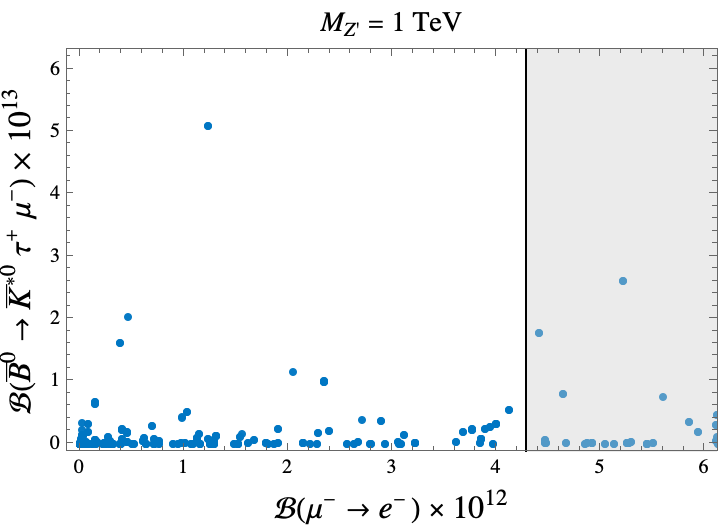} \hskip 0.4cm 
\includegraphics[width = 0.4\textwidth]{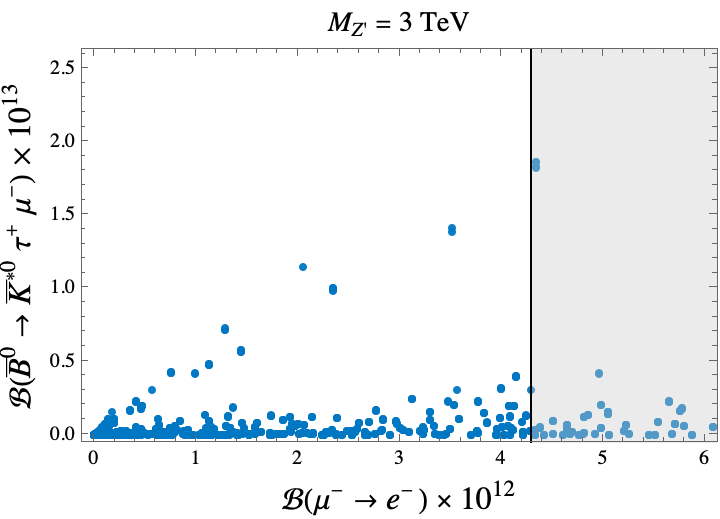}
\caption{\baselineskip 10pt  \small  Correlation between  ${\bar {\cal B}}(B_s \to \tau^+ \mu^-)$ and ${\cal B}(\mu^- \to e^- )$ due to conversion in nuclei (top panels) and 
 between ${\cal B}(\bar B^0 \to \bar K^{*0} \tau^+ \mu^-)$ and ${\cal B}(\mu^- \to e^- )$ due to conversion in nuclei  (bottom panels), for
$M_{Z^\prime}=1$ TeV (left panel) and $M_{Z^\prime}=3$ TeV (right panel).  The gray band corresponds to the upper bound ${\cal B}(\mu^- \to e^- )<4.3 \times 10^{-12} \,\, (\rm{at}\,  90\% \, {\rm CL})$\cite{Kaulard:1998rb}.
}\label{BsvsMuConv}
\end{center}
\end{figure}

%%%%%%%%%%%%%%%
%
 The processes $\mu^- \to e^- e^+ e^-$ and $\mu^- \to e^-$ conversion in nuclei play the major role in constraining LFV rare $B_{(s)}$ decays. 
Fig.~\ref{BsvsMu3e} shows that the upper limit ${\cal B}(\mu^- \to e^- e^+ e^-)<1 \times 10^{-12}\,\, (\rm{at}\,  90\% \, {\rm CL})$ \cite{SINDRUM1988,Navas:PDG}  significantly reduces   ${\bar {\cal B}}(B_s \to \tau^+ \mu^-)$ and ${\cal B}(\bar B^0 \to \bar K^{*0} \tau^+ \mu^-)$  both for $M_{Z^\prime}=1$ TeV   and for $M_{Z^\prime}=3$ TeV. After imposing this bound their largest allowed values  are  ${\cal O}(10^{-11})$,  see Table~\ref{tabLFV-leptons}.
Muon conversion in nuclei provides the most stringent constraint, as displayed in Fig.~\ref{BsvsMuConv}. The largest allowed value for  ${\bar {\cal B}}(B_s \to \tau^+ \mu^-)$ and ${\cal B}(\bar B^0 \to \bar K^{*0} \tau^+ \mu^-)$ are reduced to  ${\cal O}(10^{-13})$,  see Table~\ref{tabLFV-leptons}.

\begin{figure}[h]
\begin{center}
\includegraphics[width =0.4 \textwidth]{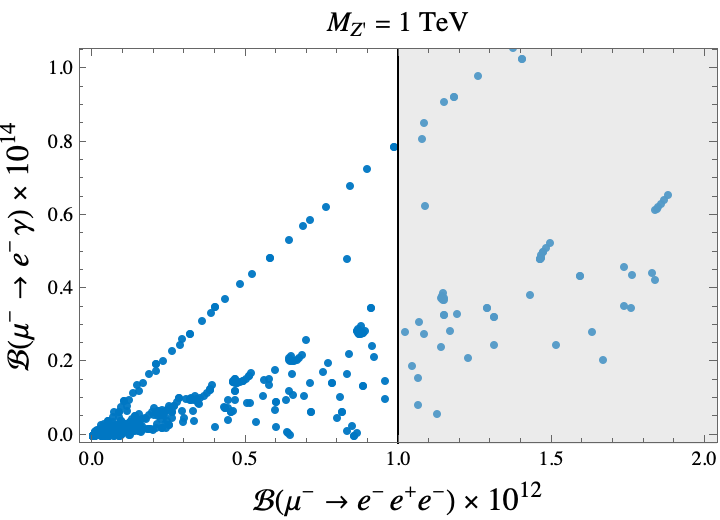} \hskip 0.4cm 
\includegraphics[width =0.4 \textwidth]{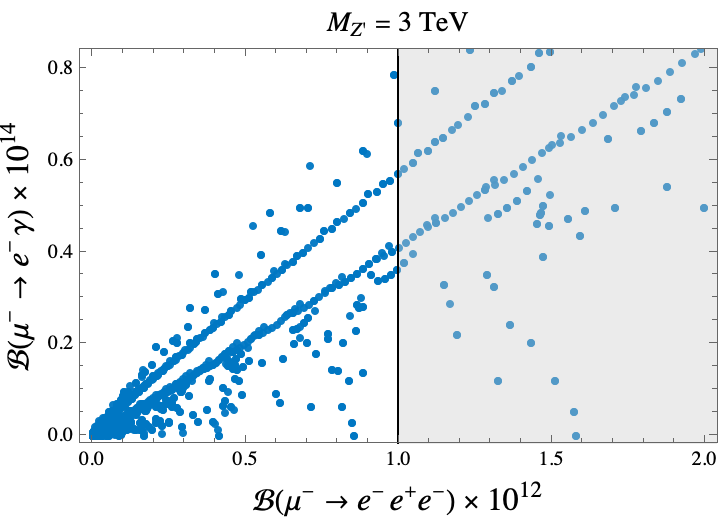} \\ \vskip 0.4cm 
\includegraphics[width = 0.4\textwidth]{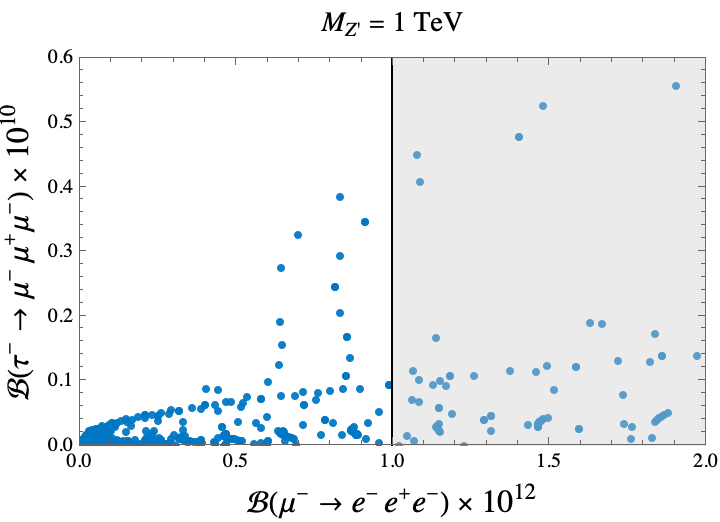} \hskip 0.4cm 
\includegraphics[width = 0.4\textwidth]{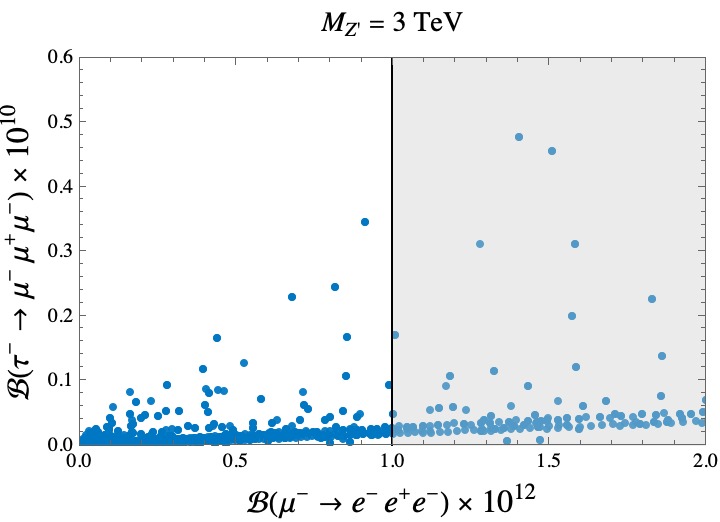}
\caption{\baselineskip 10pt  \small  Correlation between  ${\cal B}(\mu^- \to e^- e^+ e^-)$ and ${\cal B}(\mu^- \to e^- \gamma)$ (top panels) and 
 between ${\cal B}(\mu^- \to e^- e^+ e^-)$ and ${\cal B}(\tau^- \to \mu^- \mu^+ \mu^-)$  (bottom panels), for
$M_{Z^\prime}=1$ TeV (left panel) and $M_{Z^\prime}=3$ TeV (right panel).  The gray band corresponds to the upper bound ${\cal B}(\mu^- \to e^- e^+ e^-)<1 \times 10^{-12}\,\, (\rm{at}\,  90\% \, {\rm CL})$\cite{SINDRUM1988,Navas:PDG}.
}\label{leptondecaysvsmu3e}
\end{center}
\end{figure}

\begin{figure}[h]
\begin{center}
\includegraphics[width =0.45 \textwidth]{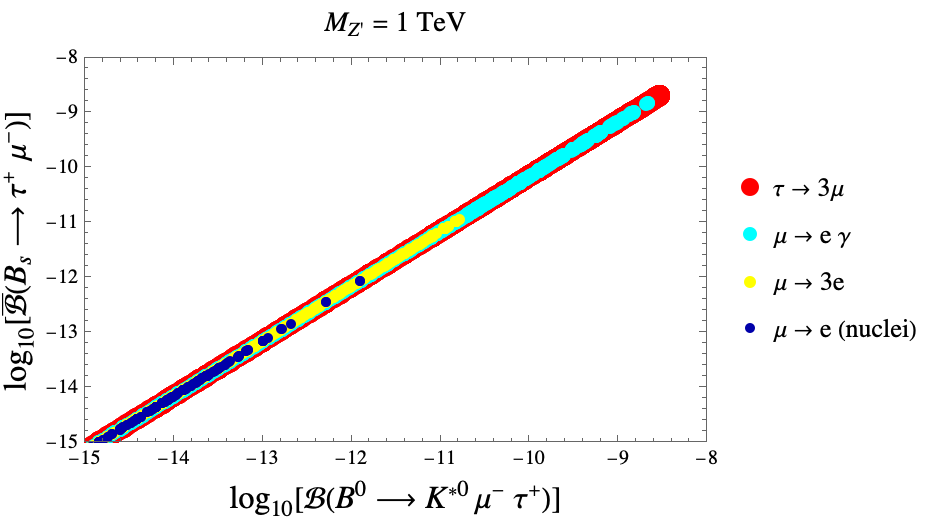} \hskip 0.4cm 
\includegraphics[width =0.45 \textwidth]{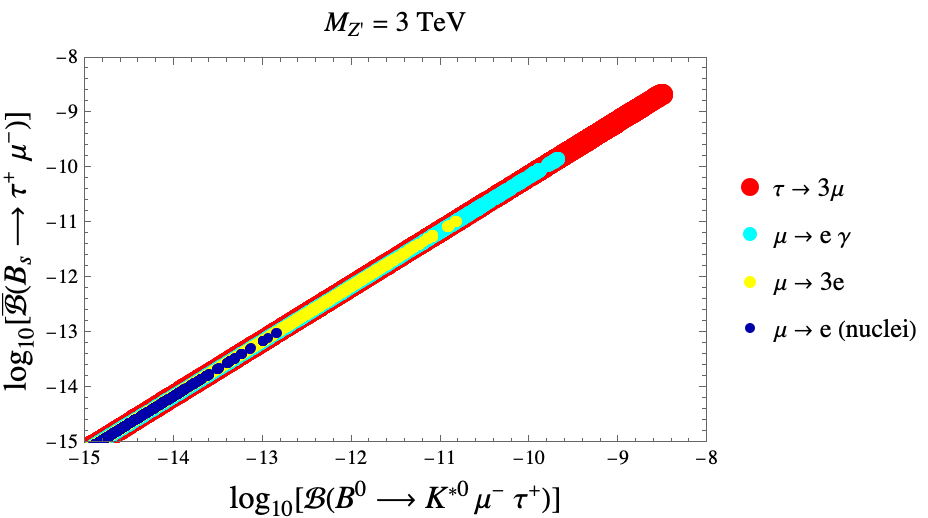} \\ 
\caption{\baselineskip 10pt  \small  Correlation between  ${\bar {\cal B}}(B_s \to \tau^+ \mu^-)$ and ${\cal B}(\bar B^0 \to \bar K^{*0} \tau^+ \mu^-)$ when constraints from LFV charged lepton decay modes are subsequently imposed as specified in the legenda, for
$M_{Z^\prime}=1$ TeV (left panel) and $M_{Z^\prime}=3$ TeV (right panel).  
}\label{correlationfinal}
\end{center}
\end{figure}

A summary of the impact of the lepton decays on ${\bar {\cal B}}(B_s \to \tau^+ \mu^-)$ and ${\cal B}(\bar B^0 \to \bar K^{*0} \tau^+ \mu^-)$ is in Fig.~\ref{correlationfinal}, where we plot their correlation   when subsequent constraints from LFV charged lepton decay modes are imposed. 
%using a logarithmic scale  to appreciate the variation.
%
It is   instructive to examine the mutual impact of different LFV lepton decays. As an example,  Fig.~\ref{leptondecaysvsmu3e} shows the constraints imposed by the experimental upper bound for ${\cal B}(\mu^- \to e^- e^+ e^-)$ on ${\cal B}(\mu^- \to e^- )$ and ${\cal B}(\tau^- \to \mu^- \mu^+ \mu^-)$: in both cases the allowed ranges are substantially reduced.
%Our analysis has shown that purely leptonic LFV processes impose stringent constraints on LFV B decays. In contrast, since the deviations observed in SM-allowed processes are comparatively small, only a moderate impact is expected in this case. We have  verified that they do not suppress the expected order of magnitude of the branching fractions of LFC modes. 

The scrutiny of $\mu^- \to e^- e^+ e^-$ and $\mu^- \to e^-$ conversion  will be the focus of future experimental programs, in particular at the COMET \cite{COMET:2018auw}, Mu2e \cite{Diociaiuti:2024stz} and Mu3e \cite{Amarinei:2025ntv} facilities that aim at an unprecedented sensitivity. The experimental bounds used in our study are based on measurements performed more than two decades ago, and have been obtained relying on theoretical formulae and approximations that need to be reconsidered. For example, for  $\mu^- \to e^- e^+ e^-$ the quoted upper  bound has been obtained  assuming a constant decay amplitude over the entire phase space. Such assumptions will  be refined in the future studies.

%
%%%%%%%%%%%%%
%
%
%%%%%%%%%%%%%
%

\subsubsection
{ Rare $B_{(s)}$ decays versus $\Delta a_\mu$}
\begin{figure}[t]
\begin{center}
\includegraphics[width =0.4 \textwidth]{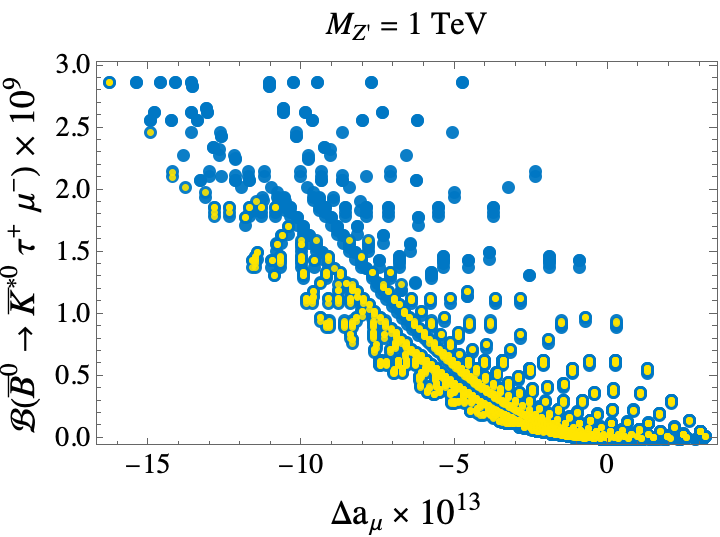} \hskip 0.4cm 
\includegraphics[width =0.4 \textwidth]{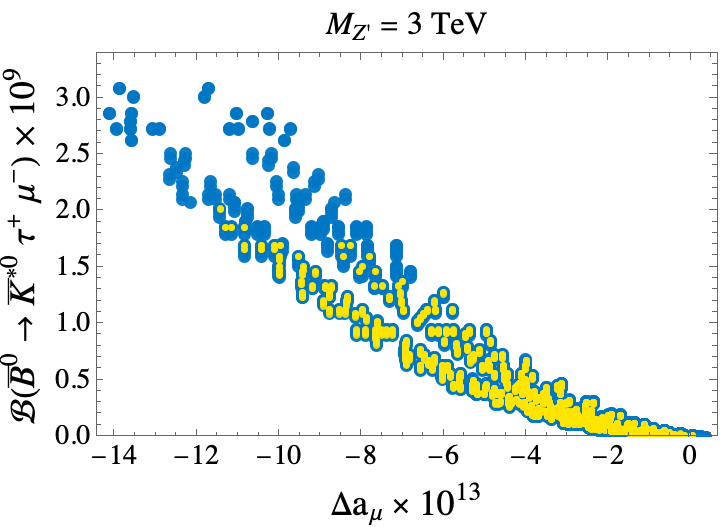}
\caption{\baselineskip 10pt  \small  Correlation between  ${\cal B}(\bar B^0 \to \bar K^{*0} \tau^+ \mu^-)$ and $\Delta a_\mu$ for
$M_{Z^\prime}=1$ TeV (left panel) and $M_{Z^\prime}=3$ TeV (right panel). 
The light yellow points  are selected requiring that  the LFC   ${\cal B}(\bar B^0 \to \bar K^{*0} \mu^+ \mu^-)$ agrees  with experiment  at $1\sigma$. }\label{gmeno2}
\end{center}
\end{figure}

%%%%%%%%%%%%%%%
%
As for the $Z^\prime$ contribution to the muon  anomalous magnetic moment,  $\Delta a_\mu$ vs ${\cal B}(\bar B^0 \to \bar K^{*0} \tau^+ \mu^-)$ is  shown in  Fig.~\ref{gmeno2}.   $\Delta a_\mu$ turns out to be negative,  it reaches at most $\Delta a_\mu \sim -10^{-12}$, a size similar to subleading hadron contributions such as the hadronic light-by-light contribution  \cite{Aoyama:2020ynm}. 

\section{Conclusions}
The search for BSM is facing the evidence that tensions with the SM predictions are spotted,  but  they are small. For example, the flavour anomalies are  below the  level of significance to claim the evidence of new physics.  Being confident on NP,   we have attempted to justify  the absence of large deviations: We have  scrutinised channels where anomalies have been detected together with  processes strictly forbidden in  SM, the charged lepton flavour violating modes, and the correlations among them.
 In a simple extension of the SM,  the ABCD model,  the new gauge boson $Z^\prime$ has 
  flavour conserving and flavour violating generation-dependent  couplings to fermions.  The  model is based on a special solution of the gauge anomaly cancellation equations which relates the quark and lepton couplings to $Z^\prime$. The flavour observables in the two sectors are intertwined,  producing nontrivial correlations.

As  noticed when  the model has been formulated \cite{Aebischer:2019blw}, quark and lepton sectors mutually act to prevent large deviations from  SM,  a welcome feature considering the small discrepancies with respect to  SM emerged by measurements of  flavour observables.
 Moreover, the possibility of having lepton flavour violation  implies that modes such as those induced by $b \to s \ell_1^- \ell_2^+$ with $\ell_1 \neq \ell_2$, can occur and are correlated both to purely leptonic LFV processes, $\mu^- \to e^- \gamma$, $\tau^- \to \mu^- \mu^+ \mu^-$, $ \mu^- \to e^- e^+ e^-$ and $\mu^- \to e^-$ conversion in nuclei,  as well as to the LFC  $b \to s \ell^- \ell^+$ transitions.

To summarize our main findings, we remark that:
\begin{itemize}
\item
Imposing $\Delta F=2$ constraints, the NP contributions to the Wilson coefficients ${\rm Re}(C_9^{\rm NP})_{\ell \ell}$ and ${\rm Re}(C_{10}^{\rm NP})_{\ell \ell}$, $\ell=\mu,\,\tau$, are found to be at the  $\simeq 10\%$ level of the corresponding SM coefficients. The analogous LFV coefficients ${\rm Re}(C_9^{\rm NP})_{\mu \tau}$ and ${\rm Re}(C_{10}^{\rm NP})_{\mu \tau}$ are  of the comparable  size to ${\rm Re}(C_{9(10)}^{\rm NP})_{\mu \mu}$, see Figs.~\ref{Wilson1000}-\ref{Wilsonmix}. 
\item
The LFV $B_{(s)} $ decays are constrained by LFC modes and by purely leptonic LFV processes in a hierarchical way;
 the weakest bounds arise from the LFC decays,  see Figs.~\ref{BsLFV}-\ref{BsphiLFV},
then  by $\tau^- \to \mu^- \mu^+ \mu^-$, $\mu^- \to e^- \gamma$ and by $ \mu^- \to e^- e^+ e^-$ and $\mu^- \to e^-$ conversion in nuclei, Figs.~\ref{Bsvsmueg1000}-\ref{Kstarvsmueg1000}. 
\item 
The role of quark and lepton observables to mutually constrain each other is visible in the results in Table \ref{tabLFV1000} and in Fig. \ref{correlationfinal}.
\end{itemize}
The above results are relevant for the NP searches   at the  new  facilities planned within the   future strategy for particle physics.

\section*{Acknowledgements}
F.D.F. thanks the coauthors of \cite{Aebischer:2019blw}.
We thank A.J. Buras for very useful discussions.
This study has  been carried out within the INFN project (Iniziativa Specifica)  SPIF.
The research has been partly funded by the European Union – Next Generation EU through the research Grant No. P2022Z4P4B “SOPHYA - Sustainable Optimised PHYsics Algorithms: fundamental physics to build an advanced society" under the program PRIN 2022 PNRR of the Italian Ministero dell’Universit\'a e Ricerca (MUR).

%%%%%%%%%%%%%%%%%%%%%%%%%%%%%%%%%%%%%%%%%%%%%%%%%%%%%%%%%%%%%%%%%%%%%%
\newpage
\appendix
\numberwithin{equation}{section}
\section{$B_q \to  V$ local form factors}\label{appA}

We use the standard parametrization of the $B_q \to V$  ($q=d,s$, $V=K^*,\,\phi$)   matrix elements of local currents in terms of form factors:

\bea
\langle V(p^\prime,\epsilon)|{\bar s} \gamma_\mu (1-\gamma_5) b| B_q(p) \rangle&=&
 \epsilon_{\mu \nu \alpha \beta} \epsilon^{* \nu} p^\alpha p^{\prime \beta}
\frac{ 2 V(q^2)}{ m_{B_q} + m_V}  \nn \\ 
&&
- i \left [ \epsilon^*_\mu (m_{B_q} + m_V) A_1(q^2) -
(\epsilon^* \cdot q) (p+p^\prime)_\mu \frac{A_2(q^2)}{m_{B_q} + m_V }
\right. \nn\\
&& - \left. (\epsilon^* \cdot q) \frac{2 m_V}{ q^2}
\big(A_3(q^2) - A_0(q^2)\big) q_\mu \right ] \,\, , \,\,\,  \hspace*{0.5cm} \label{a1}
\eea
%
 %%%%%%%%%%%%%%%%%
\begin{table}[h!]
\center{\begin{tabular}{| c | c | c | c |c |}
\hline
{\rm \footnotesize $B \to K^*$ } & {\footnotesize $m_R \,  ({\rm GeV}) $} &{\footnotesize $\alpha_0 $} & {\footnotesize $\alpha_1 $} & {\footnotesize$\alpha_2 $} \\
\hline
{\footnotesize $V$}&    {\footnotesize $5.415$}  &{\footnotesize $0.34 \pm 0.04$}   &   {\footnotesize $-1.05 \pm 0.24$ }  &  {\footnotesize $ 2.37 \pm 1.39$} \\
{\footnotesize $A_1$} & {\footnotesize  $5.829$}   &{\footnotesize $0.27 \pm 0.03$ }  &  {\footnotesize  $0.30 \pm 0.19$ }  & {\footnotesize $ -0.11 \pm 0.48$} \\
{\footnotesize $A_0$} & {\footnotesize  $5.366$}   &{\footnotesize  $0.36 \pm 0.05$}   & {\footnotesize $-1.04 \pm 0.27$ }   & {\footnotesize $1.12 \pm 1.35$} \\
{\footnotesize $A_{12}$} &{\footnotesize  $5.829$ }   & {\footnotesize $0.26 \pm 0.03$}   &{\footnotesize  $0.60 \pm 0.20$ }  &{\footnotesize  $0.12 \pm 0.84$} \\
{\footnotesize $T_1$} & {\footnotesize $5.415$}  & {\footnotesize $0.28 \pm 0.03$}    & {\footnotesize $-0.89 \pm 0.19$}  &    {\footnotesize $1.95 \pm 1.10$} \\
{\footnotesize $T_2$} &  {\footnotesize $5.829$}  &   {\footnotesize $0.28 \pm 0.03$}  & {\footnotesize $0.40 \pm 0.18$ }  & {\footnotesize $0.36 \pm 0.51$} \\
{\footnotesize $T_{23}$} & {\footnotesize $5.829$}   & {\footnotesize $0.67 \pm 0.08$}    &{\footnotesize $ 1.48 \pm 0.49$ } & {\footnotesize $1.92 \pm 1.96$}\\
 \hline
 \hline
{\rm \footnotesize $B_s \to \phi$ } & {\footnotesize $m_R \,  ({\rm GeV}) $} &{\footnotesize  $\alpha_0 $} & {\footnotesize $\alpha_1 $} & {\footnotesize $\alpha_2 $} \\
\hline
 {\footnotesize $V$} & {\footnotesize  $5.415$}  &{\footnotesize $0.39 \pm 0.03$}   & {\footnotesize  $-1.03 \pm 0.25$}   &{\footnotesize  $ 3.50 \pm 1.55$} \\
 {\footnotesize $A_1$} & {\footnotesize $5.829$ }  & {\footnotesize $0.30 \pm 0.03$ }  &  {\footnotesize $0.48 \pm 0.19$ }  & {\footnotesize$ 0.29 \pm 0.65$} \\
{\footnotesize  $A_0$ }&  {\footnotesize $5.366$ }  &  {\footnotesize $0.39 \pm 0.05$ }  & {\footnotesize $-0.78 \pm 0.26$ }   & {\footnotesize $2.41 \pm 1.48$} \\
{\footnotesize $A_{12}$} & {\footnotesize $5.829$ }   & {\footnotesize $0.25 \pm 0.03$ }   &  {\footnotesize $0.76 \pm 0.20$}   &  {\footnotesize $0.71\pm 0.96$} \\
{\footnotesize  $T_1$} & {\footnotesize $5.415$}   & {\footnotesize $0.31 \pm 0.03$ }    & {\footnotesize $-0.87 \pm 0.19$ }  &   {\footnotesize $2.75 \pm 1.19$} \\
{\footnotesize $T_2$} & {\footnotesize $5.829$ } &  {\footnotesize   $0.31 \pm 0.03$ } & {\footnotesize  $0.58 \pm 0.19$ }   & {\footnotesize $0.89 \pm 0.71$} \\
{\footnotesize $T_{23}$} & {\footnotesize $5.829$}   &  {\footnotesize $0.68 \pm 0.07$}    & {\footnotesize $ 2.11 \pm 0.46$}  &  {\footnotesize $4.94 \pm 2.25$} \\
 \hline
\end{tabular}  }
\caption { \baselineskip 10pt  \small  Parameters of the $B_{(s)} \to V$ form factors  in Eq.~\eqref{z-exp} determined in \cite{Bharucha:2015bzk}. }
\label{tabFF}
\end{table}
\bea
\langle V(p^\prime,\epsilon)|{\bar s} \sigma_{\mu \nu} q^\nu
(1+\gamma_5) b |B_q(p) \rangle&=& 
 i\epsilon_{\mu \nu \alpha
\beta} \epsilon^{* \nu} p^\alpha p^{\prime \beta}
\; 2 \; T_1(q^2)   \hspace*{0.5cm}  \nn \\
&&  \hspace*{-0.2cm}
+ \Big[ \epsilon^*_\mu (m_{B_q}^2 - m^2_V)  -
(\epsilon^* \cdot q) (p+p')_\mu \Big] \; T_2(q^2) \hspace*{0.7cm}  \nn \\
&& \hspace*{-0.2cm} + (\epsilon^* \cdot q) \left [ q_\mu - \frac{q^2}{ m_{B_q}^2 -
m^2_V} (p + p')_\mu \right ] \; T_3(q^2)  \, , \,\,\, \,\,\, 
\label{t1}
\eea
with  $q=p-p^\prime$ and $\epsilon$ the $V$ polarization vector.
The form factors are not all independent:  $A_3$ is given by
\be
A_3(q^2) =\frac {m_{B_q} + m_V }{ 2 m_V}  A_1(q^2) - \frac{m_{B_q} -
m_V }{2 m_V}  A_2(q^2)  
\ee
with $A_3(0) = A_0(0)$.  $T_1(0) = T_2(0)$ follows from the  identity 
$\displaystyle
\sigma_{\mu \nu} \gamma_5 = - \frac{i }{2}  \epsilon_{\mu \nu \alpha \beta} \sigma^{\alpha \beta}$
(with $\epsilon_{0 1 2 3}=+1$).

In our analysis we use the  determination  in \cite{Bharucha:2015bzk} based on light-cone QCD sum rules (see also  \cite{Gubernari:2023puw}).  Defining the "conformal" variable 
\be
z(t)=\frac{\sqrt{t_+-t}-\sqrt{t_+-t_0}}{\sqrt{t_+-t}+\sqrt{t_+-t_0}} \,   \label{zvar}
\ee
with $t_\pm=(m_{B_q} \pm m_V)^2$ and $t_0=t_+ \left(1-\sqrt{1-\frac{t_-}{t_+}}\right)$,
a  generic form factor $F_i$ is written as the sum of the first three terms of the $z-$expansion
\be
F_i(q^2)=\frac{1}{1-\frac{q^2}{m_{R,i}^2}}\sum_{k=0}^{2}\alpha_k^i \left[z(q^2)-z(0) \right]^k \,\, , \label{z-exp}
\ee
 with parameters $\alpha_k^i$ and   $m_{R,i}$.   The value of  $m_{R,i}$ is close to the mass of the lightest t-channel resonance involved in the various terms in  \eqref{a1}, \eqref{t1}. 
 The parameters  are determined for  $V,\,A_1,\,A_0,\,A_{12}$ and $T_1,\,T_2$, $T_{23}$ in \cite{Bharucha:2015bzk}, defining
\bea
A_{12}(q^2)&=& \frac{(m_{B_q}+m_V)^2(m_{B_q}^2-m_V^2-q^2)A_1(q^2)-\lambda(m_{B_q}^2,m_V^2,q^2)A_2(q^2)}{16m_{B_q} m_V^2(m_{B_q}+m_V)}\nn \\
T_{23}(q^2)&=& 
 \frac{(m_{B_q}^2-m_V^2)(m_{B_q}^2+3m_V^2-q^2)T_2(q^2)-\lambda(m_{B_q}^2,m_V^2,q^2)T_3(q^2)}{8m_{B_q} m_V^2(m_{B_q}-m_V)}\,\,  .
 \eea
  To make easier to reproduce our results, we quote the values of the parameters in Table \ref{tabFF}, while their correlation matrix is in \cite{Bharucha:2015bzk}.

 \numberwithin{equation}{section}
\section{Angular coefficient functions for $B \to K^*(K \pi) \ell_1^- \ell_2^+$ }\label{appAngularCoeff}
The coefficient functions $I_i^{(a)}$, for $i=1,\dots 9$ and $a=s,\,c$ in \eqref{d4G}, can be written in terms of the  transversity amplitudes
\bea
A_{\perp L,R}(q^2)  &= & N(q^2) \sqrt{2} \lambda_H^{1/2} \bigg[ 
\left[ (C_9^{\rm eff} + C_9^{{\rm eff}\prime}) \mp (C_{10}^{\rm eff}+ C_{10}^{{\rm eff}\prime}) \right] \frac{ V(q^2) }{ m_B + m_{K^*}} \nn \\
& +& \frac{2(m_b+m_s)}{q^2} (C_7^{\rm eff} + C_7^{{\rm eff}\prime}) T_1(q^2)
\bigg], 
\nn
\\
A_{\parallel L,R}(q^2)  & =& - N(q^2) \sqrt{2}(m_B^2 - m_{K^*}^2) \bigg[ \left[ (C_9^{\rm eff} - C_9^{{\rm eff}\prime}) \mp (C_{10}^{\rm eff}- C_{10}^{{\rm eff}\prime}) \right] 
\frac{A_1(q^2)}{m_B-m_{K^*}}
\nonumber\\
&+&\frac{2 (m_b-m_s)}{q^2} (C_7^{\rm eff} - C_7^{{\rm eff}\prime}) T_2(q^2)
\bigg],  \nn \\
A_{0L,R}(q^2)  &=&  - \frac{N(q^2)}{2 m_{K^*} \sqrt{q^2}}  \bigg\{ 
 \left[ (C_9^{\rm eff} - C_9^{{\rm eff}\prime}) \mp (C_{10}^{\rm eff} - C_{10}^{{\rm eff}\prime}) \right]
\\
 &\times & 
\bigg[ (m_B^2 - m_{K^*}^2 - q^2) ( m_B + m_{K^*}) A_1(q^2) 
 -\lambda_H \frac{A_2(q^2)}{m_B + m_{K^*}}
\bigg] 
\nonumber\\
&+& {2  (m_b-m_s)}(C_7^{\rm eff}- C_7^{{\rm eff}\prime}) \bigg[
 (m_B^2 + 3 m_{K^*}^2 - q^2) T_2(q^2)
-\frac{\lambda_H}{m_B^2 - m_{K^*}^2} T_3(q^2) \bigg]
\bigg\},\nn \\
 A_{t \, L,R}(q^2)  &=&- \frac{N(q^2)}{\sqrt{q^2}}\lambda_H^{1/2}  \big[ (C_9^{\rm eff} - C_9^{{\rm eff}\prime})  \mp (C_{10}^{\rm eff} - C_{10}^{{\rm eff}\prime})  \Big] A_0(q^2) .\nn 
 \eea
The factor $N(q^2)$ reads
\be
N(q^2)=\lambda_{ts}^*\Big[\frac{G_F^2 \alpha^2}{3 \times 2^{10} \pi^5 m_B^3}\lambda_H^{1/2}\lambda_L^{1/2} \Big]^{1/2} \,\, ,
\ee
with 
$\lambda_H=\lambda(q^2,m_B^2,m_{K^*}^2)$ and $\lambda_L=\lambda(q^2,m_1^2,m_2^2)$.
The expressions of the coefficient functions are:
\bea
  I_1^s & = &\frac{\lambda_L +2[q^4-(m_1^2-m_2^2)^2] }{4q^4} \left[|A_{\perp L}|^2 + |A_{\parallel L}|^2 + (L\to R) \right] 
            + \frac{4m_1 m_2}{q^2}  {\rm Re}\left[A_{\perp L}A_{\perp R}^* + A_{\parallel L}A_{\parallel R}^*\right], 
            \nn 
\\
  I_1^c & = &\frac{q^4-(m_1^2-m_2^2)^2 }{q^4} \big[|A_{0 L}|^2 +|A_{0 R}|^2 \big]
   + \frac{8m_1 m_2}{q^2}  {\rm Re}\big[A_{0 L}A_{0 R}^* - A_{t\, L}A_{t\, R}^*\big] \nn \\
   && -2\frac{(m_1^2-m_2^2)^2-q^2(m_1^2+m_2^2)}{q^4}
               \left[|A_{t \,L}|^2 +|A_{t \,R}|^2 \right]  ,
               \nn
\\
  I_2^s & =& \frac{\lambda_L}{4q^4}\left[ |A_{\perp L}|^2+ |A_{\parallel L}|^2 + (L\to R)\right],
  \nn
\\
  I_2^c & =& -\frac{\lambda_L}{q^4}\left[|A_{0 L}|^2 + (L\to R)\right],
  \nn
\\
  I_3 & =& \frac{\lambda_L}{2q^4}\left[ |A_{\perp L}|^2 - |A_{\parallel L}|^2  + (L\to R)\right],
  \nn
\\
  I_4 & =&\frac{\lambda_L}{\sqrt{2}q^4}\left[ {\rm Re} (A_{0 L}A_{\parallel L}^*) + (L\to R)\right],
\label{angcoef}
\\
  I_5 & = &\frac{\sqrt{2}\lambda_L^{1/2}}{q^2}\left[ {\rm Re}[A_{0 L}A_{\perp L}^* - (L\to R)] -\frac{m_1^2-m_2^2}{q^2}{\rm Re}[A_{t\, L}A_{\parallel L}^* + (L\to R)]
\right],
\nn
\\
  I_6^s  & = &\frac{2\lambda_L^{1/2}}{q^2}\left[ {\rm Re} (A_{\parallel L}A_{\perp L}^*) - (L\to R) \right],
  \nn
\\
   I_6^c  &  =&\frac{4\lambda_L^{1/2}}{q^2}\frac{m_1^2-m_2^2}{q^2}\left[ {\rm Re}(A_{0 L}A_{t \, L}^*) + (L\to R)\right],
 \nn
\\
%\eea
%\bea
  I_7 & =&\frac{\sqrt{2}\lambda_L^{1/2}}{q^2} \left[ {\rm Im} [A_{0 L}A_{\parallel L}^* - (L\to R)]+ \frac{m_1^2-m_2^2}{q^2}{\rm Im}[A_{\perp L}A_{t\, L}^* + (L\to R)]
\right],
\nn
\\
  I_8 & =& \frac{\lambda_L}{\sqrt{2}q^4}\left[ {\rm Im}(A_{0 L}A_{\perp L}^*) + (L\to R)\right],
  \nn
\\
  I_9 & =&\frac{\lambda_L}{q^4}\left[ {\rm Im} (A_{\parallel L}^{*}A_{\perp L}) + (L\to R)\right].
\nn
\eea
The above expressions are modified if other BSM operators  are included in the low-energy Hamiltonian,  besides the ones  in \eqref{hamil}.

\newpage
\section{Parametrization of the CKM and PMNS matrices} \label{AppCKM} 
We  use the standard parametrization of the CKM matrix
 \be
V_{\rm CKM} =
\left(
\begin{array}{ccc}
c_{12}c_{13}&s_{12}c_{13}&s_{13}e^{-i\delta}\\ -s_{12}c_{23}
-c_{12}s_{23}s_{13}e^{i\delta} &c_{12}c_{23}-s_{12}s_{23}s_{13}e^{i\delta}&
s_{23}c_{13}\\
s_{12}s_{23}-c_{12}c_{23}s_{13}e^{i\delta}&-s_{23}c_{12} -s_{12}c_{23}s_{13}e^{i\delta}&c_{23}c_{13}
\end{array}
\right)\,.
\label{CKMmat}
\ee
As four independent parameters we choose $V_{us}$,  $|V_{cb}|$,  $|V_{ub}|$ and the phase $\gamma$,  so that
 \bea
 s_{13}&=&|V_{ub}|  \,\, , \quad  s_{12}=\frac{V_{us}}{c_{13}} \,\, , \quad  s_{23}=\frac{|V_{cb}|}{c_{13}} \,\, , \quad   \delta=\gamma \,\, , \nn \\
\quad c_{ij}&=&\sqrt{1-s_{ij}^2}  \,\,. \label{CKMpar} 
 \eea
 $V_{us}$ and $\gamma$  are  set to $V_{us}=0.2253$ and $\gamma=64.6^\circ$. 
  $|V_{cb}|$ and $|V_{ub}|$ are varied in the ranges spanned by the exclusive-inclusive measurements \eqref{VcbVubRanges},
 with 
 $|V_{cb}|_{exc}=(39.1 \pm 0.5)\times 10^{-3}$,  $|V_{cb}|_{inc}=(42.19 \pm 0.78)\times 10^{-3}$,  
  $|V_{ub}|_{exc}=(3.51 \pm 0.12)\times 10^{-3}$,  $|V_{ub}|_{inc}=(4.19 \pm 0.17)\times 10^{-3}$  \cite{HeavyFlavorAveragingGroupHFLAV:2024ctg}.
 
 The standard parametrization is also used for the PMNS matrix
 \be
U_{\rm PMNS} =
\left(
\begin{array}{ccc}
k_{12}k_{13}&x_{12}k_{13}&x_{13}e^{-i\delta_{\rm PMNS}}\\ -x_{12}k_{23}
-k_{12}x_{23}x_{13}e^{i\delta_{\rm PMNS}} &k_{12}k_{23}-x_{12}x_{23}x_{13}e^{i\delta_{\rm PMNS}}&
x_{23}k_{13}\\
x_{12}x_{23}-k_{12}k_{23}x_{13}e^{i\delta_{\rm PMNS}}&-x_{23}k_{12} -x_{12}k_{23}x_{13}e^{i\delta_{\rm PMNS}}&k_{23}k_{13}
\end{array}
\right)\,,
\label{PMNSmat}
\ee
where $x_{ij}=\sin \theta_{ij} $ and $k_{ij}=\cos \theta_{ij}$. A global fit of the parameters  in \eqref{PMNSmat}  gives values which depend on the assumed hierarchy among the neutrino masses  \cite{Esteban:2024eli}:
for normal ordering 
 \bea
 \sin^2 \theta_{12}&=& 0.308^{+0.012}_{-0.011} \,\, ,
\quad
  \sin^2 \theta_{23}= 0.470^{+0.017}_{-0.013} \,\, , 
\quad
  \sin^2 \theta_{13}= 0.02215^{+0.00056}_{-0.00058} \,\, , 
 \nn \\
 \delta_{\rm PMNS}&=&(212 ^{+26}_{-41})^\circ \,\, ,\label{NO}
 \eea
for  inverted ordering 
  \bea
 \sin^2 \theta_{12}&=& 0.308^{+0.012}_{-0.011} \,\, , 
\quad 
  \sin^2 \theta_{23}= 0.550^{+0.012}_{-0.015} \,\, , 
\quad
  \sin^2 \theta_{13}= 0.02231^{+0.00056}_{-0.00056} \,\, , 
 \nn \\
 \delta_{\rm PMNS}&=&(274^{+22}_{-25})^\circ \,\,.
 \label{IO}
 \eea
 The  global fit   in  \cite{Capozzi:2025wyn} gives consistent results.
 In the numerical analysis we  use the parameters for normal ordering  \eqref{NO}.

\newpage
\bibliographystyle{JHEP}
\bibliography{refDFP}

@article{SINDRUM:1987nra,
    author = "Bellgardt, U. and others",
    collaboration = "SINDRUM",
    title = "{Search for the Decay $\mu^+ \to e^+ e^+ e^-$}",
    reportNumber = "SIN-PR-87-09",
    doi = "10.1016/0550-3213(88)90462-2",
    journal = "Nucl. Phys. B",
    volume = "299",
    pages = "1--6",
    year = "1988"
}

@article{Holdom:1985ag,
    author = "Holdom, Bob",
    title = "{Two U(1)'s and Epsilon Charge Shifts}",
    reportNumber = "UTPT-85-30",
    doi = "10.1016/0370-2693(86)91377-8",
    journal = "Phys. Lett. B",
    volume = "166",
    pages = "196--198",
    year = "1986"
}

@article{Babu:1996vt,
    author = "Babu, K. S. and Kolda, Christopher F. and March-Russell, John",
    title = "{Leptophobic U(1) $s$ and the R($b$) - R($c$) crisis}",
    eprint = "hep-ph/9603212",
    archivePrefix = "arXiv",
    reportNumber = "IASSNS-HEP-96-20",
    doi = "10.1103/PhysRevD.54.4635",
    journal = "Phys. Rev. D",
    volume = "54",
    pages = "4635--4647",
    year = "1996"
}

@article{COMET:2018auw,
    author = "Abramishvili, R. and others",
    collaboration = "COMET",
    title = "{COMET Phase-I Technical Design Report}",
    eprint = "1812.09018",
    archivePrefix = "arXiv",
    primaryClass = "physics.ins-det",
    doi = "10.1093/ptep/ptz125",
    journal = "PTEP",
    volume = "2020",
    number = "3",
    pages = "033C01",
    year = "2020"
}

@article{Diociaiuti:2024stz,
    author = "Diociaiuti, Eleonora",
    collaboration = "Mu2e",
    title = "{Status and perspectives of cLFV at Mu2e}",
    reportNumber = "FERMILAB-CONF-24-0806-V",
    doi = "10.22323/1.457.0017",
    journal = "PoS",
    volume = "WIFAI2023",
    pages = "017",
    year = "2024"
}

@article{Amarinei:2025ntv,
    author = "Amarinei, Robert Mihai",
    title = "{The Mu3e Experiment: Status and Short-Term Plans}",
    eprint = "2501.14667",
    archivePrefix = "arXiv",
    primaryClass = "hep-ex",
    month = "1",
    year = "2025"
}

@article{Wintz:1998rp,
    author = "Wintz, P.",
    editor = "Klapdor-Kleingrothaus, H. V. and Krivosheina, I. V.",
    title = "{Results of the SINDRUM-II experiment}",
    journal = "Conf. Proc. C",
    volume = "980420",
    pages = "534--546",
    year = "1998"
}

@article{SINDRUMII:2006dvw,
    author = "Bertl, Wilhelm H. and others",
    collaboration = "SINDRUM II",
    title = "{A Search for muon to electron conversion in muonic gold}",
    doi = "10.1140/epjc/s2006-02582-x",
    journal = "Eur. Phys. J. C",
    volume = "47",
    pages = "337--346",
    year = "2006"
}

@article{Hisano:1995cp,
    author = "Hisano, J. and Moroi, T. and Tobe, K. and Yamaguchi, Masahiro",
    title = "{Lepton flavor violation via right-handed neutrino Yukawa couplings in supersymmetric standard model}",
    eprint = "hep-ph/9510309",
    archivePrefix = "arXiv",
    reportNumber = "TIT-HEP-304, NSF-ITP-95-127, KEK-TH-450, LBL-37816, UT-727, TU-491",
    doi = "10.1103/PhysRevD.53.2442",
    journal = "Phys. Rev. D",
    volume = "53",
    pages = "2442--2459",
    year = "1996"
}

@article{Kitano:2002mt,
    author = "Kitano, Ryuichiro and Koike, Masafumi and Okada, Yasuhiro",
    title = "{Detailed calculation of lepton flavor violating muon electron conversion rate for various nuclei}",
    eprint = "hep-ph/0203110",
    archivePrefix = "arXiv",
    reportNumber = "KEK-TH-808",
    doi = "10.1103/PhysRevD.76.059902",
    journal = "Phys. Rev. D",
    volume = "66",
    pages = "096002",
    year = "2002",
    note = "[Erratum: Phys.Rev.D 76, 059902 (2007)]"
}

@article{Bernabeu:1993ta,
    author = "Bernabeu, J. and Nardi, E. and Tommasini, D.",
    title = "{$\mu$ - $e$ conversion in nuclei and $Z^\prime$ physics}",
    eprint = "hep-ph/9306251",
    archivePrefix = "arXiv",
    reportNumber = "UM-TH-93-08, FTUV-93-14",
    doi = "10.1016/0550-3213(93)90446-V",
    journal = "Nucl. Phys. B",
    volume = "409",
    pages = "69--86",
    year = "1993"
}

@article{Suzuki:1987jf,
    author = "Suzuki, T. and Measday, David F. and Roalsvig, J. P.",
    title = "{Total Nuclear Capture Rates for Negative Muons}",
    reportNumber = "TRI-PP-87-5",
    doi = "10.1103/PhysRevC.35.2212",
    journal = "Phys. Rev. C",
    volume = "35",
    pages = "2212",
    year = "1987"
}

@article{SINDRUM1988,
  title={Search for the Decay $\mu^+ \to e^+ e^+ e^-$},
  author={Bellgardt et al.},
  journal={Nucl. Phys. B},
  volume={299},
  year={1988},
  pages={1--6},
  doi={10.1016/0550-3213(88)90462-2}
}

@article{Kaulard:1998rb,
      author         = "Kaulard, J. and others",
      title          = "{Improved limit on the branching ratio of $\mu \to e$
                        conversion on titanium}",
      collaboration  = "SINDRUM II",
      journal        = "Phys.~Lett.",
      volume         = "B422",
      pages          = "334-338",
      doi            = "10.1016/S0370-2693(97)01423-8",
      year           = "1998",
      SLACcitation   = "%%CITATION = PHLTA,B422,334;%%",
}

@article{DelAguila:1996fw,
    author = "Del Aguila, F. and Masip, M. and Perez-Victoria, M.",
    editor = "Staruszkiewicz, A.",
    title = "{Search for new neutral bosons at future colliders}",
    eprint = "hep-ph/9603347",
    archivePrefix = "arXiv",
    reportNumber = "UG-FT-55-96",
    journal = "Acta Phys. Polon. B",
    volume = "27",
    pages = "1469--1478",
    year = "1996"
}

@article{Blanke:2018cya,
    author = "Blanke, Monika and Buras, Andrzej J.",
    title = "{Emerging $\Delta M_{d}$ -anomaly from tree-level determinations of $|V_{cb}|$ and the angle $\gamma $}",
    eprint = "1812.06963",
    archivePrefix = "arXiv",
    primaryClass = "hep-ph",
    reportNumber = "TTP18-044, AJB-18-10",
    doi = "10.1140/epjc/s10052-019-6667-x",
    journal = "Eur. Phys. J. C",
    volume = "79",
    number = "2",
    pages = "159",
    year = "2019"
}

@article{Crivellin:2015lwa,
    author = "Crivellin, Andreas and D'Ambrosio, Giancarlo and Heeck, Julian",
    title = "{Addressing the LHC flavor anomalies with horizontal gauge symmetries}",
    eprint = "1503.03477",
    archivePrefix = "arXiv",
    primaryClass = "hep-ph",
    reportNumber = "CERN-PH-TH-2015-046, ULB-TH-15-03",
    doi = "10.1103/PhysRevD.91.075006",
    journal = "Phys. Rev. D",
    volume = "91",
    pages = "075006",
    year = "2015"
}

@article{Crivellin:2015era,
    author = "Crivellin, Andreas and Hofer, Lars and Matias, Joaquim and Nierste, Ulrich and Pokorski, Stefan and Rosiek, Janusz",
    title = "{Lepton-flavour violating $B$ decays in generic $Z'$ models}",
    eprint = "1504.07928",
    archivePrefix = "arXiv",
    primaryClass = "hep-ph",
    reportNumber = "CERN-PH-TH-2015-091, TTP15-018",
    doi = "10.1103/PhysRevD.92.054013",
    journal = "Phys. Rev. D",
    volume = "92",
    pages = "054013",
    year = "2015"
}

@article{Muong-2:2025xyk,
    author = "Aguillard, D. P. and others",
    collaboration = "Muon g-2",
    title = "{Measurement of the Positive Muon Anomalous Magnetic Moment to 127 ppb}",
    eprint = "2506.03069",
    archivePrefix = "arXiv",
    primaryClass = "hep-ex",
    reportNumber = "FERMILAB-PUB-25-0364-PPD",
    month = "6",
    year = "2025"
}

@article{Muong-2:2006rrc,
    author = "Bennett, G. W. and others",
    collaboration = "Muon g-2",
    title = "{Final Report of the Muon E821 Anomalous Magnetic Moment Measurement at BNL}",
    eprint = "hep-ex/0602035",
    archivePrefix = "arXiv",
    doi = "10.1103/PhysRevD.73.072003",
    journal = "Phys. Rev. D",
    volume = "73",
    pages = "072003",
    year = "2006"
}

@article{Muong-2:2023cdq,
    author = "Aguillard, D. P. and others",
    collaboration = "Muon g-2",
    title = "{Measurement of the Positive Muon Anomalous Magnetic Moment to 0.20~ppm}",
    eprint = "2308.06230",
    archivePrefix = "arXiv",
    primaryClass = "hep-ex",
    reportNumber = "FERMILAB-PUB-23-385-AD-CSAID-PPD",
    doi = "10.1103/PhysRevLett.131.161802",
    journal = "Phys. Rev. Lett.",
    volume = "131",
    pages = "161802",
    year = "2023"
}

@article{Ciuchini:2022wbq,
    author = "Ciuchini, Marco and Fedele, Marco and Franco, Enrico and Paul, Ayan and Silvestrini, Luca and Valli, Mauro",
    title = "{Constraints on lepton universality violation from rare B decays}",
    eprint = "2212.10516",
    archivePrefix = "arXiv",
    primaryClass = "hep-ph",
    reportNumber = "P3H-22-127, TTP22-073, YITP-SB-22-42",
    doi = "10.1103/PhysRevD.107.055036",
    journal = "Phys. Rev. D",
    volume = "107",
    pages = "055036",
    year = "2023"
}

@article{Capdevila:2023yhq,
    author = "Capdevila, Bernat and Crivellin, Andreas and Matias, Joaquim",
    title = "{Review of semileptonic B anomalies}",
    eprint = "2309.01311",
    archivePrefix = "arXiv",
    primaryClass = "hep-ph",
    reportNumber = "PSI-PR-23-33, ZU-TH 50/23",
    doi = "10.1140/epjs/s11734-023-01012-2",
    journal = "Eur. Phys. J. ST",
    volume = "1",
    pages = "20",
    year = "2023"
}

@article{Belle-II:2025hpl,
    author = "Adachi, I. and others",
    collaboration = "Belle-II, Belle",
    title = "{Search for lepton flavor-violating decay modes $B^0 \to K^{\ast 0}\tau^\pm\ell^\mp$ ($\ell = e,\mu$) with hadronic B-tagging at Belle and Belle II}",
    eprint = "2505.08418",
    archivePrefix = "arXiv",
    primaryClass = "hep-ex",
    reportNumber = "Belle II preprint: 2025-014, KEK preprint: 2025-13",
    month = "5",
    year = "2025"
}

@article{LHCb:2022wrs,
    author = "Aaij, R. and others",
    collaboration = "LHCb",
    title = "{Search for the lepton-flavour violating decays $B^0 \to K^{*0} \tau^\pm \mu^\mp$}",
    eprint = "2209.09846",
    archivePrefix = "arXiv",
    primaryClass = "hep-ex",
    reportNumber = "LHCb-PAPER-2022-021, CERN-EP-2022-154",
    doi = "10.1007/JHEP06(2023)143",
    journal = "JHEP",
    volume = "06",
    pages = "143",
    year = "2023"
}

@article{Beneke:2001at,
    author = "Beneke, M. and Feldmann, T. and Seidel, D.",
    title = "{Systematic approach to exclusive $B \to  V l^+ l^-$, $V \gamma$ decays}",
    eprint = "hep-ph/0106067",
    archivePrefix = "arXiv",
    reportNumber = "PITHA-01-05",
    doi = "10.1016/S0550-3213(01)00366-2",
    journal = "Nucl. Phys. B",
    volume = "612",
    pages = "25--58",
    year = "2001"
}

@article{Colangelo:1989gi,
    author = "Colangelo, P. and Nardulli, G. and Paver, N. and Riazuddin",
    title = "{Long Distance Effects in $b \to s$ Exclusive Decays}",
    reportNumber = "BARI-TH-89-52",
    doi = "10.1007/BF01556270",
    journal = "Z. Phys. C",
    volume = "45",
    pages = "575",
    year = "1990"
}

@article{Bobeth:2017vxj,
    author = "Bobeth, Christoph and Chrzaszcz, Marcin and van Dyk, Danny and Virto, Javier",
    title = "{Long-distance effects in $B\rightarrow K^*\ell \ell $ from analyticity}",
    eprint = "1707.07305",
    archivePrefix = "arXiv",
    primaryClass = "hep-ph",
    reportNumber = "EOS-2017-01, MIT-CTP-4918, TUM-HEP-1087-17, ZU-TH-17-17",
    doi = "10.1140/epjc/s10052-018-5918-6",
    journal = "Eur. Phys. J. C",
    volume = "78",
    pages = "451",
    year = "2018"
}

@article{Bordone:2024hui,
    author = {Bordone, Marzia and Isidori, Gino and M\"achler, Sandro and Tinari, Arianna},
    title = "{Short- vs. long-distance physics in $B\rightarrow K^{(*)} \ell ^+\ell ^-$: a data-driven analysis}",
    eprint = "2401.18007",
    archivePrefix = "arXiv",
    primaryClass = "hep-ph",
    reportNumber = "CERN-TH-2024-017",
    doi = "10.1140/epjc/s10052-024-12869-5",
    journal = "Eur. Phys. J. C",
    volume = "84",
    pages = "547",
    year = "2024"
}

@article{Khodjamirian:2012rm,
    author = "Khodjamirian, A. and Mannel, Th. and Wang, Y. M.",
    title = "{$B \to K \ell^{+}\ell^{-}$ decay at large hadronic recoil}",
    eprint = "1211.0234",
    archivePrefix = "arXiv",
    primaryClass = "hep-ph",
    reportNumber = "SI-HEP-2012-10",
    doi = "10.1007/JHEP02(2013)010",
    journal = "JHEP",
    volume = "02",
    pages = "010",
    year = "2013"
}

@article{Khodjamirian:2010vf,
    author = "Khodjamirian, A. and Mannel, Th. and Pivovarov, A. A. and Wang, Y. -M.",
    title = "{Charm-loop effect in $B \to K^{(*)} \ell^{+} \ell^{-}$ and $B\to K^*\gamma$}",
    eprint = "1006.4945",
    archivePrefix = "arXiv",
    primaryClass = "hep-ph",
    reportNumber = "SI-HEP-2010-08",
    doi = "10.1007/JHEP09(2010)089",
    journal = "JHEP",
    volume = "09",
    pages = "089",
    year = "2010"
}

@article{Altmannshofer:2008dz,
    author = "Altmannshofer, Wolfgang and Ball, Patricia and Bharucha, Aoife and Buras, Andrzej J. and Straub, David M. and Wick, Michael",
    title = "{Symmetries and Asymmetries of $B \to K^{*} \mu^{+} \mu^{-}$ Decays in the Standard Model and Beyond}",
    eprint = "0811.1214",
    archivePrefix = "arXiv",
    primaryClass = "hep-ph",
    reportNumber = "IPPP-08-58, DCPT-08-116, TUM-HEP-696-08",
    doi = "10.1088/1126-6708/2009/01/019",
    journal = "JHEP",
    volume = "01",
    pages = "019",
    year = "2009"
}

@article{EOSAuthors:2021xpv,
    author = "van Dyk, Danny and others",
    collaboration = "EOS Authors",
    title = "{EOS: a software for flavor physics phenomenology}",
    eprint = "2111.15428",
    archivePrefix = "arXiv",
    primaryClass = "hep-ph",
    reportNumber = "EOS-2021-04, TUM-HEP 1371/21, P3H-21-094, SI-HEP-2021-32",
    doi = "10.1140/epjc/s10052-022-10177-4",
    journal = "Eur. Phys. J. C",
    volume = "82",
    pages = "569",
    year = "2022"
}

@article{LHCb:2024wve,
    author = "Aaij, Roel and others",
    collaboration = "LHCb",
    title = "{Search for the lepton-flavor violating decay $B_s^0 \to \phi \mu^\pm \tau^\mp$}",
    eprint = "2405.13103",
    archivePrefix = "arXiv",
    primaryClass = "hep-ex",
    reportNumber = "LHCb-PAPER-2024-006, CERN-EP-2024-114",
    doi = "10.1103/PhysRevD.110.072014",
    journal = "Phys. Rev. D",
    volume = "110",
    pages = "072014",
    year = "2024"
}

@article{MEG:2016leq,
    author = "Baldini, A. M. and others",
    collaboration = "MEG",
    title = "{Search for the lepton flavour violating decay $\mu ^+ \rightarrow \mathrm {e}^+ \gamma $ with the full dataset of the MEG experiment}",
    eprint = "1605.05081",
    archivePrefix = "arXiv",
    primaryClass = "hep-ex",
    doi = "10.1140/epjc/s10052-016-4271-x",
    journal = "Eur. Phys. J. C",
    volume = "76",
    pages = "434",
    year = "2016"
}

@article{MEGnew,
    author = "Afanaciev, K. and others",
    collaboration = "MEG",
    title = "{New limit on $\mu ^+ \rightarrow \mathrm {e}^+ \gamma $ with the  MEG II experiment}",
    eprint = "2504.15711",
    archivePrefix = "arXiv",
    primaryClass = "hep-ex",
    doi = "10.48550/arXiv.2504.15711",
    year = "2025"
}

@article{Lindner:2016bgg,
    author = "Lindner, Manfred and Platscher, Moritz and Queiroz, Farinaldo S.",
    title = "{A Call for New Physics: The Muon Anomalous Magnetic Moment and Lepton Flavor Violation}",
    eprint = "1610.06587",
    archivePrefix = "arXiv",
    primaryClass = "hep-ph",
    doi = "10.1016/j.physrep.2017.12.001",
    journal = "Phys. Rept.",
    volume = "731",
    pages = "1--82",
    year = "2018"
}

@article{Altmannshofer:2014rta,
    author = "Altmannshofer, Wolfgang and Straub, David M.",
    title = "{New physics in $b\rightarrow s$ transitions after LHC run 1}",
    eprint = "1411.3161",
    archivePrefix = "arXiv",
    primaryClass = "hep-ph",
    doi = "10.1140/epjc/s10052-015-3602-7",
    journal = "Eur. Phys. J. C",
    volume = "75",
    pages = "382",
    year = "2015"
}

@article{Capdevila:2017bsm,
    author = "Capdevila, Bernat and Crivellin, Andreas and Descotes-Genon, S\'ebastien and Matias, Joaquim and Virto, Javier",
    title = "{Patterns of New Physics in $b\to s\ell^+\ell^-$ transitions in the light of recent data}",
    eprint = "1704.05340",
    archivePrefix = "arXiv",
    primaryClass = "hep-ph",
    reportNumber = "PSI-PR-17-05, LPT-ORSAY-17-19",
    doi = "10.1007/JHEP01(2018)093",
    journal = "JHEP",
    volume = "01",
    pages = "093",
    year = "2018"
}

@article{Alguero:2021anc,
    author = "Alguer\'o, Marcel and Capdevila, Bernat and Descotes-Genon, S\'ebastien and Matias, Joaquim and Novoa-Brunet, Mart\'\i{}n",
    title = "{$b\rightarrow s\ell ^+\ell ^-$ global fits after $R_{K_S}$ and $R_{K^{*+}}$}",
    eprint = "2104.08921",
    archivePrefix = "arXiv",
    primaryClass = "hep-ph",
    reportNumber = "BARI-TH/21-732",
    doi = "10.1140/epjc/s10052-022-10231-1",
    journal = "Eur. Phys. J. C",
    volume = "82",
    pages = "326",
    year = "2022"
}

@article{Alguero:2023jeh,
    author = "Alguer\'o, Marcel and Biswas, Aritra and Capdevila, Bernat and Descotes-Genon, S\'ebastien and Matias, Joaquim and Novoa-Brunet, Mart\'\i{}n",
    title = "{To (b)e or not to (b)e: no electrons at LHCb}",
    eprint = "2304.07330",
    archivePrefix = "arXiv",
    primaryClass = "hep-ph",
    reportNumber = "BARI-TH/23-747",
    doi = "10.1140/epjc/s10052-023-11824-0",
    journal = "Eur. Phys. J. C",
    volume = "83",
    pages = "648",
    year = "2023"
}

@article{ODonnell:1991cdx,
    author = "O'Donnell, Patrick J. and Tung, Humphrey K. K.",
    title = "{Resonance contributions to the decay $b \to s \ell^+ \ell^-$}",
    reportNumber = "UTPT-90-21",
    doi = "10.1103/PhysRevD.43.R2067",
    journal = "Phys. Rev. D",
    volume = "43",
    pages = "2067--2069",
    year = "1991"
}

@article{Lim:1988yu,
    author = "Lim, C. S. and Morozumi, T. and Sanda, A. I.",
    title = "{A Prediction for $ d \Gamma (b \to s \ell \bar \ell )/d q^2$ Including the Long Distance Effects}",
    reportNumber = "KEK-TH-214, KEK-Preprint-88-79, DOE/ER/40325-50",
    doi = "10.1016/0370-2693(89)91593-1",
    journal = "Phys. Lett. B",
    volume = "218",
    pages = "343--347",
    year = "1989"
}

@article{Deshpande:1988bd,
    author = "Deshpande, N. G. and Trampetic, Josip and Panose, Kuriakose",
    title = "{Resonance Background to the Decays $b \to s \ell^+ \ell^-, B \to K^* \ell^+ \ell^-$ and $B \to K \ell^+ \ell^-$}",
    reportNumber = "OITS-394",
    doi = "10.1103/PhysRevD.39.1461",
    journal = "Phys. Rev. D",
    volume = "39",
    pages = "1461",
    year = "1989"
}

@article{Paver:1991tn,
    author = "Paver, Nello and Riazuddin",
    title = "{On the interference between short distance and long distance contributions to $b \to  s \ell^+ \ell^-$}",
    reportNumber = "PRINT-91-0453 (TRIESTE)",
    doi = "10.1103/PhysRevD.45.978",
    journal = "Phys. Rev. D",
    volume = "45",
    pages = "978--979",
    year = "1992"
}

@article{BaBar:2006tnv,
    author = "Aubert, Bernard and others",
    collaboration = "BaBar",
    title = "{Measurements of branching fractions, rate asymmetries, and angular distributions in the rare decays $B \to K \ell^{+} \ell^{-}$ and $B \to K^{*} \ell^{+} \ell^{-}$}",
    eprint = "hep-ex/0604007",
    archivePrefix = "arXiv",
    reportNumber = "SLAC-PUB-11799, BABAR-PUB-06-015",
    doi = "10.1103/PhysRevD.73.092001",
    journal = "Phys. Rev. D",
    volume = "73",
    pages = "092001",
    year = "2006"
}

@article{Belle:2009zue,
    author = "Wei, J. -T. and others",
    collaboration = "Belle",
    title = "{Measurement of the Differential Branching Fraction and Forward-Backward Asymmetry for $B \to K^{(*)}\ell^+\ell^-$}",
    eprint = "0904.0770",
    archivePrefix = "arXiv",
    primaryClass = "hep-ex",
    reportNumber = "BELLE-PREPRINT-2009-7, KEK-PREPRINT-2008-56",
    doi = "10.1103/PhysRevLett.103.171801",
    journal = "Phys. Rev. Lett.",
    volume = "103",
    pages = "171801",
    year = "2009"
}

@article{BaBar:2015wkg,
    author = "Lees, J. P. and others",
    collaboration = "BaBar",
    title = "{Measurement of angular asymmetries in the decays $B \to K^* \mu^+ \mu^-$}",
    eprint = "1508.07960",
    archivePrefix = "arXiv",
    primaryClass = "hep-ex",
    reportNumber = "BABAR-PUB-15-002, SLAC-PUB-16384",
    doi = "10.1103/PhysRevD.93.052015",
    journal = "Phys. Rev. D",
    volume = "93",
    pages = "052015",
    year = "2016"
}

@article{CDF:2011tds,
    author = "Aaltonen, T. and others",
    collaboration = "CDF",
    title = "{Measurements of the Angular Distributions in the Decays $B \to K^{(*)} \mu^+ \mu^-$ at CDF}",
    eprint = "1108.0695",
    archivePrefix = "arXiv",
    primaryClass = "hep-ex",
    reportNumber = "FERMILAB-PUB-11-364-PPD",
    doi = "10.1103/PhysRevLett.108.081807",
    journal = "Phys. Rev. Lett.",
    volume = "108",
    pages = "081807",
    year = "2012"
}

@article{CMS:2015bcy,
    author = "Khachatryan, Vardan and others",
    collaboration = "CMS",
    title = "{Angular analysis of the decay $B^0 \to K^{*0} \mu^+ \mu^-$ from pp collisions at $\sqrt  s = 8$ TeV}",
    eprint = "1507.08126",
    archivePrefix = "arXiv",
    primaryClass = "hep-ex",
    reportNumber = "CMS-BPH-13-010, CERN-PH-EP-2015-178",
    doi = "10.1016/j.physletb.2015.12.020",
    journal = "Phys. Lett. B",
    volume = "753",
    pages = "424--448",
    year = "2016"
}

@article{CMS:2017rzx,
    author = "Sirunyan, Albert M and others",
    collaboration = "CMS",
    title = "{Measurement of angular parameters from the decay $\mathrm{B}^0 \to \mathrm{K}^{*0} \mu^+ \mu^-$ in proton-proton collisions at $\sqrt{s} = $ 8 TeV}",
    eprint = "1710.02846",
    archivePrefix = "arXiv",
    primaryClass = "hep-ex",
    reportNumber = "CMS-BPH-15-008, CERN-EP-2017-240",
    doi = "10.1016/j.physletb.2018.04.030",
    journal = "Phys. Lett. B",
    volume = "781",
    pages = "517--541",
    year = "2018"
}

@article{Belle:2016fev,
    author = "Wehle, S. and others",
    collaboration = "Belle",
    title = "{Lepton-Flavor-Dependent Angular Analysis of $B\to K^\ast \ell^+\ell^-$}",
    eprint = "1612.05014",
    archivePrefix = "arXiv",
    primaryClass = "hep-ex",
    reportNumber = "BELLE-PREPRINT-2016-15, KEK-PREPRINT-2016-54",
    doi = "10.1103/PhysRevLett.118.111801",
    journal = "Phys. Rev. Lett.",
    volume = "118",
    pages = "111801",
    year = "2017"
}

@article{ATLAS:2018gqc,
    author = "Aaboud, Morad and others",
    collaboration = "ATLAS",
    title = "{Angular analysis of $B^0_d \rightarrow K^{*}\mu^+\mu^-$ decays in $pp$ collisions at $\sqrt{s}= 8$ TeV with the ATLAS detector}",
    eprint = "1805.04000",
    archivePrefix = "arXiv",
    primaryClass = "hep-ex",
    reportNumber = "CERN-EP-2017-161",
    doi = "10.1007/JHEP10(2018)047",
    journal = "JHEP",
    volume = "10",
    pages = "047",
    year = "2018"
}

@article{LHCb:2013zuf,
    author = "Aaij, R. and others",
    collaboration = "LHCb",
    title = "{Differential branching fraction and angular analysis of the decay $B^{0} \to K^{*0} \mu^{+}\mu^{-}$}",
    eprint = "1304.6325",
    archivePrefix = "arXiv",
    primaryClass = "hep-ex",
    reportNumber = "CERN-PH-EP-2013-074, LHCB-PAPER-2013-019",
    doi = "10.1007/JHEP08(2013)131",
    journal = "JHEP",
    volume = "08",
    pages = "131",
    year = "2013"
}

@article{LHCb:2013ghj,
    author = "Aaij, R and others",
    collaboration = "LHCb",
    title = "{Measurement of Form-Factor-Independent Observables in the Decay $B^{0} \to K^{*0} \mu^+ \mu^-$}",
    eprint = "1308.1707",
    archivePrefix = "arXiv",
    primaryClass = "hep-ex",
    reportNumber = "LHCB-PAPER-2013-037, CERN-PH-EP-2013-146",
    doi = "10.1103/PhysRevLett.111.191801",
    journal = "Phys. Rev. Lett.",
    volume = "111",
    pages = "191801",
    year = "2013"
}

@article{LHCb:2015svh,
    author = "Aaij, Roel and others",
    collaboration = "LHCb",
    title = "{Angular analysis of the $B^{0} \to K^{*0} \mu^{+} \mu^{-}$ decay using 3 fb$^{-1}$ of integrated luminosity}",
    eprint = "1512.04442",
    archivePrefix = "arXiv",
    primaryClass = "hep-ex",
    reportNumber = "CERN-PH-EP-2015-314, LHCB-PAPER-2015-051",
    doi = "10.1007/JHEP02(2016)104",
    journal = "JHEP",
    volume = "02",
    pages = "104",
    year = "2016"
}

@article{LHCb:2020lmf,
    author = "Aaij, Roel and others",
    collaboration = "LHCb",
    title = "{Measurement of $CP$-Averaged Observables in the $B^{0}\rightarrow K^{*0}\mu^{+}\mu^{-}$ Decay}",
    eprint = "2003.04831",
    archivePrefix = "arXiv",
    primaryClass = "hep-ex",
    reportNumber = "LHCb-PAPER-2020-002, CERN-EP-2020-027",
    doi = "10.1103/PhysRevLett.125.011802",
    journal = "Phys. Rev. Lett.",
    volume = "125",
    pages = "011802",
    year = "2020"
}

@article{LHCb:2020gog,
    author = "Aaij, Roel and others",
    collaboration = "LHCb",
    title = "{Angular Analysis of the  $B^{+}\rightarrow K^{\ast+}\mu^{+}\mu^{-}$ Decay}",
    eprint = "2012.13241",
    archivePrefix = "arXiv",
    primaryClass = "hep-ex",
    reportNumber = "LHCb-PAPER-2020-041, CERN-EP-2020-239",
    doi = "10.1103/PhysRevLett.126.161802",
    journal = "Phys. Rev. Lett.",
    volume = "126",
    pages = "161802",
    year = "2021"
}

@article{LHCb:2023gpo,
    author = "Aaij, Roel and others",
    collaboration = "LHCb",
    title = "{Amplitude Analysis of the $B^0 \to K^{*0} \mu^+ \mu^-$ Decay}",
    eprint = "2312.09115",
    archivePrefix = "arXiv",
    primaryClass = "hep-ex",
    reportNumber = "LHCb-PAPER-2023-033, CERN-EP-2023-273",
    doi = "10.1103/PhysRevLett.132.131801",
    journal = "Phys. Rev. Lett.",
    volume = "132",
    pages = "131801",
    year = "2024"
}

@article{LHCb:2024onj,
    author = "Aaij, Roel and others",
    collaboration = "LHCb",
    title = "{Comprehensive analysis of local and nonlocal amplitudes in the $B^{0} \to  K^{*0} \mu^+ \mu^-$ decay}",
    eprint = "2405.17347",
    archivePrefix = "arXiv",
    primaryClass = "hep-ex",
    reportNumber = "LHCb-PAPER-2024-011, CERN-EP-2024-122",
    doi = "10.1007/JHEP09(2024)026",
    journal = "JHEP",
    volume = "09",
    pages = "026",
    year = "2024"
}

@article{Ali:1991is,
    author = "Ali, Ahmed and Mannel, T. and Morozumi, T.",
    title = "{Forward backward asymmetry of dilepton angular distribution in the decay $b \to s \ell^+ \ell^-$}",
    reportNumber = "DESY-91-097, RU-91-8-B",
    doi = "10.1016/0370-2693(91)90306-B",
    journal = "Phys. Lett. B",
    volume = "273",
    pages = "505--512",
    year = "1991"
}

@article{Ali:1999mm,
    author = "Ali, Ahmed and Ball, Patricia and Handoko, L. T. and Hiller, G.",
    title = "{A Comparative study of the decays $B \to$ ($K$, $K^{*)} \ell^+ \ell^-$ in standard model and supersymmetric theories}",
    eprint = "hep-ph/9910221",
    archivePrefix = "arXiv",
    reportNumber = "SLAC-PUB-8269, DESY-99-146, CERN-TH-99-298, LNF-99-026-P",
    doi = "10.1103/PhysRevD.61.074024",
    journal = "Phys. Rev. D",
    volume = "61",
    pages = "074024",
    year = "2000"
}

@article{Ali:1994bf,
    author = "Ali, Ahmed and Giudice, G. F. and Mannel, T.",
    title = "{Towards a model independent analysis of rare $B$ decays}",
    eprint = "hep-ph/9408213",
    archivePrefix = "arXiv",
    reportNumber = "CERN-TH-7346-94",
    doi = "10.1007/BF01624585",
    journal = "Z. Phys. C",
    volume = "67",
    pages = "417--432",
    year = "1995"
}

@article{Liu:1994cfa,
    author = "Liu, Dong-sheng",
    title = "{Estimates for forward - backward asymmetry in $B \to K^*(892) \ell^+ \ell^-$}",
    eprint = "hep-ph/9501276",
    archivePrefix = "arXiv",
    reportNumber = "UTAS-PHYS-94-27",
    doi = "10.1016/0370-2693(95)00007-8",
    journal = "Phys. Lett. B",
    volume = "346",
    pages = "355--362",
    year = "1995"
}

@article{Colangelo:1995jv,
    author = "Colangelo, P. and De Fazio, F. and Santorelli, Pietro and Scrimieri, E.",
    title = "{QCD sum rule analysis of the decays $B \to K \ell^{+} \ell^{-}$ and $B \to K^{*} \ell^{+} \ell^{-}$}",
    eprint = "hep-ph/9510403",
    archivePrefix = "arXiv",
    reportNumber = "BARI-TH-95-206, DSF-T-95-42",
    doi = "10.1103/PhysRevD.53.3672",
    journal = "Phys. Rev. D",
    volume = "53",
    pages = "3672--3686",
    year = "1996",
    note = "[Erratum: Phys.Rev.D 57, 3186 (1998)]"
}

@article{Beneke:2000wa,
    author = "Beneke, M. and Feldmann, T.",
    title = "{Symmetry breaking corrections to heavy to light B meson form-factors at large recoil}",
    eprint = "hep-ph/0008255",
    archivePrefix = "arXiv",
    reportNumber = "PITHA-00-20",
    doi = "10.1016/S0550-3213(00)00585-X",
    journal = "Nucl. Phys. B",
    volume = "592",
    pages = "3--34",
    year = "2001"
}

@article{Charles:1998dr,
    author = "Charles, J. and Le Yaouanc, A. and Oliver, L. and Pene, O. and Raynal, J. C.",
    title = "{Heavy to light form-factors in the heavy mass to large energy limit of QCD}",
    eprint = "hep-ph/9812358",
    archivePrefix = "arXiv",
    reportNumber = "LPTHE-ORSAY-98-77",
    doi = "10.1103/PhysRevD.60.014001",
    journal = "Phys. Rev. D",
    volume = "60",
    pages = "014001",
    year = "1999"
}

@article{Becirevic:2011bp,
    author = "Becirevic, Damir and Schneider, Elia",
    title = "{On transverse asymmetries in $B \to K^* \ell^+ \ell^-$}",
    eprint = "1106.3283",
    archivePrefix = "arXiv",
    primaryClass = "hep-ph",
    reportNumber = "LPT-11-50",
    doi = "10.1016/j.nuclphysb.2011.09.004",
    journal = "Nucl. Phys. B",
    volume = "854",
    pages = "321--339",
    year = "2012"
}

@article{Kruger:2005ep,
    author = "Kruger, Frank and Matias, Joaquim",
    title = "{Probing new physics via the transverse amplitudes of $B^0\to K^{*0} (\to K^- \pi^+) l^+l^-$ at large recoil}",
    eprint = "hep-ph/0502060",
    archivePrefix = "arXiv",
    reportNumber = "UAB-FT-560",
    doi = "10.1103/PhysRevD.71.094009",
    journal = "Phys. Rev. D",
    volume = "71",
    pages = "094009",
    year = "2005"
}

@article{HeavyFlavorAveragingGroupHFLAV:2024ctg,
    author = "Banerjee, Swagato and others",
    collaboration = "Heavy Flavor Averaging Group (HFLAV)",
    title = "{Averages of $b$-hadron, $c$-hadron, and $\tau$-lepton properties as of 2023}",
    eprint = "2411.18639",
    archivePrefix = "arXiv",
    primaryClass = "hep-ex",
    month = "11",
    year = "2024"
}

@article{Kitahara:2024azt,
    author = "Kitahara, Teppei",
    title = "{Theoretical point of view on Cabibbo angle anomaly}",
    eprint = "2407.00122",
    archivePrefix = "arXiv",
    primaryClass = "hep-ph",
    reportNumber = "CHIBA-EP-265",
    doi = "10.1142/S0217751X24420119",
    journal = "Int. J. Mod. Phys. A",
    volume = "39",
    pages = "2442011",
    year = "2024"
}

@article{LHCb:2014iah,
    author = "Aaij, Roel and others",
    collaboration = "LHCb",
    title = "{Precision measurement of $CP$ violation in $B_s^0 \to J/\psi K^+K^-$ decays}",
    eprint = "1411.3104",
    archivePrefix = "arXiv",
    primaryClass = "hep-ex",
    reportNumber = "LHCB-PAPER-2014-059, CERN-PH-EP-2014-271",
    doi = "10.1103/PhysRevLett.114.041801",
    journal = "Phys. Rev. Lett.",
    volume = "114",
    pages = "041801",
    year = "2015"
}

@article{Aoyama:2020ynm,
    author = "Aoyama, T. and others",
    title = "{The anomalous magnetic moment of the muon in the Standard Model}",
    eprint = "2006.04822",
    archivePrefix = "arXiv",
    primaryClass = "hep-ph",
    reportNumber = "FERMILAB-PUB-20-207-T, INT-PUB-20-021, KEK Preprint 2020-5,
  MITP/20-028, KEK Preprint 2020-5, MITP/20-028, CERN-TH-2020-075, IFT-UAM/CSIC-20-74, LMU-ASC 18/20, LTH 1234,
  LU TP 20-20, LTH 1234, LU TP 20-20, MAN/HEP/2020/003, PSI-PR-20-06, UWThPh 2020-14, ZU-TH 18/20",
    doi = "10.1016/j.physrep.2020.07.006",
    journal = "Phys. Rept.",
    volume = "887",
    pages = "1--166",
    year = "2020"
}

@article{Aliberti:2025beg,
    author = "Aliberti, R. and others",
    title = "{The anomalous magnetic moment of the muon in the Standard Model: an update}",
    eprint = "2505.21476",
    archivePrefix = "arXiv",
    primaryClass = "hep-ph",
    reportNumber = "CERN-TH-2025-101, FERMILAB-PUB-25-0344-T, INT-PUB-25-015,
  IPARCOS-UCM-25-029, KEK Preprint 2025-22, LTH 1403, MITP-25-037, UWThPh
  2025-15, ZU-TH 37/25",
    month = "5",
    year = "2025"
}

@article{Borsanyi:2020mff,
    author = "Borsanyi, Sz. and others",
    title = "{Leading hadronic contribution to the muon magnetic moment from lattice QCD}",
    eprint = "2002.12347",
    archivePrefix = "arXiv",
    primaryClass = "hep-lat",
    doi = "10.1038/s41586-021-03418-1",
    journal = "Nature",
    volume = "593",
    number = "7857",
    pages = "51--55",
    year = "2021"
}

@article{CMD-3:2023alj,
    author = "Ignatov, F. V. and others",
    collaboration = "CMD-3",
    title = "{Measurement of the $e^+ e^- \to \pi^+ \pi^-$ cross section from threshold to 1.2~GeV with the CMD-3 detector}",
    eprint = "2302.08834",
    archivePrefix = "arXiv",
    primaryClass = "hep-ex",
    doi = "10.1103/PhysRevD.109.112002",
    journal = "Phys. Rev. D",
    volume = "109",
    pages = "112002",
    year = "2024"
}

@article{Cornella:2021sby,
    author = "Cornella, Claudia and Faroughy, Darius A. and Fuentes-Martin, Javier and Isidori, Gino and Neubert, Matthias",
    title = "{Reading the footprints of the B-meson flavor anomalies}",
    eprint = "2103.16558",
    archivePrefix = "arXiv",
    primaryClass = "hep-ph",
    doi = "10.1007/JHEP08(2021)050",
    journal = "JHEP",
    volume = "08",
    pages = "050",
    year = "2021"
}

@article{Bordone:2021usz,
    author = "Bordone, Marzia and Rahimi, Muslem and Vos, K. Keri",
    title = "{Lepton flavour violation in rare $\Lambda _b$ decays}",
    eprint = "2106.05192",
    archivePrefix = "arXiv",
    primaryClass = "hep-ph",
    reportNumber = "P3H-21-039, SI-HEP-2021-17, Nikhef-2021-012",
    doi = "10.1140/epjc/s10052-021-09531-9",
    journal = "Eur. Phys. J. C",
    volume = "81",
    pages = "756",
    year = "2021"
}

@article{Panda:2024ygr,
    author = "Panda, Dhiren and Mohapatra, Manas Kumar and Mohanta, Rukmani",
    title = "{Exploring the lepton flavor violating decay modes  $b \to s \mu^\pm \tau^\mp$ in SMEFT approach}",
    eprint = "2403.09393",
    archivePrefix = "arXiv",
    primaryClass = "hep-ph",
    doi = "10.1016/j.nuclphysb.2024.116720",
    journal = "Nucl. Phys. B",
    volume = "1008",
    pages = "116720",
    year = "2024"
}

@article{Becirevic:2024vwy,
    author = "Be\v{c}irevi\'c, Damir and Jaffredo, Florentin and Pinheiro, J. Paulo and Sumensari, Olcyr",
    title = "{Lepton flavor violation in exclusive $b \to d \ell_i \ell_j$ and $b \to s \ell_i \ell_j$ decay modes}",
    eprint = "2407.19060",
    archivePrefix = "arXiv",
    primaryClass = "hep-ph",
    doi = "10.1103/PhysRevD.110.075004",
    journal = "Phys. Rev. D",
    volume = "110",
    pages = "075004",
    year = "2024"
}

@article{Buras:2021nns,
    author = "Buras, Andrzej J. and Venturini, Elena",
    title = "{Searching for New Physics in Rare $K$ and $B$ Decays without $|V_{cb}|$ and $|V_{ub}|$ Uncertainties}",
    eprint = "2109.11032",
    archivePrefix = "arXiv",
    primaryClass = "hep-ph",
    reportNumber = "AJB-21-7, TUM-HEP-1364/21",
    doi = "10.5506/APhysPolB.53.6-A1",
    journal = "Acta Phys. Polon. B",
    volume = "53",
    number = "6",
    pages = "6-A1",
    month = "9",
    year = "2021"
}

@article{Buras:2022wpw,
    author = "Buras, Andrzej J. and Venturini, Elena",
    title = "{The exclusive vision of rare K and B decays and of the quark mixing in the standard model}",
    eprint = "2203.11960",
    archivePrefix = "arXiv",
    primaryClass = "hep-ph",
    reportNumber = "AJB-22-6, TUM-HEP-1393/22",
    doi = "10.1140/epjc/s10052-022-10583-8",
    journal = "Eur. Phys. J. C",
    volume = "82",
    pages = "615",
    year = "2022"
}

@article{Colangelo:2016ymy,
    author = "Colangelo, Pietro and De Fazio, Fulvia",
    title = "{Tension in the inclusive versus exclusive determinations of $|V_{cb}|$: a possible role of new physics}",
    eprint = "1611.07387",
    archivePrefix = "arXiv",
    primaryClass = "hep-ph",
    reportNumber = "BARI-TH-709-2016",
    doi = "10.1103/PhysRevD.95.011701",
    journal = "Phys. Rev. D",
    volume = "95",
    pages = "011701",
    year = "2017"
}

@article{Colangelo:2024ped,
    author = "Colangelo, Pietro and De Fazio, Fulvia and Loparco, Francesco and Losacco, Nicola",
    title = "{Flavour anomalies, correlations, hadronic uncertainties, and all that}",
    eprint = "2401.02796",
    archivePrefix = "arXiv",
    primaryClass = "hep-ph",
    reportNumber = "BARI-TH/751-23",
    doi = "10.1393/ncc/i2024-24145-5",
    journal = "Nuovo Cim. C",
    volume = "47",
    pages = "145",
    year = "2024"
}

@article{Matias:2012xw,
    author = "Matias, Joaquim and Mescia, Federico and Ramon, Marc and Virto, Javier",
    title = "{Complete Anatomy of $\bar{B}_d \to \bar{K}^{* 0} (\to K \pi) \ell^+ \ell^-$ and its angular distribution}",
    eprint = "1202.4266",
    archivePrefix = "arXiv",
    primaryClass = "hep-ph",
    reportNumber = "UAB-FT-706, ICCUB-12-076, ECM-UB-68",
    doi = "10.1007/JHEP04(2012)104",
    journal = "JHEP",
    volume = "04",
    pages = "104",
    year = "2012"
}

@article{DeBruyn:2012wj,
    author = "De Bruyn, Kristof and Fleischer, Robert and Knegjens, Robert and Koppenburg, Patrick and Merk, Marcel and Tuning, Niels",
    title = "{Branching Ratio Measurements of $B_s$ Decays}",
    eprint = "1204.1735",
    archivePrefix = "arXiv",
    primaryClass = "hep-ph",
    reportNumber = "NIKHEF-2012-005",
    doi = "10.1103/PhysRevD.86.014027",
    journal = "Phys. Rev. D",
    volume = "86",
    pages = "014027",
    year = "2012"
}

@article{Aebischer:2019blw,
	archiveprefix = {arXiv},
	author = {Aebischer, Jason and Buras, Andrzej J. and Cerd\`a-Sevilla, Maria and De Fazio, Fulvia},
	doi = {10.1007/JHEP02(2020)183},
	eprint = {1912.09308},
	journal = {JHEP},
	pages = {183},
	primaryclass = {hep-ph},
	title = {{Quark-lepton connections in $Z^\prime$ mediated FCNC processes: gauge anomaly cancellations at work}},
	volume = {02},
	year = {2020}}

@article{Bharucha:2015bzk,
    author = "Bharucha, Aoife and Straub, David M. and Zwicky, Roman",
    title = "{$B \to V\ell^+\ell^-$ in the Standard Model from light-cone sum rules}",
    eprint = "1503.05534",
    archivePrefix = "arXiv",
    primaryClass = "hep-ph",
    reportNumber = "TUM-HEP-957-14, CP3-Origins-2015-010, DIAS-2015-10",
    doi = "10.1007/JHEP08(2016)098",
    journal = "JHEP",
    volume = "08",
    pages = "098",
    year = "2016"
}

@article{Gubernari:2023puw,
    author = "Gubernari, Nico and Reboud, M\'eril and van Dyk, Danny and Virto, Javier",
    title = "{Dispersive analysis of $B \to K^{(*)}$ and $B_{s} \to \phi$ form factors}",
    eprint = "2305.06301",
    archivePrefix = "arXiv",
    primaryClass = "hep-ph",
    reportNumber = "EOS-2023-02, IPPP/23/22, P3H-23-026, SI-HEP-2023-09",
    doi = "10.1007/JHEP12(2023)153",
    journal = "JHEP",
    volume = "12",
    pages = "153",
    year = "2023",
    note = "[Erratum: JHEP 01, 125 (2025)]"
}

@article{Capozzi:2025wyn,
    author = "Capozzi, Francesco and Giar\`e, William and Lisi, Eligio and Marrone, Antonio and Melchiorri, Alessandro and Palazzo, Antonio",
    title = "{Neutrino masses and mixing: Entering the era of subpercent precision}",
    eprint = "2503.07752",
    archivePrefix = "arXiv",
    primaryClass = "hep-ph",
    month = "3",
    year = "2025"
}

@article{Esteban:2024eli,
    author = "Esteban, Ivan and Gonzalez-Garcia, M. C. and Maltoni, Michele and Martinez-Soler, Ivan and Pinheiro, Jo\~ao Paulo and Schwetz, Thomas",
    title = "{NuFit-6.0: updated global analysis of three-flavor neutrino oscillations}",
    eprint = "2410.05380",
    archivePrefix = "arXiv",
    primaryClass = "hep-ph",
    reportNumber = "IFT-UAM/CSIC-24-140, YITP-SB-2024-24, IPPP/24/64, IPPP/24/64, IFT-UAM/CSIC-24-140, YITP-SB-2024-24",
    doi = "10.1007/JHEP12(2024)216",
    journal = "JHEP",
    volume = "12",
    pages = "216",
    year = "2024"
}

@article{Leike:1998wr,
	archiveprefix = {arXiv},
	author = {Leike, A.},
	doi = {10.1016/S0370-1573(98)00133-1},
	eprint = {hep-ph/9805494},
	journal = {Phys. Rept.},
	pages = {143--250},
	reportnumber = {LMU-03-98},
	title = {{The Phenomenology of extra neutral gauge bosons}},
	volume = {317},
	year = {1999}}

@article{Langacker:2000ju,
	archiveprefix = {arXiv},
	author = {Langacker, Paul and Plumacher, Michael},
	doi = {10.1103/PhysRevD.62.013006},
	eprint = {hep-ph/0001204},
	journal = {Phys. Rev. D},
	pages = {013006},
	reportnumber = {UPR-0870-T},
	title = {{Flavor changing effects in theories with a heavy $Z^\prime$ boson with family nonuniversal couplings}},
	volume = {62},
	year = {2000}}

@inproceedings{Rizzo:2006nw,
	archiveprefix = {arXiv},
	author = {Rizzo, Thomas G.},
	booktitle = {{Theoretical Advanced Study Institute in Elementary Particle Physics}: {Exploring New Frontiers Using Colliders and Neutrinos}},
	eprint = {hep-ph/0610104},
	month = {10},
	pages = {537--575},
	reportnumber = {SLAC-PUB-12129},
	title = {{$Z^\prime$ phenomenology and the LHC}},
	year = {2006}}

@article{Appelquist:2002mw,
	archiveprefix = {arXiv},
	author = {Appelquist, Thomas and Dobrescu, Bogdan A. and Hopper, Adam R.},
	doi = {10.1103/PhysRevD.68.035012},
	eprint = {hep-ph/0212073},
	journal = {Phys. Rev. D},
	pages = {035012},
	reportnumber = {YCTP-11-02, FERMILAB-PUB-02-307-T},
	title = {{Nonexotic Neutral Gauge Bosons}},
	volume = {68},
	year = {2003}}

@article{Allanach:2018vjg,
	archiveprefix = {arXiv},
	author = {Allanach, B. C. and Davighi, Joe and Melville, Scott},
	doi = {10.1007/JHEP02(2019)082},
	eprint = {1812.04602},
	journal = {JHEP},
	note = {[Erratum: JHEP 08, 064 (2019)]},
	pages = {082},
	primaryclass = {hep-ph},
	reportnumber = {DAMTP-2018-41},
	title = {{An Anomaly-free Atlas: charting the space of flavour-dependent gauged $U(1)$ extensions of the Standard Model}},
	volume = {02},
	year = {2019}}

@article{Carena:2004xs,
	archiveprefix = {arXiv},
	author = {Carena, Marcela and Daleo, Alejandro and Dobrescu, Bogdan A. and Tait, Timothy M. P.},
	doi = {10.1103/PhysRevD.70.093009},
	eprint = {hep-ph/0408098},
	journal = {Phys. Rev.},
	pages = {093009},
	primaryclass = {hep-ph},
	reportnumber = {FERMILAB-PUB-04-129-T},
	slaccitation = {%%CITATION = HEP-PH/0408098;%%},
	title = {{$Z^\prime$ gauge bosons at the Tevatron}},
	volume = {D70},
	year = {2004}}

@article{Muong-2:2021ojo,
	archiveprefix = {arXiv},
	author = {Abi, B. and others},
	collaboration = {Muon g-2},
	doi = {10.1103/PhysRevLett.126.141801},
	eprint = {2104.03281},
	journal = {Phys. Rev. Lett.},
	pages = {141801},
	primaryclass = {hep-ex},
	reportnumber = {FERMILAB-PUB-21-132-E},
	title = {{Measurement of the Positive Muon Anomalous Magnetic Moment to 0.46 ppm}},
	volume = {126},
	year = {2021}}

@article{Crivellin:2015mga,
	archiveprefix = {arXiv},
	author = {Crivellin, Andreas and D'Ambrosio, Giancarlo and Heeck, Julian},
	doi = {10.1103/PhysRevLett.114.151801},
	eprint = {1501.00993},
	journal = {Phys. Rev. Lett.},
	pages = {151801},
	primaryclass = {hep-ph},
	reportnumber = {CERN-PH-TH-2015-001, ULB-TH-14-26},
	title = {{Explaining $h \to \mu^\pm\tau^\mp$, $B\to K^* \mu^+\mu^-$ and $B\to K \mu^+\mu^-/B\to K e^+e^-$ in a two-Higgs-doublet model with gauged $L_\mu-L_\tau$}},
	volume = {114},
	year = {2015}}

@article{Navas:PDG,
	author = {Navas, S. and Others},
	collaboration = {Particle Data Group},
	journal = {Phys. Rev. D},
	pages = {030001},
	title = {{Review of Particle Physics}},
	volume = {110},
	year = {2024}}

@article{Buras:2012jb,
	archiveprefix = {arXiv},
	author = {Buras, Andrzej J. and De Fazio, Fulvia and Girrbach, Jennifer},
	doi = {10.1007/JHEP02(2013)116},
	eprint = {1211.1896},
	journal = {JHEP},
	pages = {116},
	primaryclass = {hep-ph},
	reportnumber = {FLAVOUR(267104)-ERC-26, BARI-TH-12-665},
	title = {{The Anatomy of $Z^\prime$ and $Z$ with Flavour Changing Neutral Currents in the Flavour Precision Era}},
	volume = {02},
	year = {2013}}

@article{Buras:1993xp,
    author = "Buras, A. J. and Misiak, M. and Munz, M. and Pokorski, S.",
    title = "{Theoretical uncertainties and phenomenological aspects of $B \to X_s \gamma$ decay}",
    eprint = "hep-ph/9311345",
    archivePrefix = "arXiv",
    reportNumber = "MPI-PH-93-77, TUM-T31-50-93",
    doi = "10.1016/0550-3213(94)90299-2",
    journal = "Nucl. Phys. B",
    volume = "424",
    pages = "374--398",
    year = "1994"
}

@book{Buras:2020xsm,
    author = "Buras, Andrzej J.",
    title = "{Gauge Theory of Weak Decays}",
    doi = "10.1017/9781139524100",
    isbn = "978-1-139-52410-0, 978-1-107-03403-7",
    publisher = "Cambridge University Press",
    month = "6",
    year = "2020"
}

@article{Misiak:2018cec,
    author = "Misiak, M.",
    editor = "Je\.zabek, Marek",
    title = "{Radiative Decays of the $B$ Meson: a Progress Report}",
    doi = "10.5506/APhysPolB.49.1291",
    journal = "Acta Phys. Polon. B",
    volume = "49",
    pages = "1291--1300",
    year = "2018"
}
\end{document}